\documentclass[11pt]{article}
\usepackage[utf8]{inputenc}
\pdfoutput=1
\usepackage{amssymb,amsmath,mathrsfs,enumerate}
\usepackage{graphicx,rotate,multicol}
\usepackage{float}
\usepackage{tocloft}
\usepackage{subfig}
\usepackage[margin=10pt,labelfont=bf]{caption}
\usepackage{cite}
\usepackage{soul}
\usepackage[colorlinks=true,
            linkcolor=blue,
            urlcolor=blue,
            citecolor=teal]{hyperref}

\makeatletter
\let\Hy@linktoc\Hy@linktoc@page
\makeatother

\usepackage{color}
\definecolor{ourcolor}{rgb}{0.7, 0.25, 0.05}
\usepackage{tikz,braket}
\long\def\rpl#1!!#2!!{\textcolor{red}{#1} \textcolor{blue}{#2}}

\let\tilde=\widetilde
\let\hat=\widehat
\let\bar=\overline

\def \order(#1){{\mathcal O} \left(#1 \right)}

\evensidemargin=\oddsidemargin
\allowdisplaybreaks

\title{\color{black}{Rare Top Decays in Minimal and Non-minimal Universal Extra Dimension}}

\author {\sf Ujjal Kumar Dey,$^{a,}$\footnote{ujjaldey@prl.res.in} 
\hspace{4pt}  Tapoja Jha,$^{b,}$\footnote{tapoja.phy@gmail.com} \\[10pt]
\small\em $^a$ Theoretical Physics Division, 
		Physical Research Laboratory,
		Navrangpura, Ahmedabad 380009, India\\
\small\em $^b$Department of Physics, University of Calcutta, 92 Acharya Prafulla Chandra Road, Kolkata 700009, India\\
}

\date{}

\begin{document}

\maketitle

\begin{abstract}The flavour changing decays of the top quark are severely suppressed in the Standard Model by virtue of the Glashow-Iliopoulos-Maiani mechanism. Many beyond Standard Model extensions predict the decay rates at a level that is observable in the LHC. We perform a complete one-loop calculation of the flavour changing top quark decays $t\to c\gamma$ and $t\to ch$ in the universal extra dimensional model. Apart from considering the decay rates in the minimal version of the model, we also calculate the rates in the non-minimal scenario where the presence of boundary localised terms interestingly modify the set-up. We find that the decay rates in the minimal variant of the model do not change much from their Standard Model values. In the non-minimal version of this model, these decay rates can be higher for specific choices of the boundary localised parameters for a certain range of inverse compactification radius. But these model parameters lead to Kaluza-Klein particle masses that are in tension with various searches at the LHC.
\end{abstract}

~~~~Keywords: Beyond Standard Model; Rare decay; FCNC Interaction

\newpage

\hrule \hrule
\tableofcontents
\vskip 10pt
\hrule \hrule 

\section{Introduction}
The discovery of the Higgs boson at the LHC~\cite{Aad:2012tfa, Chatrchyan:2012xdj} marks the completion of the Standard Model (SM) of particle physics. Moreover, the SM is immensely successful in explaining almost all the observations from high energy colliders like Tevatron and the LHC. Nevertheless, there are other observations \textit{e.g.,} the existence of dark matter (DM), the confirmation of neutrino mass, etc., which demand some physics beyond the SM (BSM). One interesting possibility is to consider the existence of space-like dimensions beyond the usual three. These types of extensions of the SM are termed as extra-dimensional theories.\footnote{Extra dimensions are ubiquitous in string theories. However, the possibility of a TeV scale extra dimension was first put forward in~\cite{Antoniadis:1990ew, Antoniadis:1998ig}.} Now, many variants of these models are possible depending on the following: the number of extra spatial dimensions, the intrinsic geometry of the full space-time continuum, and obviously the possible ultraviolet completions of those theories. In this paper we take the case of the simplest of these, namely the universal extra dimension (UED).
Among many variants of extra dimensional models, UED, proposed by Appelquist \textit{et al.}~\cite{Appelquist:2000nn}, among other things, addresses the DM problem in an elegant way. In UED one considers one or more\footnote{In this work we stick to the one extra spatial dimension scenario.} extra spatial dimensions and unlike other extra dimensional models (\textit{e.g.,} ADD~\cite{ArkaniHamed:1998rs} or RS~\cite{Randall:1999ee, Randall:1999vf}) in UED the extra flat spatial dimension is open to all the SM fields. The extra dimension ($y$) is compactified on a circle ($S^{1}$) of radius $R$. The compactification gives rise to an infinite tower of Kaluza-Klein (KK) modes with increasing masses for any field, called the KK tower. However, to ensure the presence of chiral fermions in the SM one further needs to orbifold (by imposing one extra symmetry $y\leftrightarrow -y$) the extra dimension and impose appropriate boundary conditions on the fermionic fields. The resulting manifold of the extra dimension is called the $S^{1}/\mathbb{Z}_{2}$ orbifold. As it stands, the orbifolding breaks the translational invariance along the extra dimension and consequently the momentum in the fifth direction $p_{5}$, viz. the KK number is violated. Even then there remains an accidental discrete symmetry ($y\to y+\pi R$), called KK parity which for the $n$-th KK mode particle is $(-1)^{n}$. An SM field is identified as the zeroth mode of the infinite KK tower. Clearly, the conservation of KK parity ensures the stability of the lightest KK particle (LKP) which can be a good DM candidate~\cite{Servant:2002aq, Servant:2002hb, Burnell:2005hm, Kong:2005hn}. However such a set-up gives rise to an almost degenerate mass spectrum. But radiative corrections lift this degeneracy~\cite{Cheng:2002iz}. Since the five-dimensional (5D) theory is non-renormalisable, UED can be considered to be an effective field theory valid under a cut-off\footnote{Recent vacuum stability studies on Higgs boson mass and couplings in the context of minimal UED suggest $\Lambda R \sim 6$~\cite{Blennow:2011tb, Datta:2012db}.} scale $\Lambda$. Evidently the radiative corrections that lift the degeneracy in the mass spectrum  depend on $\Lambda$. This one-loop corrected UED is known in the literature as the minimal UED (mUED). Since UED is an effective theory in four dimensions (4D), one should take into account all operators that are allowed by the SM gauge symmetry and Lorentz symmetry. This type of operator can be in the bulk or localized in the fixed points or the boundary of the orbifold. Actually in mUED all the boundary localized terms (BLT) are assumed to be vanishing at the cut-off scale $\Lambda$, but these terms are generated radiatively at the low scales~\cite{Cheng:2002iz}. The set-up with the non-vanishing BLTs is termed as non-minimal UED (nmUED)~\cite{Flacke:2008ne}. In the nmUED the mass spectrum as well as the couplings are dependent on the BLT parameters. Therefore nmUED has a rich phenomenology compared to the mUED scenario thanks to the presence of the BLT parameters. Some of the recent studies on the phenomenology of the BLTs can be found in~\cite{delAguila:2003bh,delAguila:2003gu, delAguila:2003kd, delAguila:2003gv, delAguila:2006atw, Datta:2012tv, Datta:2013nua, Datta:2013yaa, Dey:2013cqa, Flacke:2014jwa, Jha:2014faa, Datta:2015aka}.
Now, an important aspect of SM is the absence of flavour changing neutral current (FCNC) interactions at the tree-level. Moreover, in the loop-level FCNC is possible, but that too is strongly suppressed by the Glashow-Iliopoulos-Maiani (GIM) mechanism. Normally these types of loop-driven processes involve two different generations of fermions in the initial and final states and all possible generation of fermions running in the loop. These FCNC processes are strongly suppressed in the SM; so the discovery of any such process would be a clear hint of some BSM physics. Clearly in the case of the BSM scenario for these types of processes no BSM particle has to be produced on-shell but their effects in the loop would be enough to look into the picture at hand. This is particularly important in a time when there is a lack of any direct evidence of new physics at the LHC.\footnote{Recent 750 GeV scalar resonance observed, although not with extreme significance limit, by ATLAS and CMS may be the first hint to physics beyond SM~\cite{ATLAS-CONF-2015-081, CMS-PAS-EXO-15-004}.} In this same vein many BSM scenarios are studied through these type of FCNC processes. One nice place to look for such FCNC processes is the rare decays of top quark in the context of some new physics model. There have been many studies to consider the rare decays of the top quark in the SM~\cite{Deshpande:1981zq, Deshpande:1982mi, DiazCruz:1989ub, Mele:1998ag, Mele:1999zk, AguilarSaavedra:2002ns, AguilarSaavedra:2004wm, Chen:2013qta, Khanpour:2014xla, Hesari:2014eua, Kim:2015oua, Hesari:2015oya, Khatibi:2015aal} as well as in various BSM scenarios, \textit{e.g.,} in supersymmetry~\cite{Guasch:1999jp, Eilam:2001dh, Frank:2005vd, Cao:2007dk, Cao:2008vk, Cao:2014udj, Dedes:2014asa, Bardhan:2016txk}, two Higgs doublet model (2HDM)~\cite{Eilam:1990zc, Iltan:2001yt, Arhrib:2005nx, Gaitan:2015hga, Abbas:2015cua}, warped extra dimension~\cite{Gao:2013fxa}, UED~\cite{GonzalezSprinberg:2007zz}, etc. A model independent effective field theory based study of FCNC top decays can be found in~\cite{CorderoCid:2004vi, Datta:2009zb}. In this paper we  consider the rare decays $t\to c\gamma$ and $t\to ch$ in the context of minimal as well as non-minimal UED.

This paper is organised as follows. In  Sec.~\ref{s:model} we briefly describe the mUED and nmUED model. The general features of the rare top quark decays to $c\gamma$ and $ch$ are discussed in Sec.~\ref{s:raretopdecay}. In Sec.~\ref{s:results} we elaborate on the results that we obtain for SM as well as in the cases of mUED and nmUED. Sec.~\ref{s:stufcnc} is dedicated to the electroweak precision and FCNC related issues. Finally, we summarise and conclude in Sec.~\ref{s:concl}. In the Appendix we list the relevant Feynman rules.

\section{Model Description} 
\label{s:model}
In this section we briefly review the model to set the notations and conventions. For completeness first we describe the general set-up of nmUED with BLTs and as we proceed we point out how to revert to mUED. For a more detailed discussion of the model see~\cite{Flacke:2008ne, delAguila:2003bh,delAguila:2003gu, delAguila:2003gv, delAguila:2006atw, Datta:2012tv, Datta:2013nua, Datta:2013yaa, Dey:2013cqa, Jha:2014faa, Datta:2015aka, delAguila:2003kd, Flacke:2014jwa}. 
%

%
\subsection{Lagrangians and Interactions}
\label{sbsc:lags}
\begin{figure}[h]
\begin{center}
\includegraphics[scale=0.45]{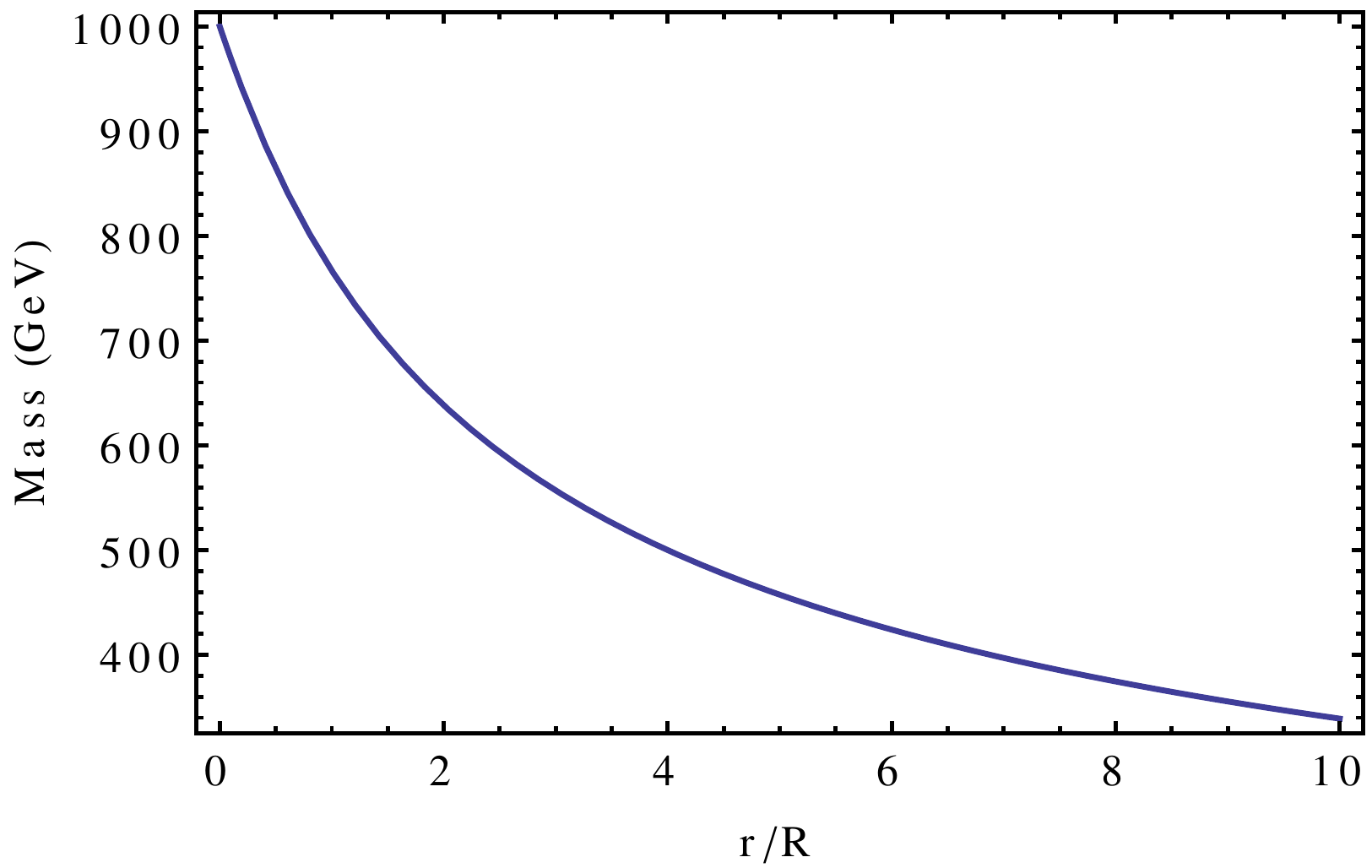}
\caption{Dependence of first KK level mass on the BLKT parameter for $1/R = 1$ TeV.}
\label{f:transcmass}
\end{center}
\end{figure}

To begin with, consider the 5D action for the quark\footnote{Leptonic fields will follow a similar procedure.} fields in the presence of boundary localised kinetic terms (BLKT),
\begin{align}
\label{eq:squark}
\mathcal{S}_{quark} = &\int d^4 x \int_{0}^{\pi R} dy \Big[\overline{Q} i\Gamma^{M} \mathcal{D}_{M} Q + r_f \{ \delta(y) + \delta(y-\pi R) \} \overline{Q} i\gamma^{\mu} \mathcal{D}_{\mu} P_L Q \nonumber \\ 
 &  + \overline{U} i\Gamma^{M} \mathcal{D}_{M} U + r_f \{ \delta(y) + \delta(y-\pi R) \} \overline{U} i\gamma^{\mu} \mathcal{D}_{\mu} P_R U \nonumber \\ 
 &  + \overline{D} i\Gamma^{M} \mathcal{D}_{M} D + r_f \{ \delta(y) + \delta(y-\pi R) \} \overline{D} i\gamma^{\mu} \mathcal{D}_{\mu} P_R D\Big]. 
\end{align}
The five-dimensional four-component quark fields ($Q,U$ and $D$) are comprised of two-component chiral spinors and their KK excitations and  they can be written as,
\begin{subequations}
\begin{gather}
Q(x,y) =   \sum^{\infty}_{n=0} \left( \begin{array}{c} Q_{L}^{n}(x) f_L^n(y) \\ Q_{R}^{n}(x) g_L^n(y)\end{array} \right) ,\\
U(x,y) = \sum^{\infty}_{n=0}  \left( \begin{array}{c} U_{L}^{n}(x) f_R^n(y) \\ U_{R}^{n}(x) g_R^n(y) \end{array} \right), \quad
D(x,y) = \sum^{\infty}_{n=0}  \left( \begin{array}{c} D_{L}^{n}(x) f_R^n(y) \\ D_{R}^{n}(x) g_R^n(y) \end{array} \right). 
\,\,\,\,  
\end{gather} 
\end{subequations}
In the effective 4D theory the zero modes of $Q$ give rise to the $SU(2)_{L}$ doublet quarks whereas the zero modes of $U$ ($D$) are identified with the up- (down-) type singlet quarks, \textit{i.e.,} after compactification and orbifolding the zero modes of $Q$ are be the left-handed doublet comprising SM $t_{L}$ and $b_{L}$, whereas $t_{R}$ and $b_{R}$ emerge from the $U$ and $D$, respectively. The latin indices in Eq.~\ref{eq:squark} run from 0 to 4 and the greek indices from 0 to 3. We use the \textit{mostly minus} metric convention, \textit{i.e.,} $g_{MN}\equiv \rm{diag}(+1,-1,-1,-1,-1)$. The covariant derivative $D_M\equiv\partial_M-i\widetilde{g}W_M^a T^a-i \widetilde{g}^\prime B_M Y$, where $\widetilde{g}$ and $\widetilde{g}^\prime$ are the five-dimensional gauge coupling constants of $SU(2)_L$ and $U(1)_Y$, respectively, and $T^a$ and $Y$ are the corresponding generators. The 5D gamma matrices are $\Gamma^{M}=(\gamma^{\mu},-i\gamma_{5})$. In Eq.~\ref{eq:squark} the terms containing the parameter $r_{f}$ are the BLKTs. Clearly in the mUED, $r_{f}$ is assumed to be vanishing. It is worth mentioning that by setting BLKT parameters to zero one can translate from nmUED to mUED. 
Now, from the variation of action and considering appropriate boundary conditions, one can obtain the $y$-dependent mode functions $f$ and $g$,
\begin{eqnarray}
\label{eq:modefunctions}
f_{L}(y) = g_{R}(y) = N_{Qn} \left\{ \begin{array}{rl}
                \displaystyle \frac{\cos[M_{Q_{n}} \left (y - \frac{\pi R}{2}\right)]}{C_{Q_{n}}}  &\mbox{for $n$ even,}\\
                \displaystyle \frac{{-}\sin[M_{Q_{n}} \left (y - \frac{\pi R}{2}\right)]}{S_{Q_{n}}} &\mbox{for $n$ odd,}
                \end{array} \right. 
\end{eqnarray}
and
\begin{eqnarray}
\label{eq:modefunctionsodd}
g_{L}(y) = f_{R}(y) = N_{Qn} \left\{ \begin{array}{rl}
                \displaystyle \frac{\sin[M_{Q_{n}} \left (y - \frac{\pi R}{2}\right)]}{C_{Q_{n}}}  &\mbox{for $n$ even,}\\
                \displaystyle \frac{\cos[M_{Q_{n}} \left (y - \frac{\pi R}{2}\right)]}{S_{Q_{n}}} &\mbox{for $n$ odd,}
                \end{array} \right. 
\end{eqnarray}
with
\begin{equation}
C_{Q_{n}} = \cos\left( \frac{M_{Q_{n}} \pi R}{2} \right)\, , 
\,\,\,\,
S_{Q_{n}} = \sin\left( \frac{M_{Q_{n}} \pi R}{2} \right).
\end{equation}
The orthonormality conditions satisfied by $f$s and $g$s are given as
\begin{equation}\label{eq:orthonorm}
\int dy \left[1 + r_{f}\{ \delta(y) + \delta(y - \pi R)\}
\right] ~k^{(m)}(y) ~k^{(n)}(y) = \delta^{mn} = \int dy ~l^{(m)}(y) ~l^{(n)}(y)
\end{equation}
where $k$ can be $f_{L}$ or $g_{R}$ and $l$ corresponds to $g_{L}$ or $f_{R}$. From the above condition one can obtain the normalization factors as
\begin{equation} \label{eq:norm}
N_{Qn} = \sqrt{\frac{2}{\pi R}}\left[ \frac{1}{\sqrt{1 + \frac{r_f^2 M_{Qn}^{2}}{4} 
+ \frac{r_f}{\pi R}}}\right].
\end{equation}
In passing note that $r_{f} = 0$ implies the usual mUED normalization $\sqrt{2/(\pi R)}$. The quantity $M_{Qn}$ in the previous equations represents the KK mass and is given by
\begin{eqnarray} \label{eq:transc}
  r_{f} M_{Qn}= \left\{ \begin{array}{rl}
         -2\tan \left(\frac{M_{Qn}\pi R}{2}\right) &\mbox{for $n$ even,}\\
          2\cot \left(\frac{M_{Qn}\pi R}{2}\right) &\mbox{for $n$ odd.}
          \end{array} \right.         
 \end{eqnarray}
Clearly for $r_{f} = 0$ we get back the mUED KK mass $n/R$. In Fig.~\ref{f:transcmass} we show the dependence of KK mass on the BLKT parameter; here we have taken $1/R$ to be 1 TeV. 
After discussing the fermions we now describe the actions for gauge and scalar fields and the Yukawa interactions. The respective actions are given by
\begin{align}
\label{eq:gaugeacn}
\mathcal{S}_{\rm gauge} &= -\frac14 \int d^{4}x \int_{0}^{\pi R}dy
             \bigg[
                  \sum_{a}\left(F^{MNa}F_{MN}^{a}
                  +r_{G}\{\delta(y)+\delta(y-\pi R)\}
                  F^{\mu \nu a}F_{\mu \nu}^{a}\right) \nonumber
                   \\
             & ~~~~~~~~~~~~~~~~~~~~~~~~~~~~~~ 
                   +B^{MN}B_{MN}+r_{G}
                   \{\delta(y)+\delta(y-\pi R)\}
                   B^{\mu \nu}B_{\mu \nu}
              \bigg],\\
\label{eq:scalacn}
\mathcal{S}_{\rm scalar} &=  \int d^{4}x \int_{0}^{\pi R}dy
             \bigg[
             \left(D^{M}\Phi\right)^{\dagger}\left(D_{M}\Phi\right)
             +r_{\Phi}\{\delta(y)+\delta(y-\pi R)\}
             \left(D^{\mu}\Phi\right)^{\dagger}\left(D_{\mu}\Phi\right)
             \bigg], \\
\label{eq:yukacn}               
\mathcal{S}_{\rm Yuk} &= - \int d^{4}x \int_{0}^{\pi R}dy 
             \bigg[
             \tilde{y}_{t}\bar{Q}\tilde{\Phi}U +
             \tilde{y}_{b}\bar{Q}\Phi D + 
             r_{Y}\{\delta(y)+\delta(y-\pi R)\} \nonumber \\
       &  ~~~~~~~~~~~~~~~~~~~~~~~~~~~~~~ 
             \times \left(
             \tilde{y}_{t}\bar{Q}_{L}\tilde{\Phi}U_{R} +
             \tilde{y}_{b}\bar{Q}_{L}\Phi D_{R} \right) + {\rm h.c.}
             \bigg],              
\end{align}
where $a$ is the gauge index, the field strength $F_{MN}^{a}\equiv \left(\partial_{M}A_{N}^{a}-\partial_{N}A_{M}^{a}+
\tilde{g}f^{abc}A_{M}^{b}A_{N}^{c}\right)$ is associated with $SU(2)_{L}$, and the field strength $B_{MN} \equiv (\partial_{M}B_{N}-\partial_{N}B_{M})$ is with $U(1)_{Y}$. The standard Higgs doublet is denoted by $\Phi$, and $\tilde{\Phi} = i\tau^{2}\Phi^{\ast}$. The BLKT parameters for gauge and scalar fields are $r_{G}$ and $r_{\Phi}$ respectively, whereas $r_{Y}$ denotes the boundary localised Yukawa parameter. The tilde in the gauge (and Yukawa) couplings labels them as 5D couplings which are actually dimensionful quantities and related to the dimensionless 4D couplings via appropriate scalings, \textit{e.g.,}
\begin{align}
g = \frac{\tilde{g}}{\sqrt{r_{G}+\pi R}}~.
\end{align}
The KK mode functions for gauge and scalar fields are similar in form to the fermionic mode functions (like $f_{L}$ or $g_{R}$). The mass of the KK excitations of gauge and scalar fields $M_{Gn}$ and $M_{\Phi n}$ follows the same transcendental equation given in Eq.~\ref{eq:transc} with $r_{f}$ replaced appropriately.
In our analysis we consider the boundary localised kinetic terms for fermions, gauge bosons and scalars as well as the boundary localised Yukawa terms . Moreover, we assume equal BLKT parameter for gauge and scalar fields, {\it i.e.,} $r_{G} = r_{\Phi}$. In a general setting one can take $r_{G} \neq r_{\Phi}$, but then the differential equation satisfied by the mode functions of the gauge bosons  contains a term proportional to the $r_{\Phi}$ after the electroweak symmetry breaking and thus the solutions of the mode functions are not of a simple form as mentioned in our Eq.~\ref{eq:modefunctions}~\cite{Datta:2014sha}. To avoid this unnecessary complication we stick to $r_{G} = r_{\Phi}$.
Also note that since the fifth component of a gauge field (\textit{e.g.,} $W_{5}$, $A_{5}$ etc.) is projected out by the $\mathbb{Z}_{2}$ symmetry, the $y$-dependent mode functions of it are given by a similar form as in Eq.~\ref{eq:modefunctionsodd}~\cite{Muck:2004zz, Datta:2013yaa}.  
We use 't Hooft--Feynman gauge in our calculation and it is important to spell out the gauge fixing action in this scenario. This action, following~\cite{Muck:2004zz, Jha:2014faa}, is given by
\begin{align}
\label{eq:gfaction}
\mathcal{S}_{\rm GF}^{W} = -\frac{1}{\xi_{y}} 
                 \int d^{4}x \int_{0}^{\pi R}dy 
                 \left|\partial_{\mu}W^{\mu +}
                 +\xi_{y}\left[\partial_{5}W^{5+}
                 -iM_{W}\phi^{+}\{1+r_{\Phi}
                 (\delta(y)+\delta(y-\pi R))\}
                 \right]\right|^{2},
\end{align}
where $M_{W}$ is the $W$-boson mass and $\xi_{y}$ is related to the physical gauge fixing parameter $\xi$ as
\begin{align}
\xi = \xi_{y}\left[1+r_{\Phi}\{\delta(y)+\delta(y-\pi R)\}\right],
\end{align}
and in 't Hooft--Feynman gauge $\xi \to 1$ whereas in Landau gauge $\xi \to 0$.
Since in this work we are interested in calculating the widths of various rare decays of the top quark we need to consider the interactions between the top quark (and final decay products) and various higher KK excitations, which show up in the loop-induced decay processes. The standard procedure to calculate the effective 4D couplings for this type of interaction is to write the original 5D interaction term and then replace each field by its corresponding KK expansions and then integrate out the extra coordinate $y$. In mUED these types of couplings are equivalent to their SM counterparts and most of the time can be read off from the Lagrangian itself. But in the case of nmUED, the couplings get modification from the overlap integrals of the form
\begin{align}
I^{ijk} = \int_{0}^{\pi R}dy~f_{\alpha}^{i}(y)
              ~f_{\beta}^{j}(y)~f_{\gamma}^{k}(y),
\end{align}      
where the greek indices (subscripts) denote the type of field and the latin indices (superscripts) refer to the KK level of the respective fields. This type of modification in coupling is characteristic to the nmUED scenario. The root of this modification lies in the fact that unlike mUED, the KK mode function in nmUED has BLT parameter dependence, explicitly in normalization factors and implicitly in KK masses. Also note that if $(i+j+k)$ is an odd integer then these overlap integrals vanish due to the conservation of KK parity.

At this point we mention about a few overlap integrals that appear in our calculations,
\begin{subequations}\label{eq:overlapints}
\begin{align}
\label{eq:ovint1}
I_{1}^{jk} &= \int_{0}^{\pi R}dy \left[
              1+r_{f}\{\delta(y)+\delta(y-\pi R)\}\right]
              f_{Qt_{L}}^{(j)}(y)
              ~f_{\Phi}^{(k)}(y)~f_{b_{L}}^{(0)}(y),\\
\label{eq:ovint2}
I_{2}^{jk} &= \int_{0}^{\pi R}dy ~f_{Qt_{R}}^{(j)}(y) 
              ~f_{W_{5}}^{(k)}(y)~f_{b_{L}}^{(0)}(y),\\
\label{eq:ovint3}
I_{3}^{k} &= \int_{0}^{\pi R}dy \left[
              1+r_{f}\{\delta(y)+\delta(y-\pi R)\}\right]
              f_{Qt_{R}}^{(0)}(y)~f_{W_{\mu}}^{(k)}(y)
              ~f_{b_{L}}^{(0)}(y).   
\end{align}
\end{subequations}
These are the overlap integrals that modify the respective couplings.
\begin{figure}[h]
\begin{center}
\includegraphics[scale=0.7]{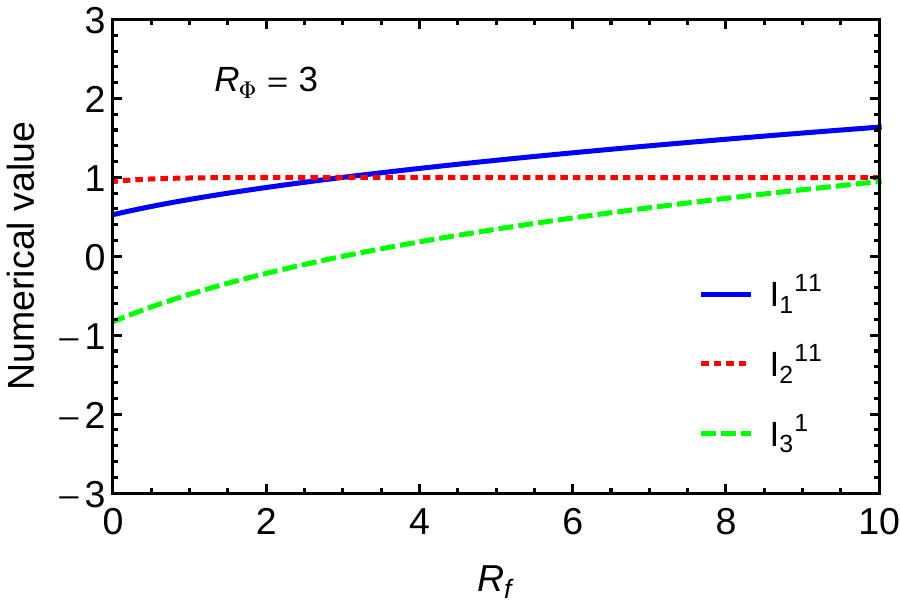}
~~~~
\includegraphics[scale=0.7]{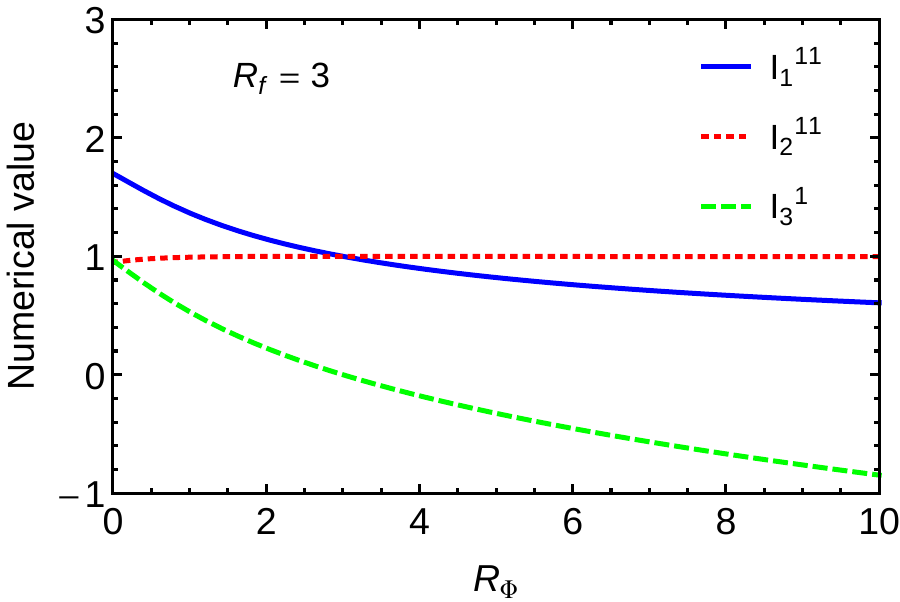} 
\caption{Characteristic dependence of overlap integrals on the BLT parameters. Here $R_{\Phi,f} = r_{\Phi,f}/R$ and we take only first KK modes into account, \textit{i.e.,} we take $j,k=1$ in Eqs.~\ref{eq:overlapints}.}
\label{f:ovint}
\end{center}
\end{figure}
In Fig.~\ref{f:ovint} we show characteristic behaviour of the overlap integrals with respect to various BLT parameters. It should be kept in mind that even though for some choice of BLT parameters the numerical value of the overlap integrals can be greater than unity, the final values of the relevant couplings remain within the perturbativity limit.

\subsection{Physical Eigenstates}
\label{sbsc:phyeign}
In the effective 4D theory, the higher KK modes of various fields mixes to give rise to physical fields. This type of mixing is present in fermionic as well as scalar/gauge sector.
In the quark sector the strength of mixing is proportional to the quark mass. Thus, it is significant for the top sector. However, to incorporate the GIM mechanism and our further analysis, we consider the mixing in the down sector too. This mixing matrix can be diagonalised by separate unitary matrices for left- and right-handed quarks,
\begin{align}
\mathscr{U}_{L}^{(n)} = \begin{pmatrix}
-\cos \alpha_{n} & \sin \alpha_{n} \\
 \sin \alpha_{n} & \cos \alpha_{n}
\end{pmatrix},~~
\mathscr{U}_{R}^{(n)} = \begin{pmatrix}
 \cos \alpha_{n} & -\sin \alpha_{n} \\
 \sin \alpha_{n} & \cos \alpha_{n}
\end{pmatrix},
\end{align}
where $\alpha_{n} = \frac12 \tan^{-1}\left(\frac{m_{b}}{M_{Qn}}\right)$ with $m_{b}$ denoting the SM bottom quark mass. In the $n$-th KK level, the mass term can be written as
\begin{equation}
\begin{pmatrix}
\bar{Q}_{j_{L}}^{(m)} & \bar{D}_{L}^{(m)}
\end{pmatrix}
\begin{pmatrix}
-M_{Qn}\delta^{mn} & m_{j} \alpha_{1}{\cal I}^{mn} \\ m_{j} \alpha_{1} & M_{Qn}\delta^{mn}
\end{pmatrix}
\begin{pmatrix}
Q_{j_{R}}^{(n)} \\ D_{R}^{(n)}
\end{pmatrix}+{\rm h.c.},
\end{equation}
where $M_{Qn}$ are the solutions of transcendental equations given in Eq.~\ref{eq:transc}. ${\cal I}^{mn}$ is an overlap integral of the form 
\begin{align*}
\int_0 ^{\pi R} \left[ 1+ r_Y \{\delta(y) + \delta(y - \pi R)\} \right] f_{L}^m (y) g_{R}^n (y)\;dy,
\end{align*}
and
\begin{align*}
\alpha_{1} =\frac{\pi R+ r_{f}}{\pi R+r_{Y}}.
\end{align*}
In general ${\cal I}^{mn}$ is non-zero whether $n=m$ or $n\neq m$. However, the $n\neq m$ case would lead to the (KK-)mode mixing among the quark of a particular flavour. An interesting point to note is that for the choice $r_f = r_Y$, ${\cal I}^{mn} = \delta_{mn}$ and obviously $\alpha_{1}=1$. Thus to get a simpler form of fermion mixing matrix and avoid the mode mixing we  stick to the choice of equal $r_Y$ and $r_f$. Taking into account these matrices one can now relate the gauge eigenstates $Q_{j}^{(n)}$ [$D^{(n)}$] and mass eigenstates $Q_{j}^{\prime (n)}$ [$D^{\prime (n)}$] as (in this notation $j$ refers to the down quark flavour),
\begin{subequations}
\begin{align}
Q_{j_{L/R}}^{(n)} &= \mp \cos \alpha_{n} Q_{j_{L/R}}^{\prime(n)} +
                      \sin \alpha_{n} D_{L/R}^{\prime(n)},\\
D_{L/R}^{(n)} &= \pm \sin \alpha_{n} Q_{j_{L/R}}^{\prime(n)} + 
                    \cos \alpha_{n} D_{L/R}^{\prime(n)}.
\end{align}
\end{subequations}
The mass eigenstates, in this case, share the same mass eigenvalue,  
\begin{align}
m_{Q_{b}^{\prime (n)}} = m_{D^{\prime (n)}} 
                          = \sqrt{m_{b}^{2}+M_{Qn}^{2}} 
                          \equiv M_{\rm bottom}.
\end{align}
A similar procedure follows for the up sector also. 
In the scalar sector, the 4D effective Lagrangian also contains bilinear terms involving KK excitations of the fifth components of $W^{\pm}$ ($Z$) bosons and the KK excitations of $\phi^{\pm}$ ($\chi^{0}$), which basically are the components of the Higgs doublet. Now, using Eqs.~\ref{eq:gaugeacn},~\ref{eq:scalacn} and~\ref{eq:gfaction} one can write, in $R_{\xi}$ gauge, the bilinear terms of $W_{5}^{\pm (n)}$ and $\phi^{\pm (n)}$ as follows: 
\begin{align}
\mathcal{L}_{W_{5}^{\pm (n)}\phi^{\mp (n)}} = 
       -\begin{pmatrix}
       W_{5}^{(n)-} & \phi^{(n)-}
       \end{pmatrix}
       \begin{pmatrix}
       M_{W}^{2}+\xi M_{\Phi}^{2}  & -i(1-\xi)M_{W}M_{\Phi n} \\
       i(1-\xi)M_{W}M_{\Phi n}   &  M_{\Phi n}^{2}+\xi M_{W}^{2}
       \end{pmatrix}
       \begin{pmatrix}
       W_{5}^{(n)+} \\
       \phi^{(n)+}
       \end{pmatrix}.
\end{align}
This mass matrix upon diagonalization gives rise to a tower of charged Goldstone bosons $G^{(n)\pm }$ and charged Higgs bosons $H^{(n)\pm }$ with masses $\sqrt{\xi(M_{\Phi n}^{2}+M_{W}^{2})}$ and $\sqrt{M_{\Phi n}^{2}+M_{W}^{2}}$ respectively. In the component form they can be written as
\begin{subequations}
\begin{align}
G^{(n)\pm } &= \frac{M_{\Phi n}W^{\pm5(n)} \mp i M_{W}\phi^{\pm(n)}}
               {\sqrt{M_{\Phi n}^{2}+M_{W}^{2}}},\\
H^{(n)\pm } &= \frac{M_{\Phi n}\phi^{\pm(n)} \mp i M_{W}W^{\pm5(n)}}
               {\sqrt{M_{\Phi n}^{2}+M_{W}^{2}}}. 
\end{align}
\end{subequations}
Thus, in the 't Hooft--Feynman gauge ($\xi \to 1$) the fields $G^{ (n)\pm}$, $H^{(n)\pm }$ and $W^{\mu(n)\pm}$ have a common mass eigenvalue $M_{Wn} = \sqrt{M_{\Phi n}^{2}+M_{W}^{2}}$. It is worth mentioning that these combinations of charged Higgs and Goldstones ensure vanishing of the $\gamma^{(0)}H^{(n)\pm}W^{(n)\mp}$ coupling, where $\gamma^{(0)}$ is the SM photon.

\section{Rare Top Decays} 
\label{s:raretopdecay}
In this section we discuss some of the rare decays of the top quark in the model presented above. The flavor changing rare decays of top quarks occur at loop level in the SM. On top of this loop suppression,  there are CKM and GIM suppression~\cite{Eilam:1990zc, AguilarSaavedra:2002ns, AguilarSaavedra:2004wm}. In the present work we consider the decays, $t\to c\gamma$ and $t\to c h$. Clearly in the present model, the higher KK mode particles contribute in these loop-driven processes. In the following first we will discuss the general Lorentz structure for each decay width and present the corresponding Feynman diagrams in this model. We use 't Hooft--Feynman gauge in our calculation as the divergences are more manageable in this gauge but at the cost of having extra diagrams with unphysical scalars. We present the important Feynman rules in the Appendix.

\subsection{$t \to c\gamma$} 
\label{sbsc:tcgam}
We are now going to lay down the details of the calculation of the decay width of $t\to c\gamma$ in this model. The most general form of the amplitude of the decay $t(p)\to c(k_{2})\gamma(k_{1})$ for on-shell quarks and real photons can be presented as~\cite{AguilarSaavedra:2002ns, Datta:2009zb},
\begin{align}
\mathcal{M}(t\to c\gamma) = \frac{i}{m_{t}+m_{c}}\bar{u}
                            (k_{2})[\sigma^{\mu \nu}
                            k_{1\nu}\left(A_{L}P_{L}+B_{R}P_{R}\right)]
                            u(p)\epsilon^{\ast}_{\mu}(k_{1}),                            
\end{align}
where $u$, $\bar{u}$ and $\epsilon_{\mu}$ are the incoming, outgoing spinors and photon polarization respectively; $P_{L,R} = (1\mp \gamma_{5})/2$ are the usual projection operators. The coefficients $A_{L}$ and $B_{R}$ contain the information about couplings, CKM matrix elements and the loop momenta integration. We ignore the effect of KK particle contribution on the CKM elements; for details see~\cite{Buras:2002ej}. Note that when writing the full amplitude for the process $t\to c\gamma$ following all the Feynman rules [in the SM or (n)mUED] one may come across terms proportional to $\gamma_{\mu}P_{L,R}$ in the amplitudes\footnote{To be precise, in the $m_{c} = 0$ case, only $\gamma_{\mu}P_{L}$ appears and for the $m_{c}\neq 0$ case both $\gamma_{\mu}P_{L,R}$ are present.} of Feynman diagrams. But when the amplitudes of all the diagrams are summed and the GIM mechanism\footnote{Basically by incorporating the GIM mechanism, we mean the utilisation of the relation, $V_{tj}^{\ast}V_{cj}[i\mathcal{M}(m_{j})] = V_{tb}^{\ast}V_{cb}[i\mathcal{M}(m_{b})-i\mathcal{M}(m_{s})]$, where $\mathcal{M}$ represents the sum of the amplitudes of all the Feynman diagrams and $m_{s}$ is the strange quark mass that we take to be zero.} is incorporated, the terms proportional to $\gamma_{\mu}P_{L,R}$ cancel. In the process all the apparent divergences that appear in the individual diagrams also get cancelled. These remarks are true irrespective of whether $m_{c}$ is taken to be zero or not.
At this stage it is worth mentioning that in the limit $m_{c}\to 0$, which is a reasonable approximation, the coefficient $A_{L}$ vanishes. In the $m_{c}\neq 0$ case, however, both $A_{L}$ and $B_{R}$ contribute. The apparent divergences in these loop-driven processes get cancelled among the triangle and self-energy-type diagrams. In the general non-vanishing $m_{c}$ case the decay width is given by
\begin{align}
\Gamma_{t\to c\gamma} = \frac{1}{16\pi}\frac{(m_{t}^{2}-m_{c}^{2})^{3}}
                        {m_{t}^{3}(m_{c}+m_{t})^{2}}
                        \left(|A_{L}|^{2} + |B_{R}|^{2}\right).
\end{align}

\begin{figure}[!htbp]
  \centering
  \subfloat[]{
    \includegraphics[scale=0.35]{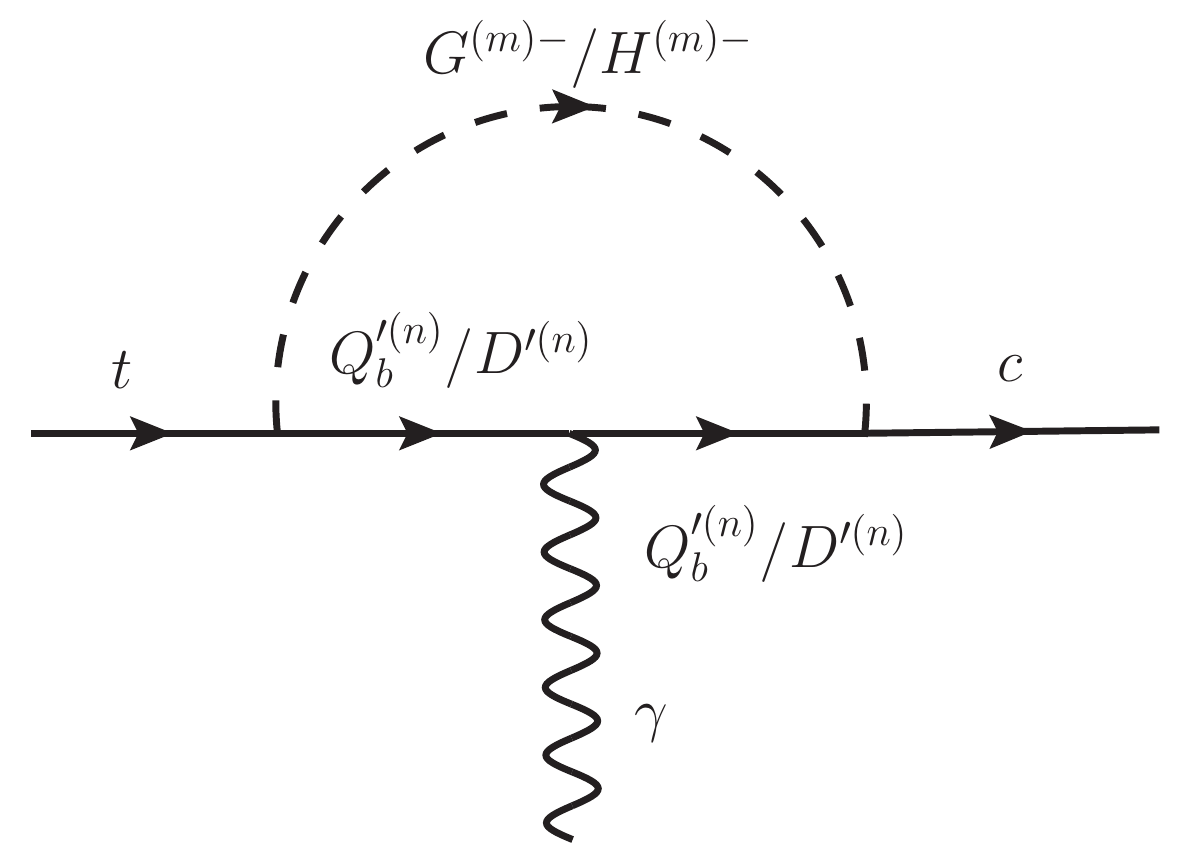}
    }
  \subfloat[]{
    \includegraphics[scale=0.35]{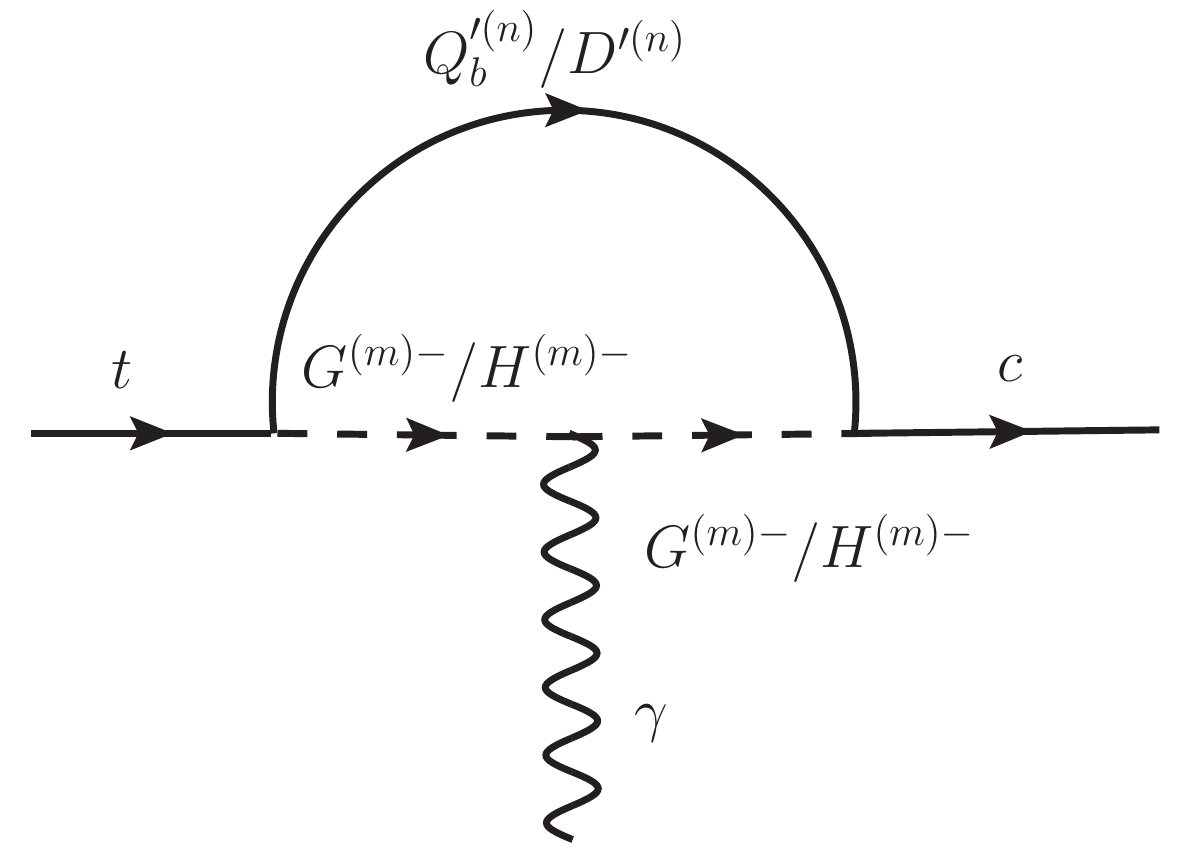}
    }
  \subfloat[]{
    \includegraphics[scale=0.35]{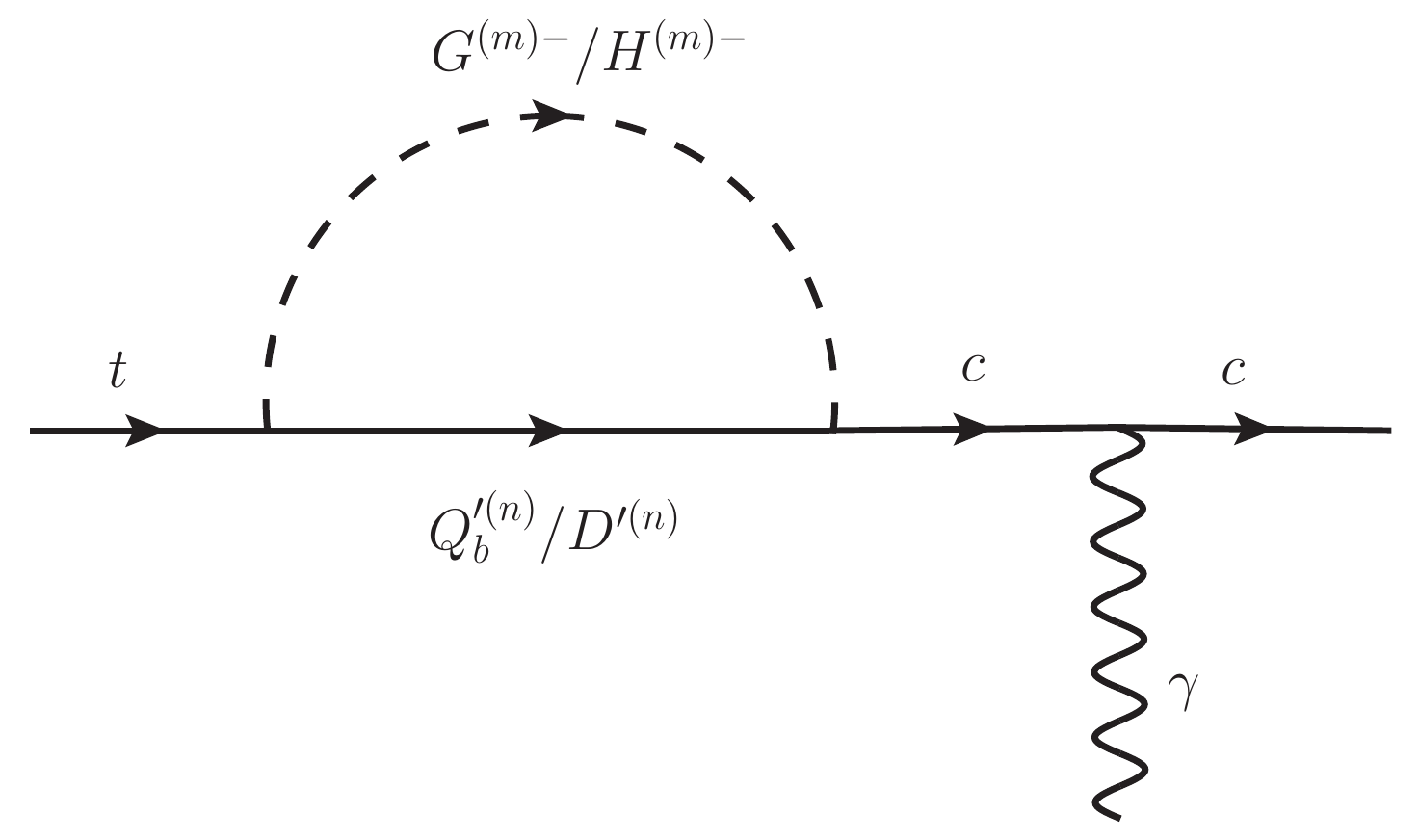}
    } \\
  \subfloat[]{
    \includegraphics[scale=0.35]{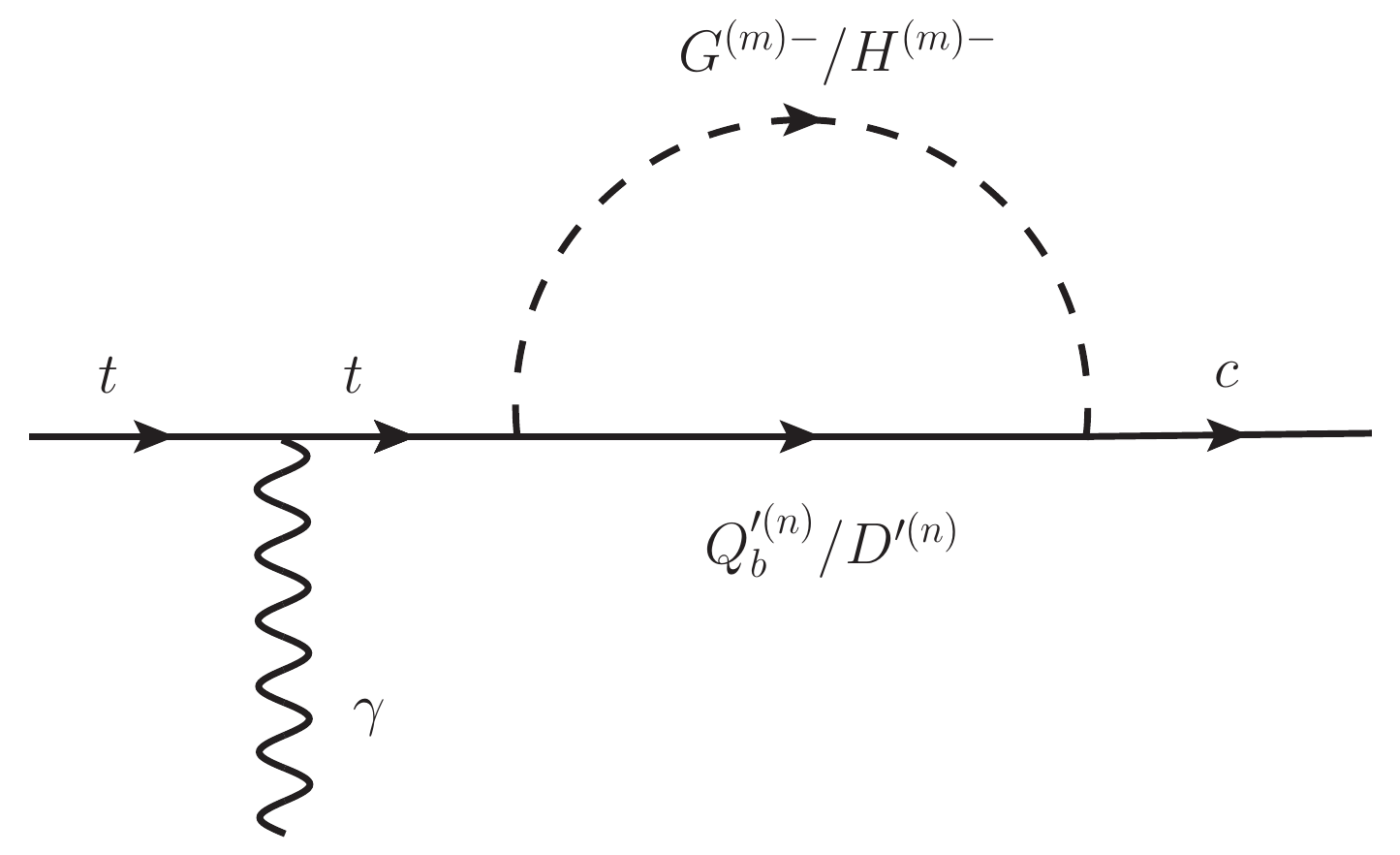}
    }
  \subfloat[]{
    \includegraphics[scale=0.35]{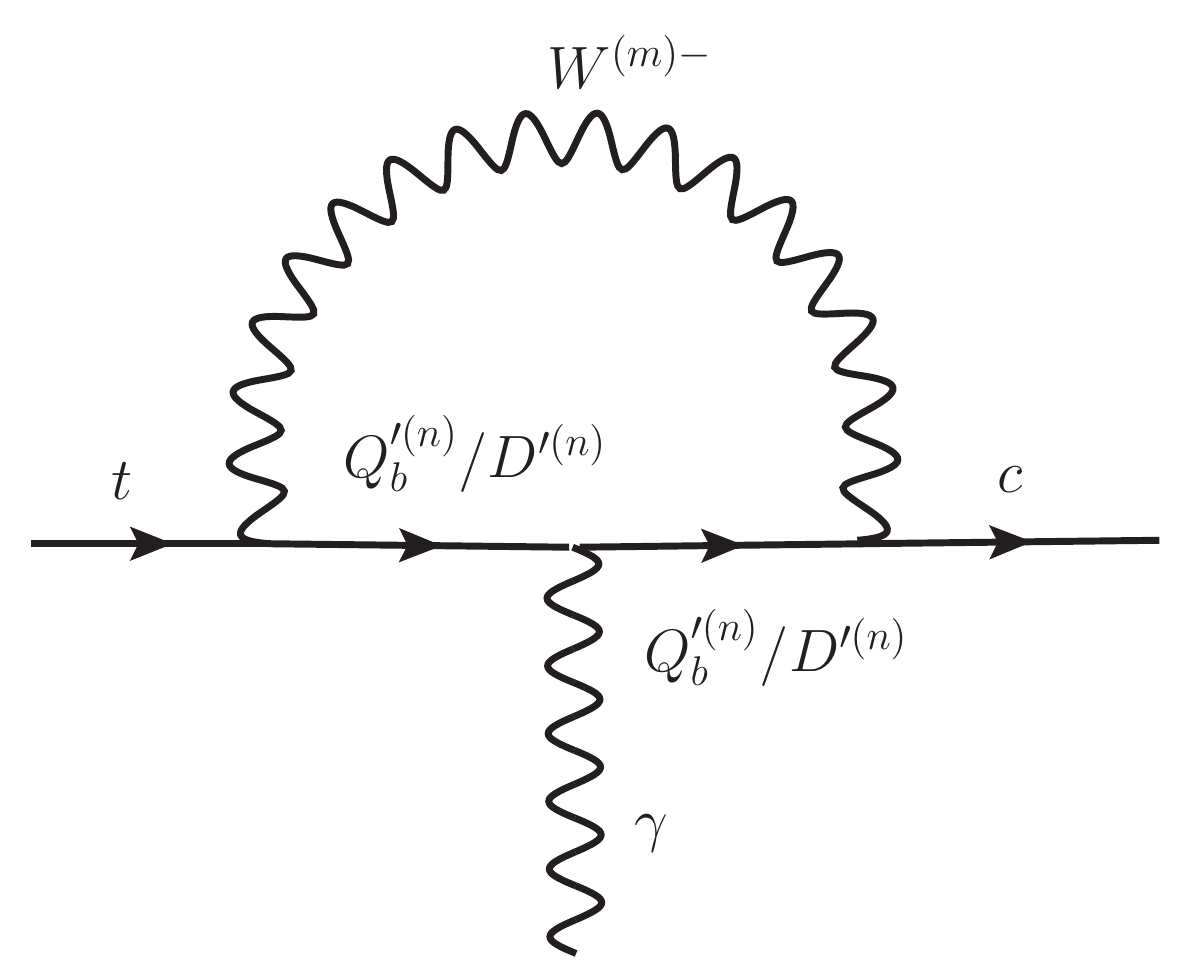}
    }
  \subfloat[]{
    \includegraphics[scale=0.35]{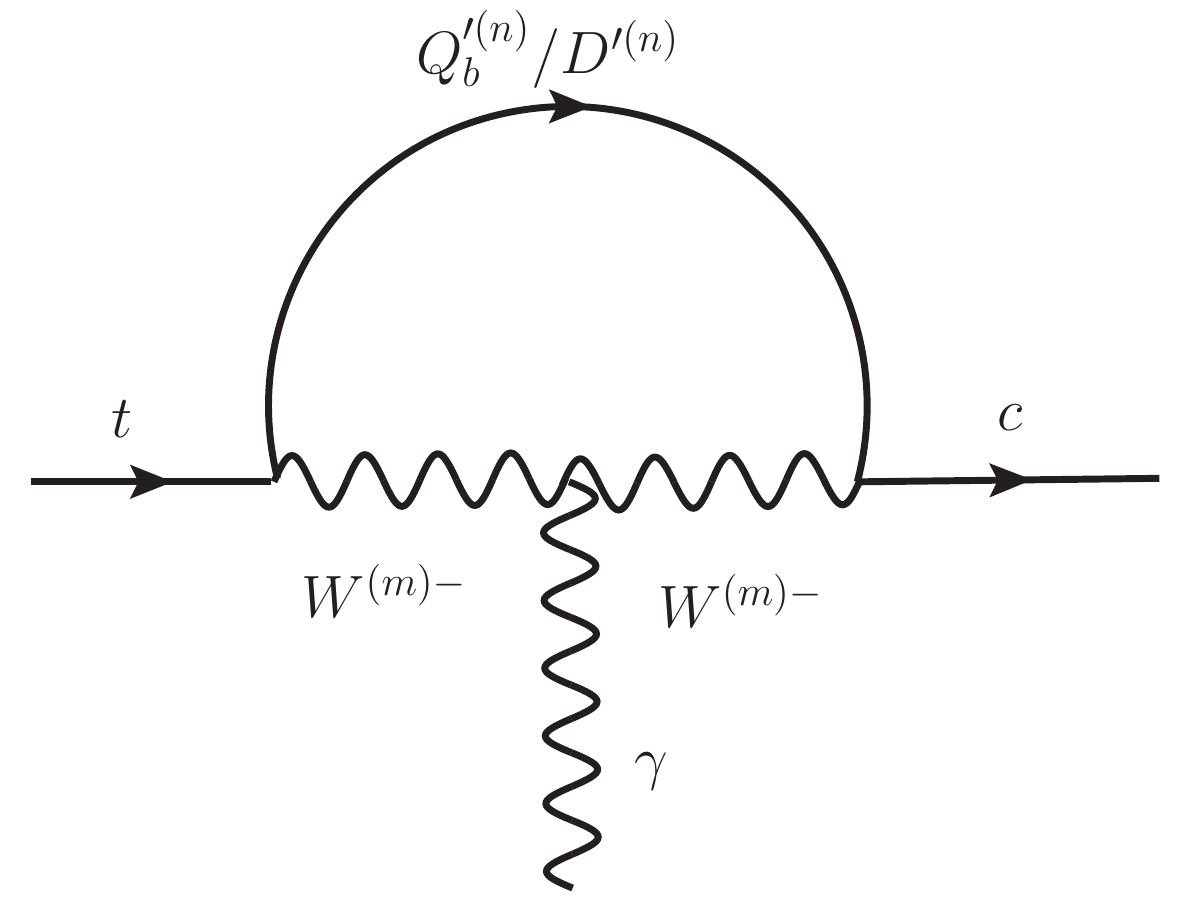}
    } \\
  \subfloat[]{
    \includegraphics[scale=0.35]{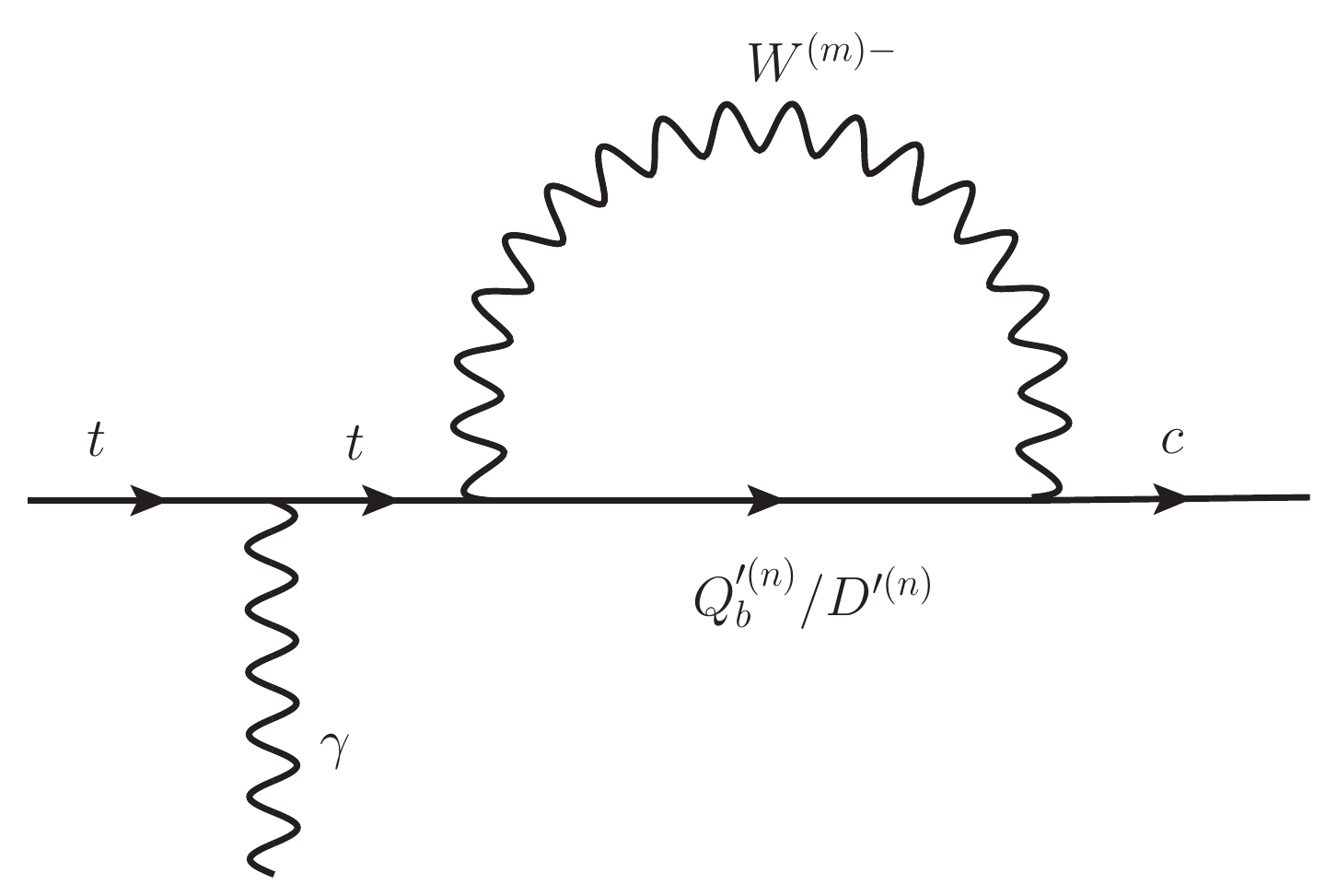}
    }
  \subfloat[]{
    \includegraphics[scale=0.35]{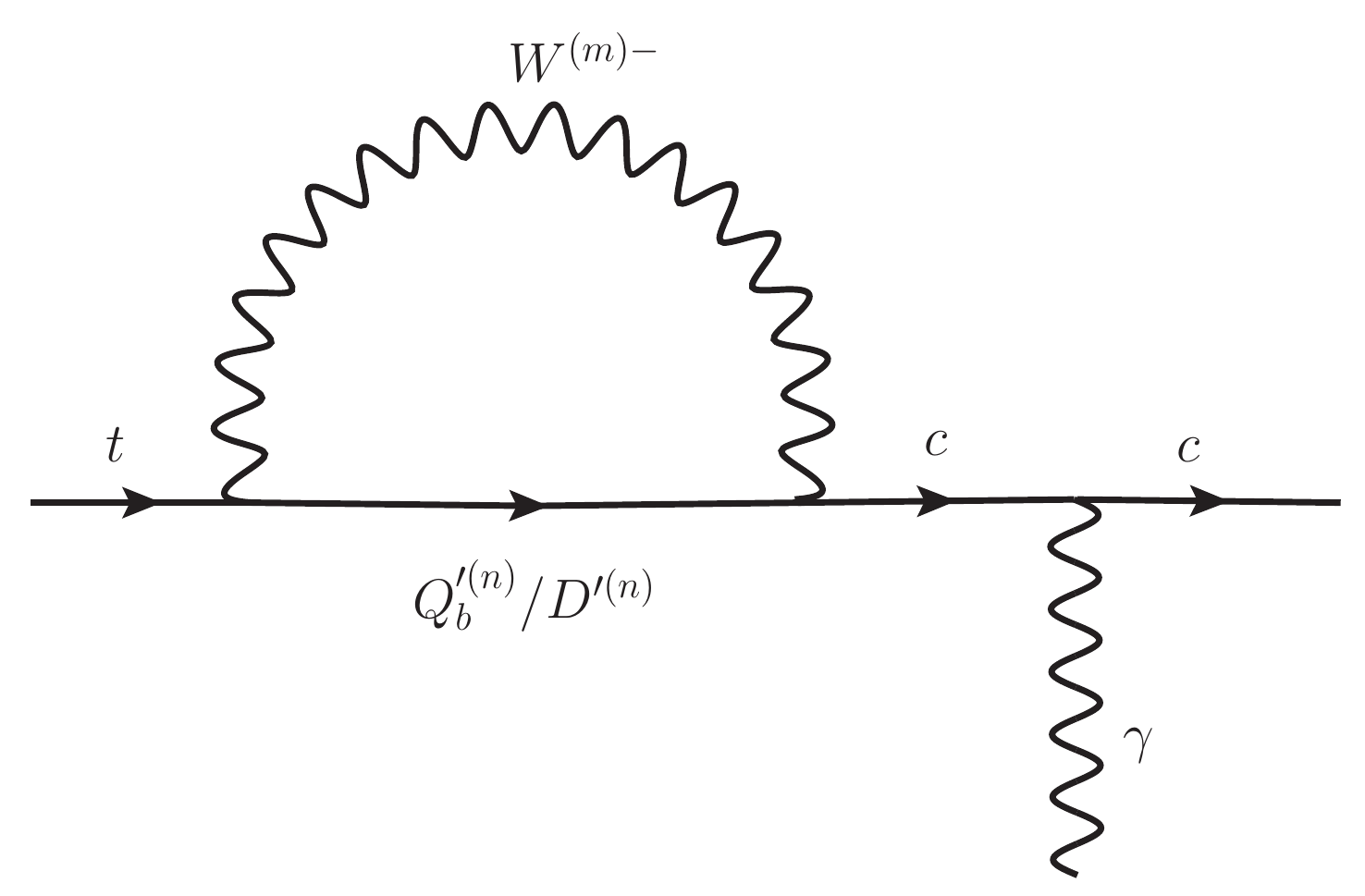}
    }
  \subfloat[]{
    \includegraphics[scale=0.35]{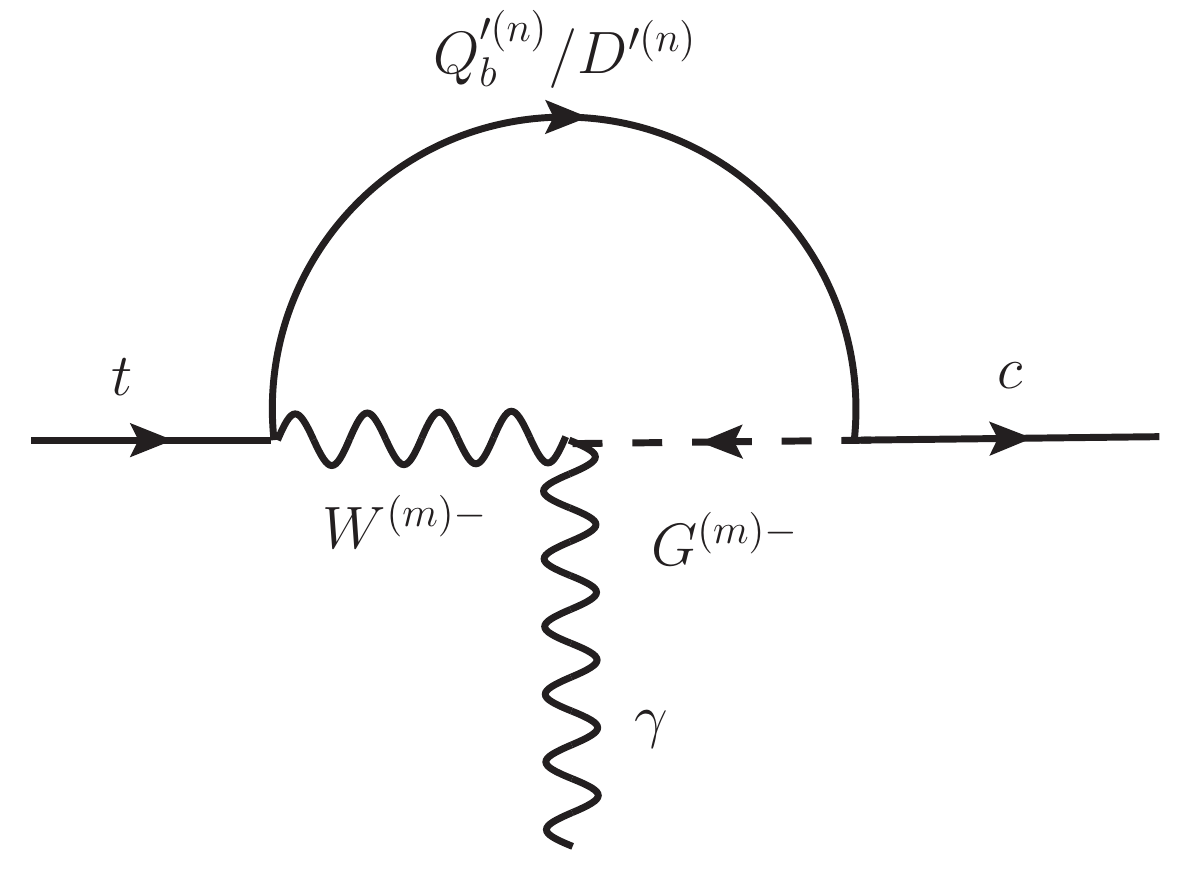}
    } \\
   \subfloat[]{
    \includegraphics[scale=0.35]{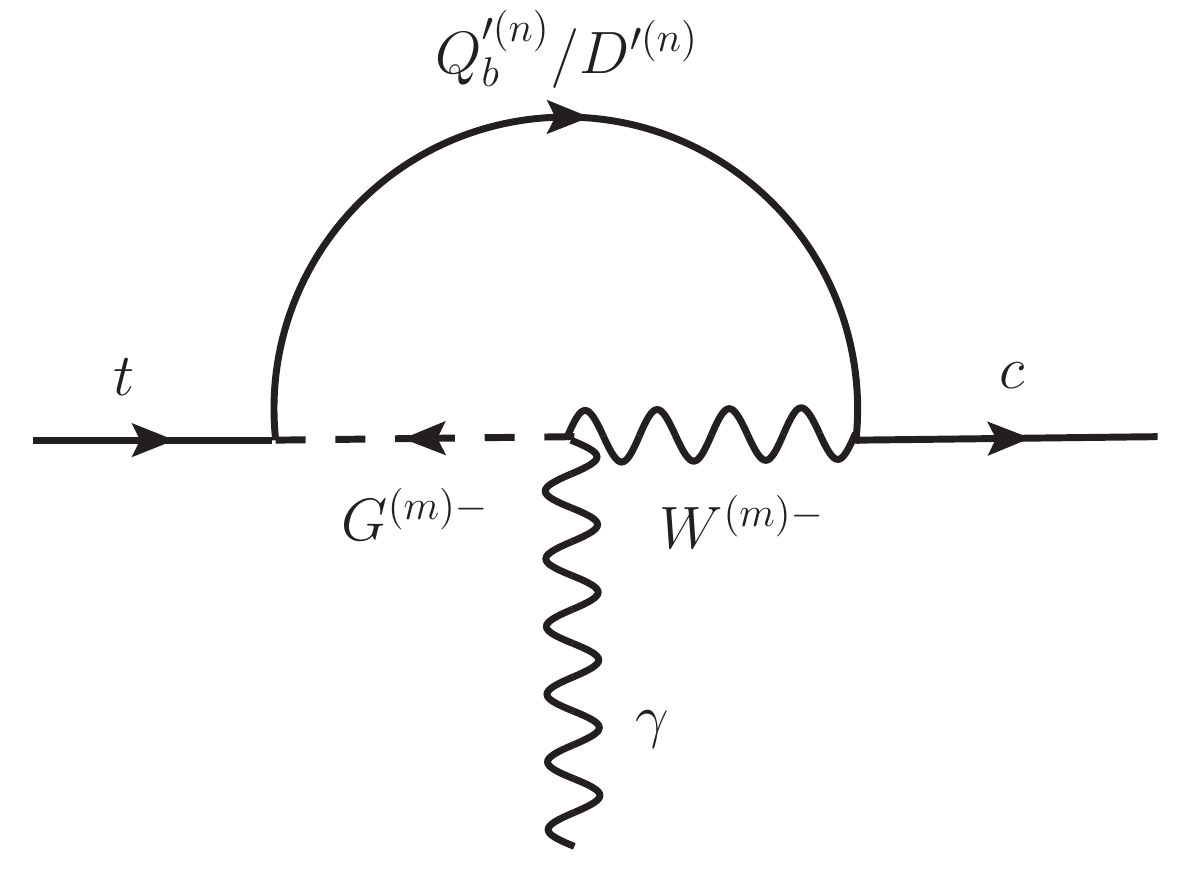}
    } 
    \caption[]{Feynman diagrams for the process $t\to c\gamma$ in the 't Hooft--Feynman gauge in nmUED. Note that the particles in the legs contain no KK indices as they represent the SM particles and their KK indices are assumed to be zero. Having said that we also emphasize that the indices $n$ can also be zero if that vertex is allowed by KK parity. We consider those types of diagrams also but do not show this explicitly in this set of diagrams to reduce cluttering.}
\label{fig:tcgam}
\end{figure}

The Feynman diagrams for this process are presented in Fig.~\ref{fig:tcgam}. In these diagrams, the superscripts $(n)$ or $(m)$ represents the $n$ (or $m$)-th KK mode of the corresponding particle. Also, since in mUED KK number is conserved in any specific vertex we always have $m=n$; but in the nmUED case $m$ and $n$ can be different, obviously satisfying the conservation of the KK parity. Evidently the quantities $A_{L}$ and $B_{R}$ contain the sum over the KK modes. In our analysis we took the KK sum up to level five (we have checked that the results of the KK sum up to level ten is almost same as that of the sum up to level five) as the contribution for higher modes decouples. Also from Fig.~\ref{f:tcgam_mued} we see that the mass of $m_{c}$ plays an insignificant role in the total decay width. Unless otherwise stated we take a vanishing $m_{c}$ in our numerical analysis.

\subsection{$t \to ch$} 
\label{sbsc:tch}
\begin{figure}[!htbp]
  \centering
  \subfloat[]{
    \includegraphics[scale=0.35]{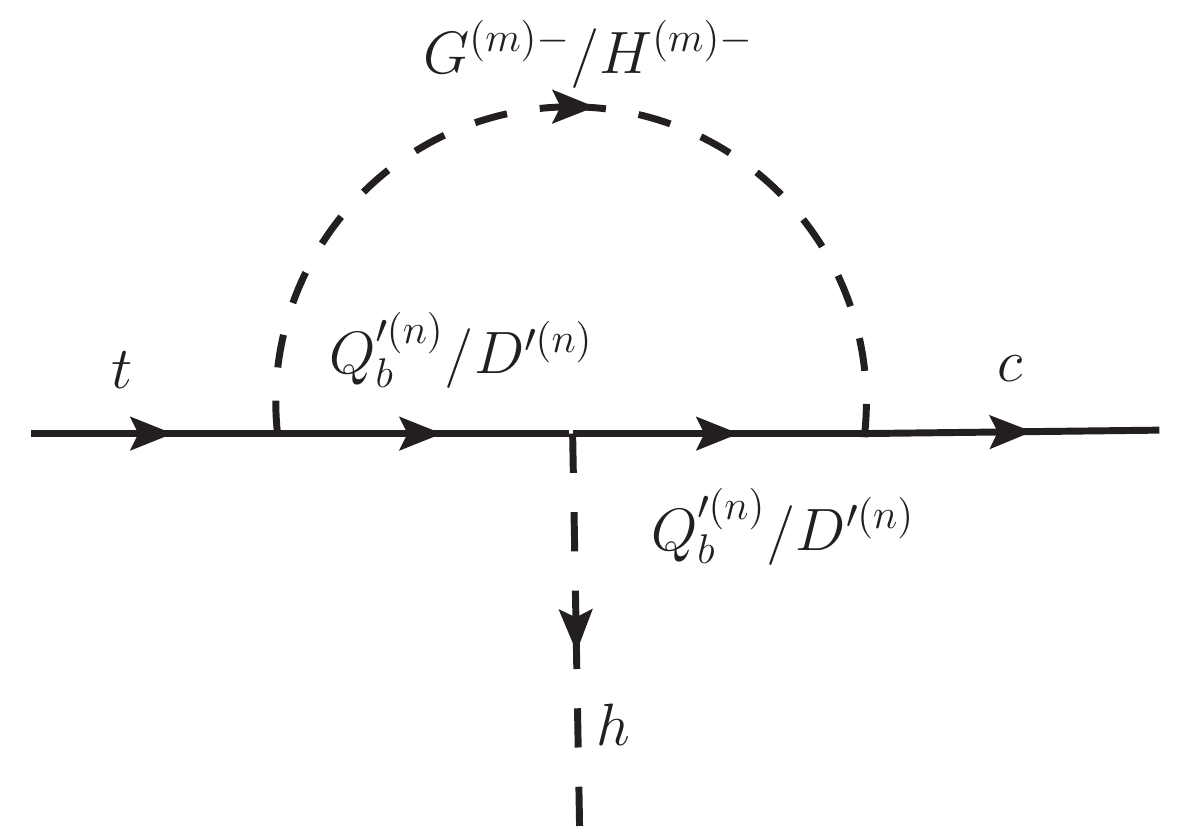}
    }
  \subfloat[]{
    \includegraphics[scale=0.35]{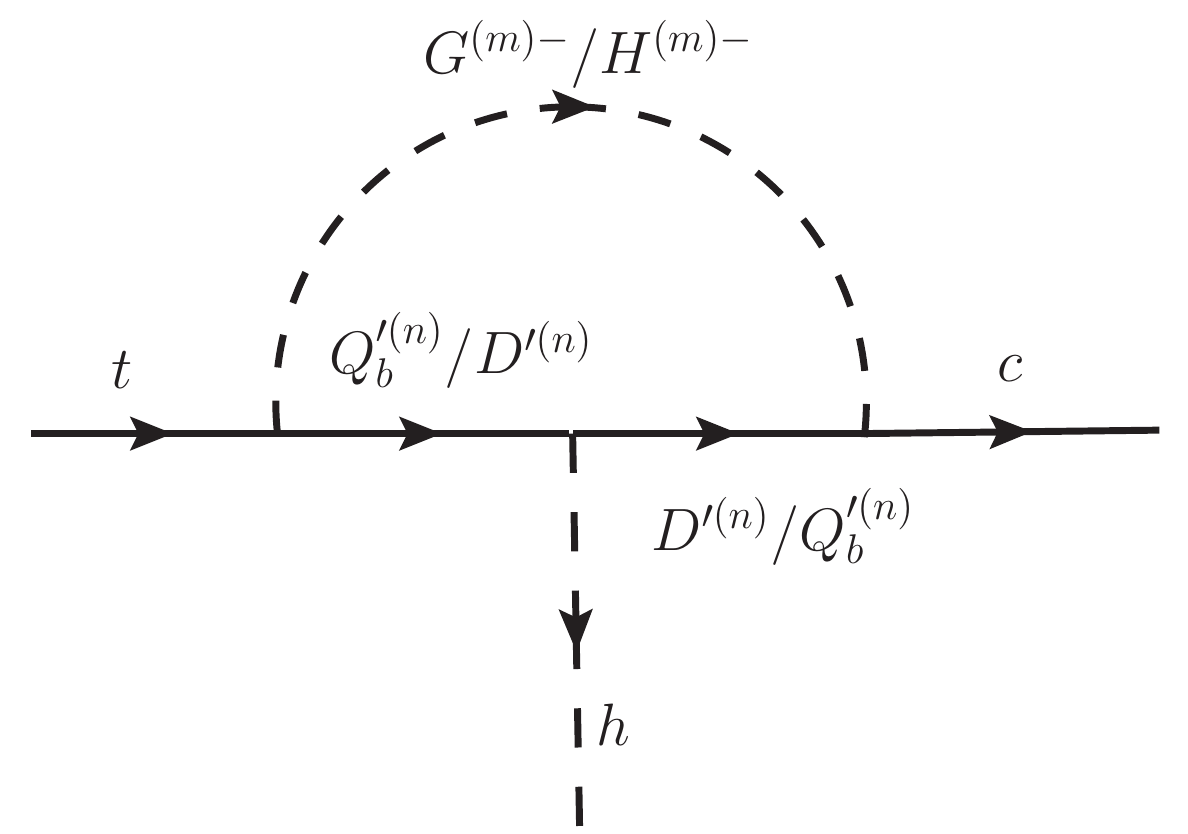}
    }
  \subfloat[]{
    \includegraphics[scale=0.35]{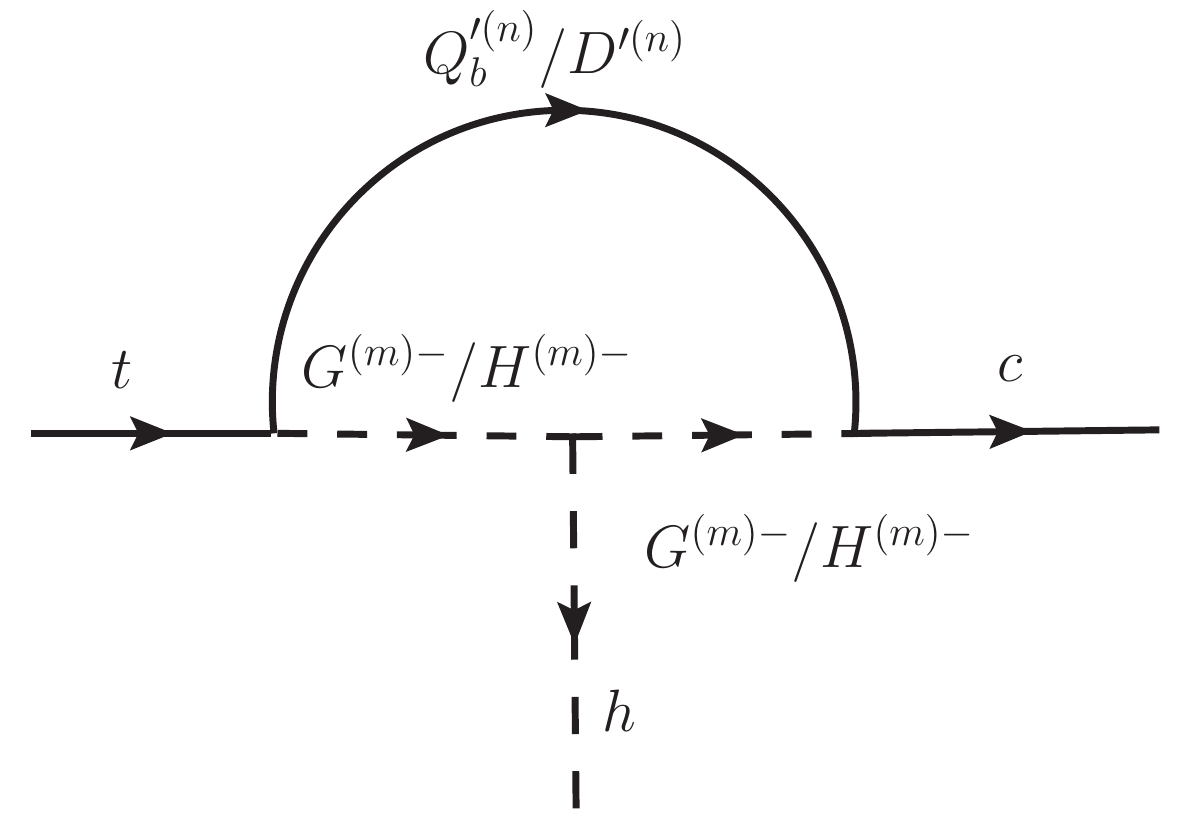}
    } \\
  \subfloat[]{
    \includegraphics[scale=0.35]{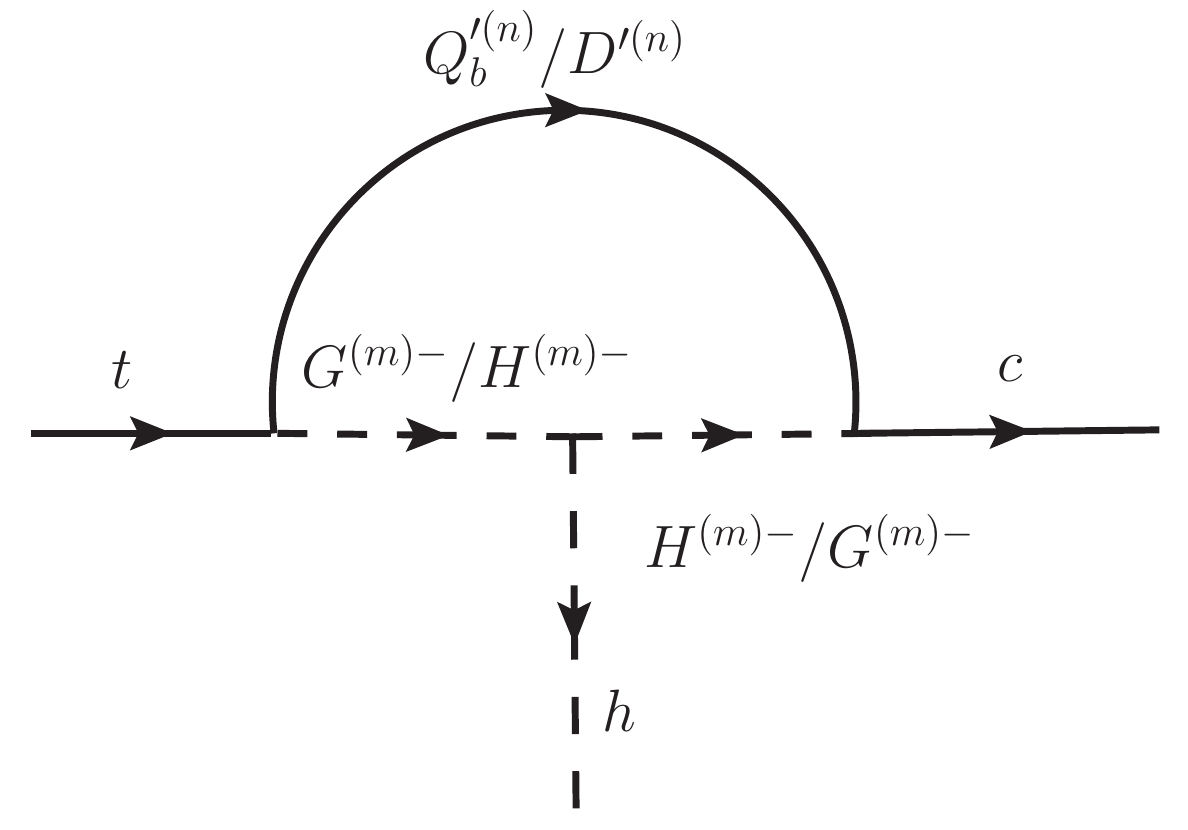}
    }    
  \subfloat[]{
    \includegraphics[scale=0.35]{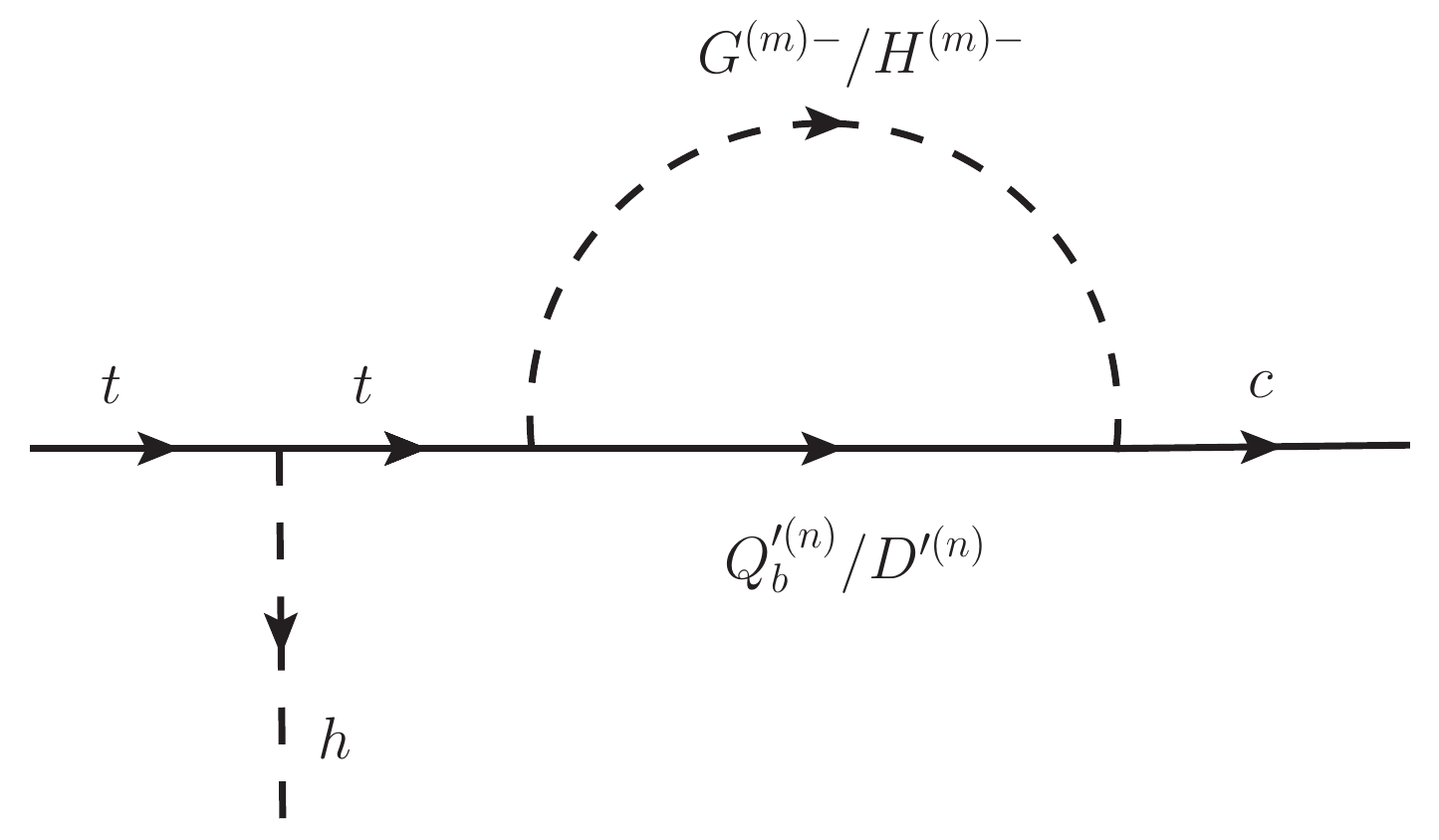}
    } 
  \subfloat[]{
    \includegraphics[scale=0.35]{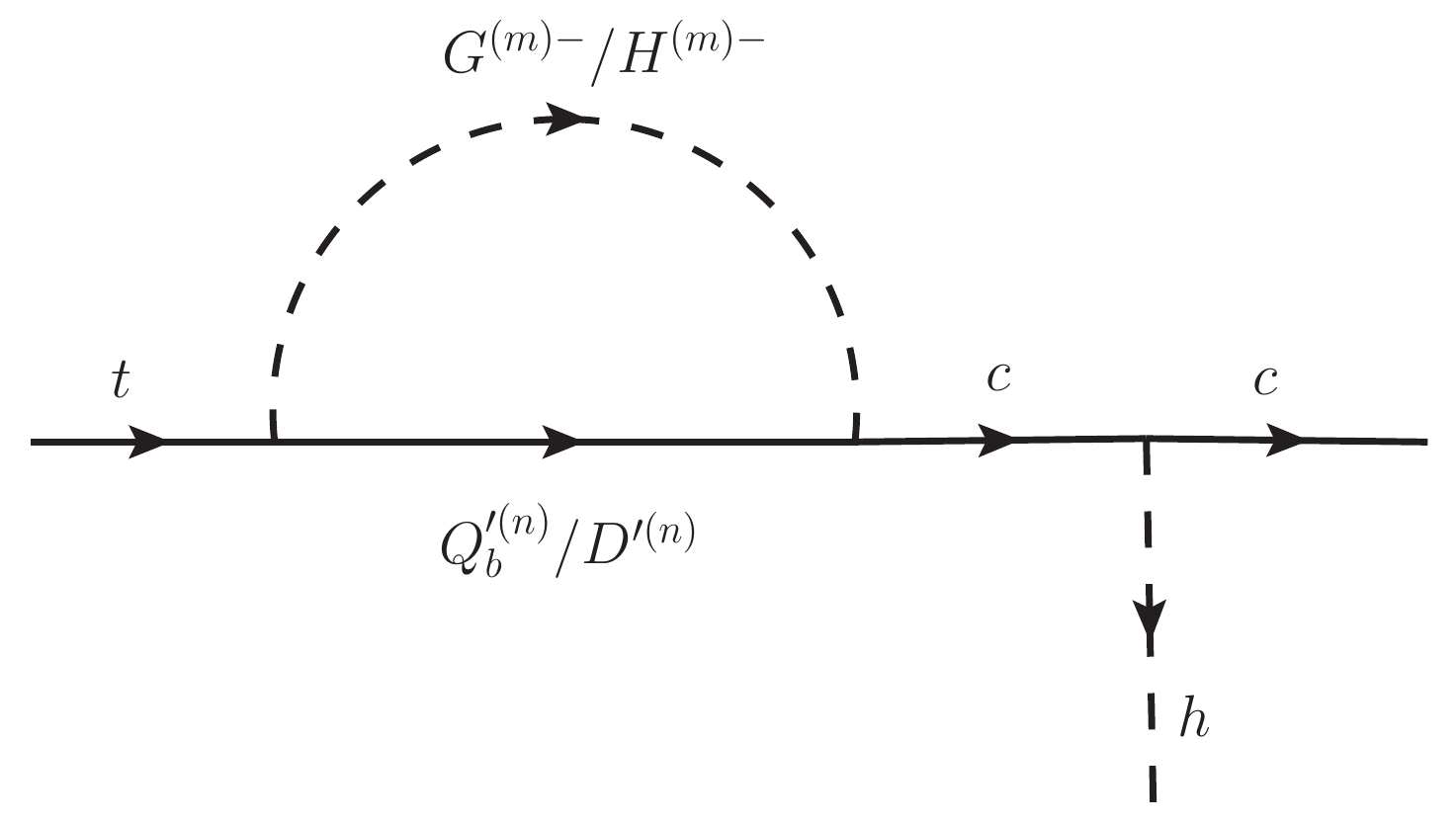}
    } \\
  \subfloat[]{
    \includegraphics[scale=0.35]{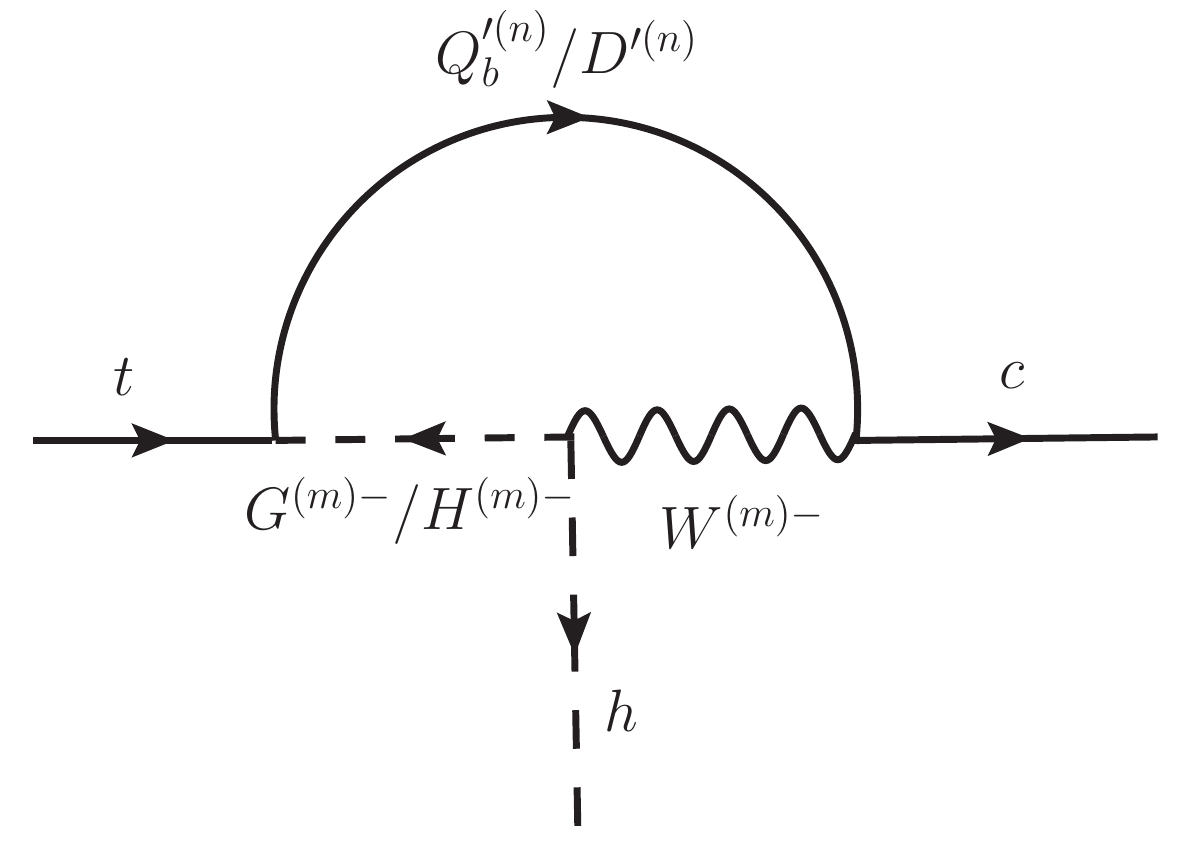}
    }
  \subfloat[]{
    \includegraphics[scale=0.35]{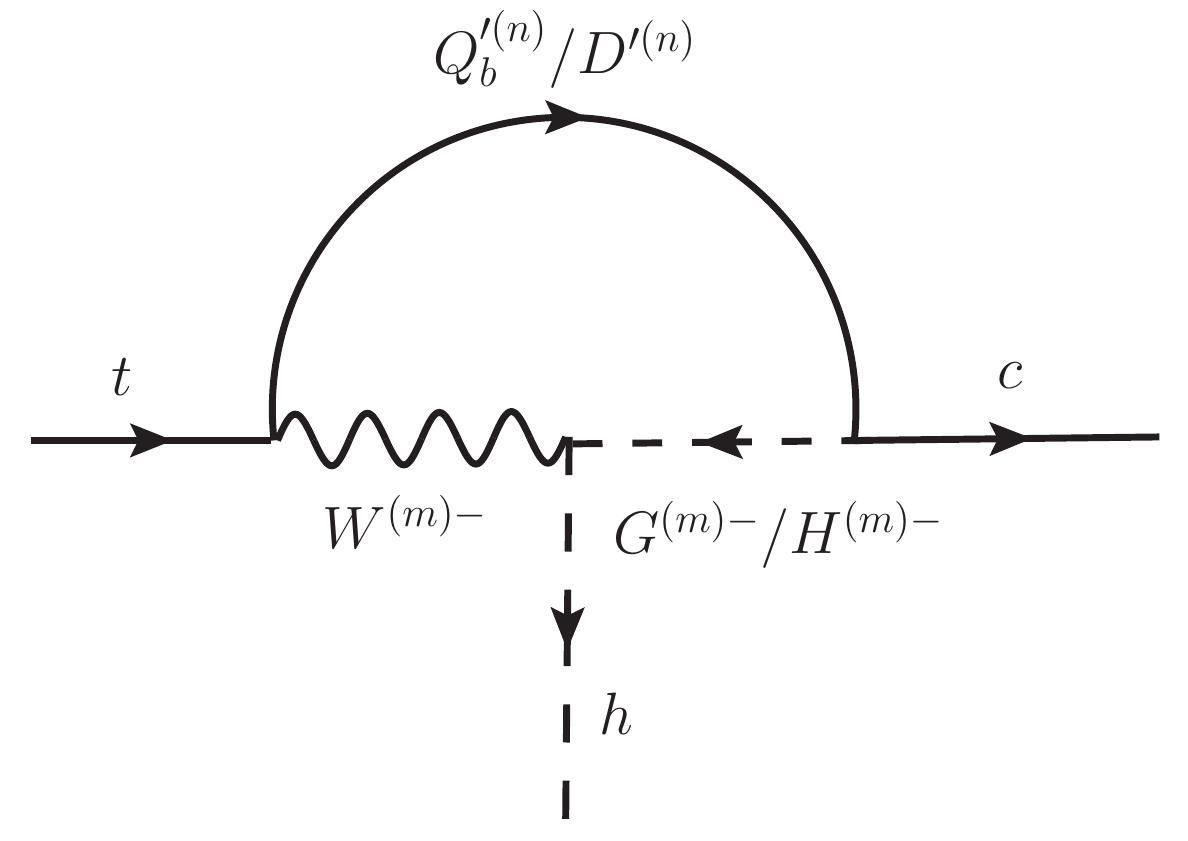}
    } 
  \subfloat[]{
    \includegraphics[scale=0.35]{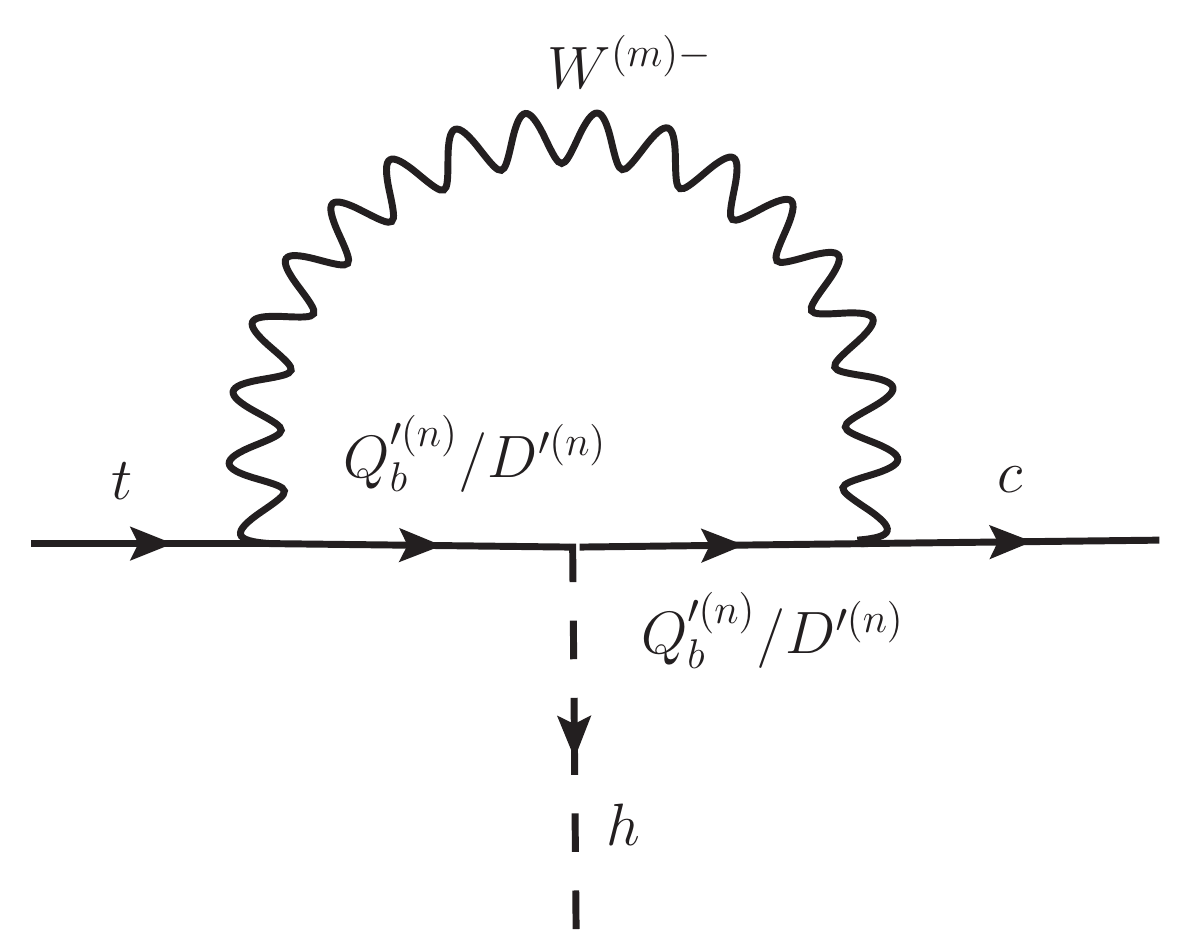}
    } \\
  \subfloat[]{
    \includegraphics[scale=0.35]{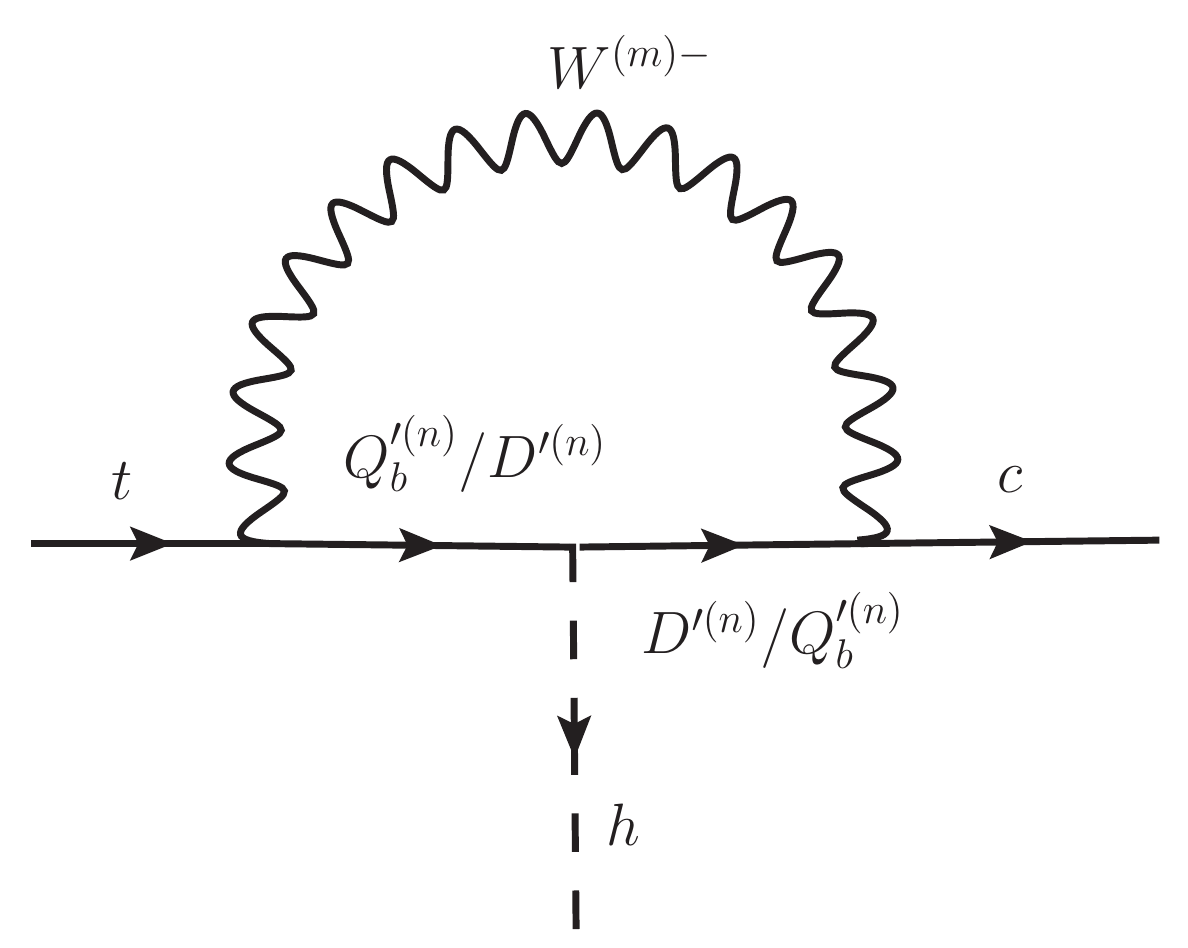}
    }    
  \subfloat[]{
    \includegraphics[scale=0.35]{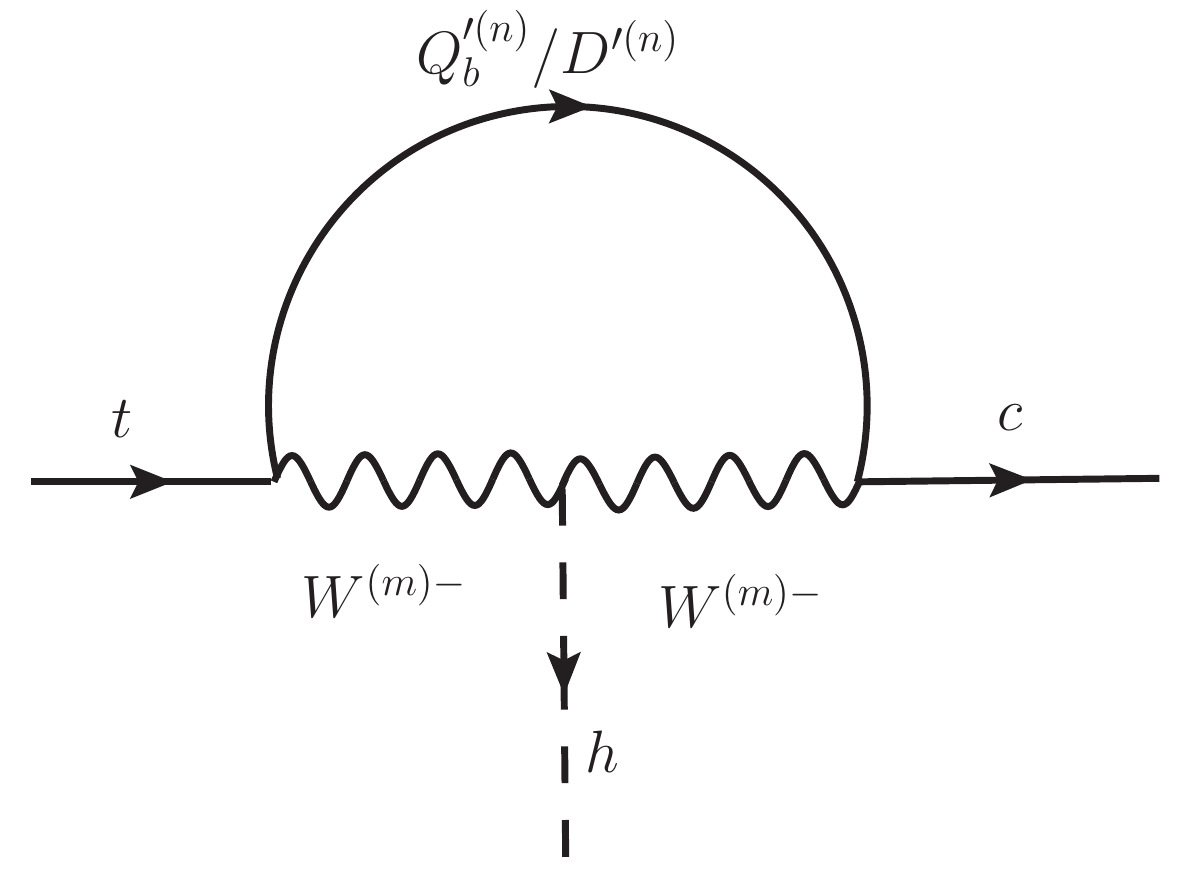}
    }
  \subfloat[]{
    \includegraphics[scale=0.35]{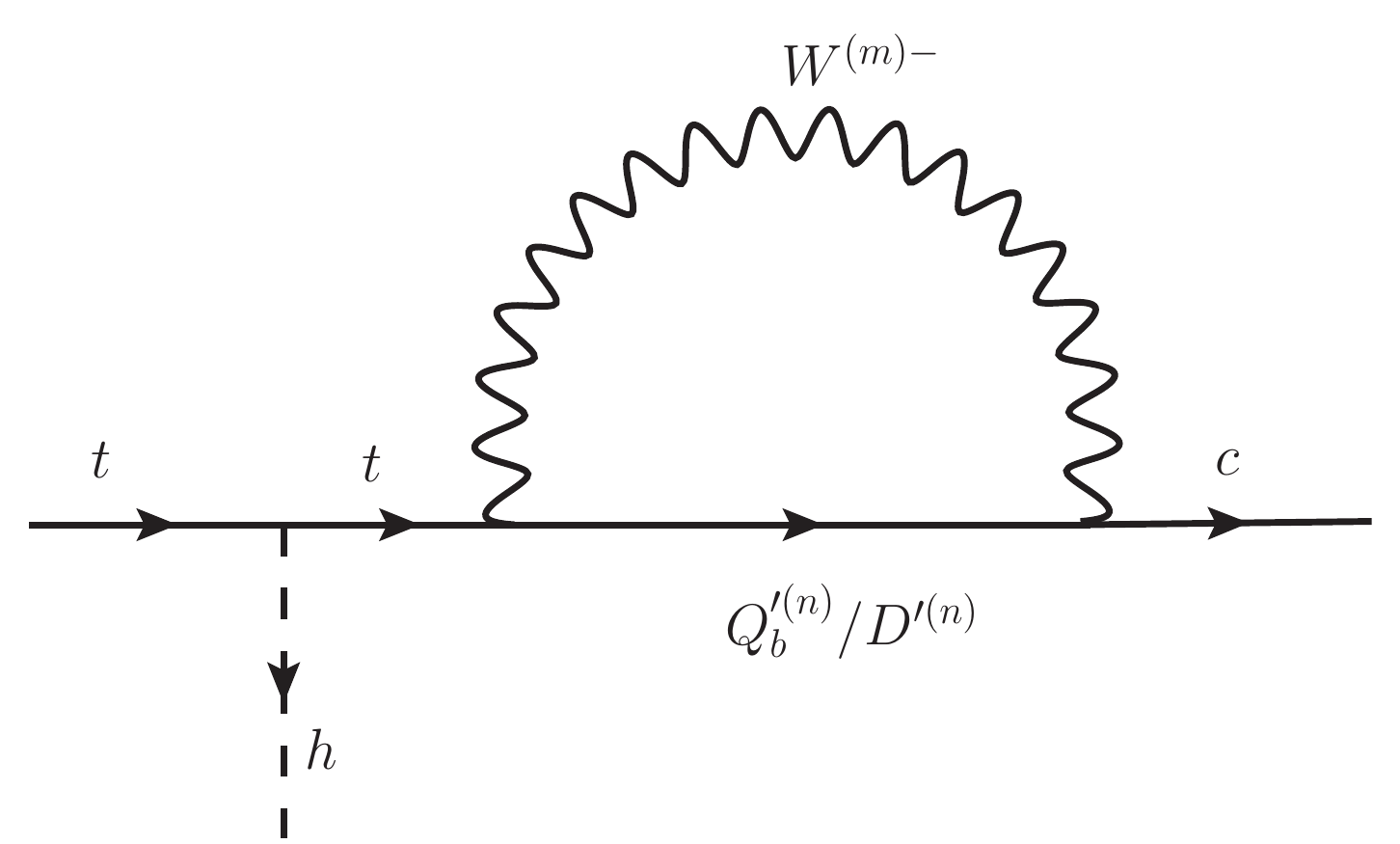}
    } \\
   \subfloat[]{
    \includegraphics[scale=0.35]{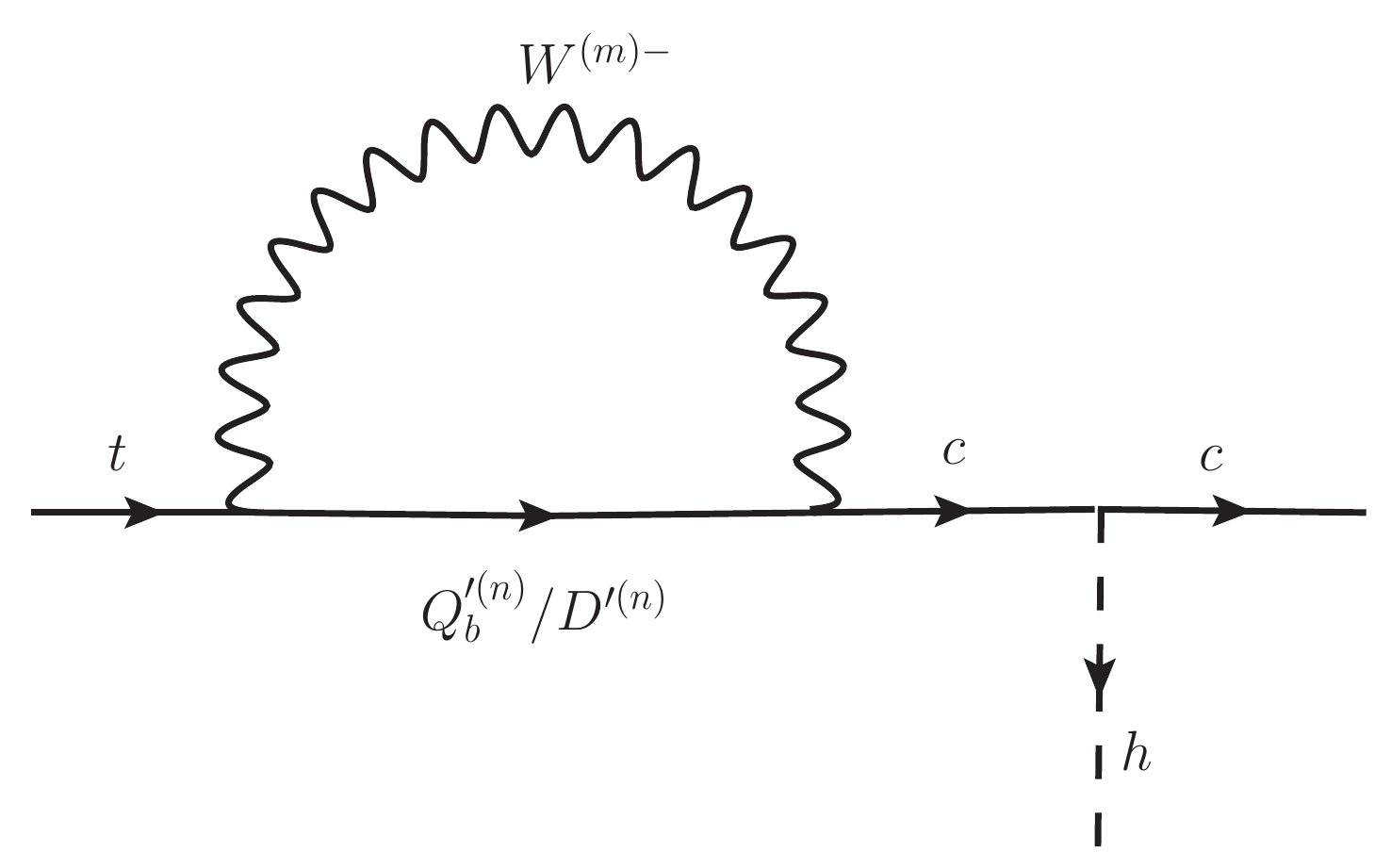}
    } 
    \caption[]{Feynman diagrams for the process $t\to ch$ in the 't Hooft--Feynman gauge in nmUED. Diagrams with vertices where $n$ is zero, \textit{i.e.,} diagrams with vertices with KK numbers $00m$ are also considered (but not shown) if they were allowed by the KK parity.}
\label{fig:tch}
\end{figure}

One of the other important rare decays of the top quark is its flavour violating decay to the charm quark and Higgs boson. The most general form of the amplitude of the decay $t(p)\to c(k_{2})h(k_{1})$ is given by
\begin{align}
\mathcal{M}(t\to ch) = \bar{u}(k_{2})\left[F_{S} + i\gamma_{5}F_{P}\right]u(p),
\end{align}
where the $F_{S}$ and $F_{P}$ are scalar and pseudo-scalar form factors, respectively. The assertions we made in the case of $t\to c\gamma$ regarding the divergence cancellation etc. hold true here also. Moreover, we keep the information of couplings, CKM elements, and loop momenta embedded in these form factors. 
It is straightforward to calculate the decay width for the process $t\to ch$ from the amplitude mentioned above. The decay width is given by
\begin{align}
\Gamma_{t\to ch} = \frac{1}{16\pi m_{t}^{3}}
                   &\sqrt{\left(m_{t}^{2}-(m_{c}+m_{h})^{2}\right)
                    \left(m_{t}^{2}-(m_{c}-m_{h})^{2}\right)} \nonumber 
                    \\
                    &\times \left(
                    \{(m_{c}+m_{t})^{2} - m_{h}^{2}\}|F_{S}|^{2} + 
                    \{(m_{c}-m_{t})^{2} - m_{h}^{2}\}|F_{P}^{\prime}|^{2}
                    \right),
\end{align}
where $F_{P}^{\prime} = iF_{P}$  and in the last piece, \textit{i.e.,} the form factor squared quantities, the KK sum is taken. Also, for $m_{c} = 0$ the two form factors are equal, \textit{i.e.,} $F_{S} = F_{P}^{\prime}$. The relevant Feynman diagrams for this process are shown in Fig.~\ref{fig:tch}. The KK indices $m$ and $n$ satisfy the same set of assertions mentioned in the previous Sec.~\ref{sbsc:tcgam}. 

\section{Results} 
\label{s:results}

\subsection{$t \to c\gamma$} 
\label{sbsc:results_tcgam}
Before presenting the mUED and nmUED results it is important to relook at the SM results. First of all the dominant decay mode of the top quark is to a bottom quark and a $W$ boson; this decay width is given by
\begin{align}
\Gamma_{t\to bW} = \frac{g^{2}}{64\pi}|V_{tb}|^{2}\left[
                    1 - 3\left(\frac{m_{W}}{m_{t}}\right)^{4} 
                      + 2\left(\frac{m_{W}}{m_{t}}\right)^{6} \right].
\end{align}
For $m_{W} = 80.39$ GeV, $m_{t} = 174.98$ GeV~\cite{Abazov:2014dpa}, $\Gamma_{t\to bW}$ is approximately $1.5$ GeV. This being the most prominent decay mode of the top quark any branching ratio can be given as
\begin{align}
\text{BR}(t\to XY) = \frac{\Gamma_{t\to XY}}{\Gamma_{t\to bW}}.
\end{align}
The SM prediction for the branching of $t\to c\gamma$ is
\begin{align}
\text{BR}(t\to c\gamma) = \left(4.6^{+1.2}_{-1.0} \pm 
                           0.4^{+1.6}_{-0.5} \right)\times 10^{-14},
\end{align}
where the first uncertainty is due to the uncertainty in bottom mass, the second from the CKM mixing angle uncertainties, and the third from the variation of the renormalization scale between $m_{Z}$ (+ve sign) and $1.5m_{t}$ (-ve sign)~\cite{AguilarSaavedra:2002ns}. Taking the pole mass of the $b$-quark to be $4.18$ GeV~\cite{Agashe:2014kda} our SM prediction for the $t\to c\gamma$ branching ratio is $2.4\times 10^{-13}$ and for the running mass $\bar{m}_{b}(m_{t}) = 2.74$ GeV the branching ratio becomes $5.18\times 10^{-14}$. Clearly the exact value of this decay width, as well as the other flavour violating decays of the top quark, is highly sensitive to the bottom quark mass, as has been pointed out in~\cite{AguilarSaavedra:2004wm}. For the numerical evaluations we have used \texttt{Package-X}~\cite{Patel:2015tea} and \texttt{LoopTools}~\cite{Hahn:1998yk}. 
From these numbers it is evident that the branching ratios for flavour changing top quark decays in the SM are exorbitantly suppressed, making the prospect of its detection at the LHC or even higher energetic FCC quite bleak. We discuss the present LHC reaches in our summary Sec.~\ref{s:concl}. On the other hand, as a positive side if any signature of these types of decays is found with a measurable amount of enhancement that must arise from some new physics beyond the SM.  

\subsubsection{mUED Results}
\label{sbsbsc:muedtcgam}
In the mUED scenario the loop-induced decay of the top quark to charm quark and photon gets additional contribution from the higher mode KK particles running in the loop. The representative Feynman diagrams are shown in Fig.~\ref{fig:tcgam}. Since in mUED, KK number is a conserved quantity the KK indices $m$ and $n$ in each vertex of the diagrams  respect this symmetry; to be more precise, in all the diagrams, for mUED at least, $m$ and $n$ should be equal. 
As has been mentioned in the model description, see Sec.~\ref{s:model}, the only relevant parameter for the mUED set-up is the inverse of the compactification radius $1/R$ and the masses of all the KK particles are dependent on this quantity. The important difference from the SM in mUED is basically the presence of KK counterparts of SM particles in the loop as well as the presence of charged KK scalars. Moreover, the mixing in the KK fermion sector plays an important role. 
\begin{figure}[H]
\begin{center}
\includegraphics[scale=0.5]{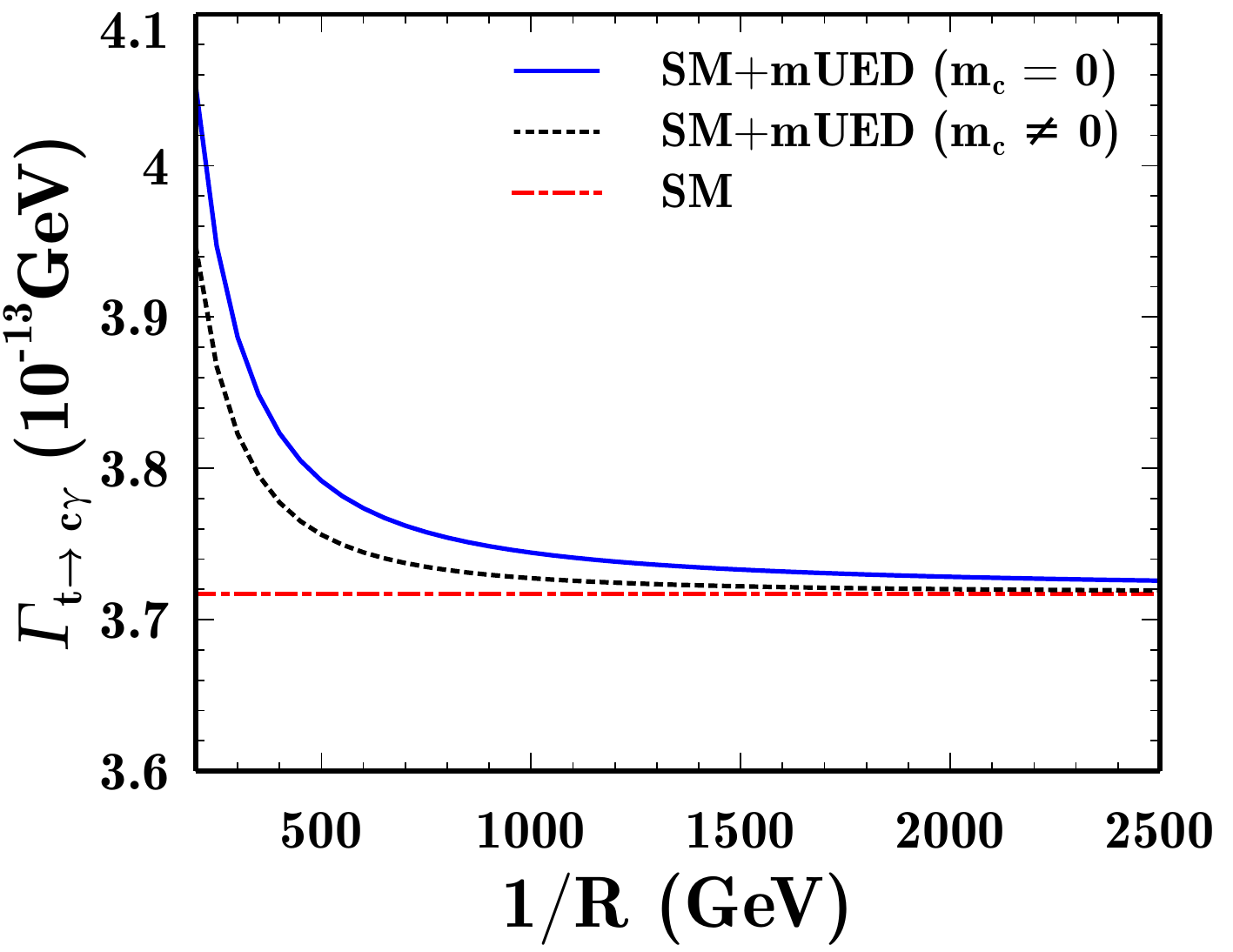}
\caption{The decay width of the process $t\to c\gamma$ as a function of the inverse compactification radius $1/R$ in the case of mUED. For the $m_{c}\neq 0$ case, the charm quark mass is taken to be 1.275 GeV.}
\label{f:tcgam_mued}
\end{center}
\end{figure}
Fig.~\ref{f:tcgam_mued} summarises the $t\to c\gamma$ decay width in mUED. We have calculated the decay width by taking the SM as well as the new physics, \textit{i.e.,} the mUED into account. Clearly for a higher value of the inverse compactification radius $1/R$ the masses of the KK modes become too heavy and they decouple, effectively making a negligible contribution. The red (dash-dotted) line shows the SM only value that we obtained using the pole mass of the $b$-quark. At the higher values of $1/R$ the convergence of the blue (solid), for $m_{c} = 0$,  and black (dotted), for $m_{c} \neq 0$, line with SM line only reflects the decoupling of the KK modes. We take $m_{c} = 1.275$ GeV~\cite{Agashe:2014kda} for the $m_{c}\neq 0$ case.
The lower values (less than $1$ TeV) of $1/R$ are disfavoured from the LHC data~\cite{Servant:2014lqa}. Moreover from Fig.~\ref{f:tcgam_mued} we see that even for the lower values of $1/R$ the order of magnitude of the decay width does not change much. Thus, one can conclude that the mUED set-up cannot enhance the branching ratio of $t\to c\gamma$ to any significant level from that of the SM value while remaining in the allowed ranges, obtained from the LHC, of the inverse compactification radius. However, the situation is different in the case of nmUED, as we see shortly. 

\subsubsection{nmUED Results}
\label{sbsbsc:nmuedtcgam}
The presence of BLKT parameters makes the situation quite different from the minimal scenario. It has already been mentioned that the BLKT parameters control the mass spectrum via the transcendental equation [see Eq.~\ref{eq:transc}], as well as the couplings via the appropriate overlap integrals. Like the mUED scenario, here also the loop-induced $t\to c\gamma$ process gets contributions from the higher KK excitations in the loop. But in this case the BLT parameters play a significant role in determining the masses of those particles running in the loops as well as the relevant couplings. One other important distinction from the mUED scenario is that in nmUED, KK number is no longer a conserved quantity, but still the conservation of KK parity holds due to the presence of the same BLKT parameters at the two orbifold fixed points, $y=0$ and $\pi R$. Consequently, unlike mUED, the couplings of particles with KK numbers 0-0-$n$, where $n$ is even, are present at tree-level. Thus, there are extra Feynman diagrams contributing in the process, \textit{e.g.,} in Fig.~\ref{fig:tcgam} the appropriate diagrams with $n$ being zero will also contribute.  
We have already discussed that in the most general scenario each field present in the model can have its own independent BLT parameters. However that would give rise to many new independent parameters. In this study, to keep the situation more tractable, we take universal BLT parameters $r_{f}$ for all the fermions and for gauge and scalar fields a common BLT parameter $r_{\Phi}$. As these BLT parameters are dimensionful parameters we use the dimensionless quantity $R_{X} = r_{X}/R$ when presenting our results. The parameters $R_{X}$ can, in principle, have small negative values but $R_{\Phi,f}~(\equiv r_{\Phi,f}/R) < -\pi$ can lead to tachyonic zero modes, as can be guessed from a cursory look at Eq.~\ref{eq:norm}. Moreover, $R_{\Phi} < -\pi$ can also lead to imaginary gauge couplings. In our study we strictly use only the positive values of these parameters.
We now discuss the results in the nmUED scenario. Note that we have two BLT parameters $R_{\Phi}$ and $R_{f}$ at our disposal. Thus we consider two cases, one being the universal BLT case, \textit{i.e.,} $R_{\Phi} = R_{f}$ making the same BLT for all types of fields and another being the case of $R_{\Phi} \neq R_{f}$. 
\begin{figure}[H]
\begin{center}
\includegraphics[scale=0.45]{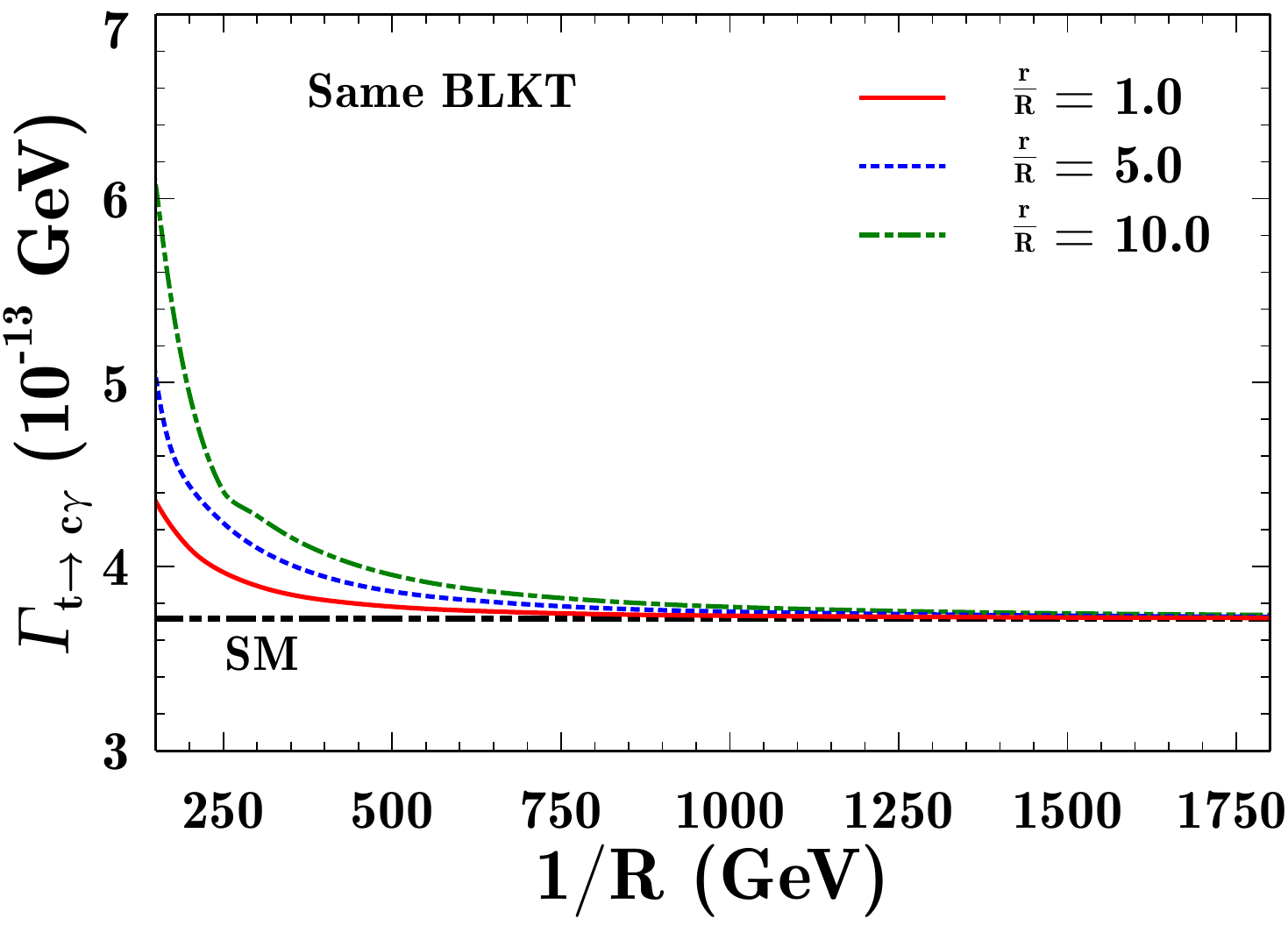}
\caption{The decay width of the process $t\to c\gamma$ as a function of the inverse compactification radius $1/R$ in the case of nmUED for different BLT parameters. In this case we consider a universal BLT parameter $r$.}
\label{f:tcgam_nmuedsame}
\end{center}
\end{figure}
First, take the case of universal BLT, \textit{i.e.,} $R_{\Phi} = R_{f}\equiv r/R$. Note that in this scenario all the overlap integrals that modify the couplings become unity by virtue of the orthonormalization conditions, see Eqs.~\ref{eq:overlapints} and \ref{eq:orthonorm}. Therefore, the effect of the common BLT parameter $r/R$ is only to determine the masses of the KK particles running in the loop. Clearly this situation is almost like the mUED but with the freedom that the KK masses can now be tuned with the BLT parameter $r/R$.  
In Fig.~\ref{f:tcgam_nmuedsame} we present the results for the same BLT case for different values of the parameter. The black (dash-dot-dot) line represents our SM value for the $t\to c\gamma$ decay width; the red (solid), blue (dotted), and green (dash-dotted) curves are for BLT parameter $r/R= 1.0$, $5.0$, and $10.0$ respectively. 
We see that the contribution from the KK particle decouples at lower values (compared to the mUED case) of the compactification radius for lower values of $r/R$. This is expected as the higher values of $r/R$ imply a lower (compared to the mUED case) KK mass for a specific value of the inverse compactification radius $1/R$. Clearly for a specific value of $1/R$, a higher value of the BLT parameter would lead to a lower KK mass than in the mUED thus making the propagator suppression less effective. Consequently it is evident that the higher value BLKT parameters will result in a decoupling for higher values of $1/R$.
However, like in mUED, in the universal BLT case also the value of the decay width does not change its order of magnitude from its SM value even for lower $1/R$. But unlike mUED a lower value of $1/R$ is not much constrained from LHC data in the nmUED scenario. To the best of our knowledge the only collider studies made in nmUED, to date are~\cite{Datta:2012tv, Datta:2013yaa}. A detailed study of nmUED in the light of LHC data is underway.
Now we take up the case of different BLTs, \textit{i.e.,} $R_{\Phi} \neq R_{f}$. Clearly in this case the KK excitations of fermions and the KK scalar/gauge bosons have different masses depending on their respective BLT parameters. Moreover, the couplings in this case get modified by the appropriate overlap integrals, mentioned in Eqs.~\ref{eq:overlapints}. The variation of the $t\to c\gamma$ decay width for various choices of BLT parameters is shown in Fig.~\ref{f:tcgam_nmueddiff}. Different choices of BLT parameters give rise to distinct features in the $1/R$ dependence of the decay width. Unlike the mUED or universal BLT scenario, here the total decay width (SM plus nmUED) can be smaller than the SM value for some choice of parameters. Moreover,  the higher values of the BLT parameters (see the figure in the lower right panel of Fig.~\ref{f:tcgam_nmueddiff}) can enhance the decay width by several orders of magnitude from the SM value, in the lower $1/R$ region. For example, for $R_{f} = 8$ and $R_{\Phi} = 15$ and $1/R = 500$ GeV the decay rate can be $\sim 4.1\times 10^{-12}$ GeV. But, for these sets of parameter values the masses of first KK level particles are less than 200 GeV, to be precise $\sim 190$ GeV for first KK level fermions and $\sim 140$ GeV for first KK level bosons; and the second levels are of $\sim 570$ GeV (fermions) and $\sim 540$ GeV (bosons). We elaborate on the implications of this in Sec.~\ref{s:stufcnc}. 
\begin{figure}[H]
\begin{center}
\includegraphics[scale=0.41]{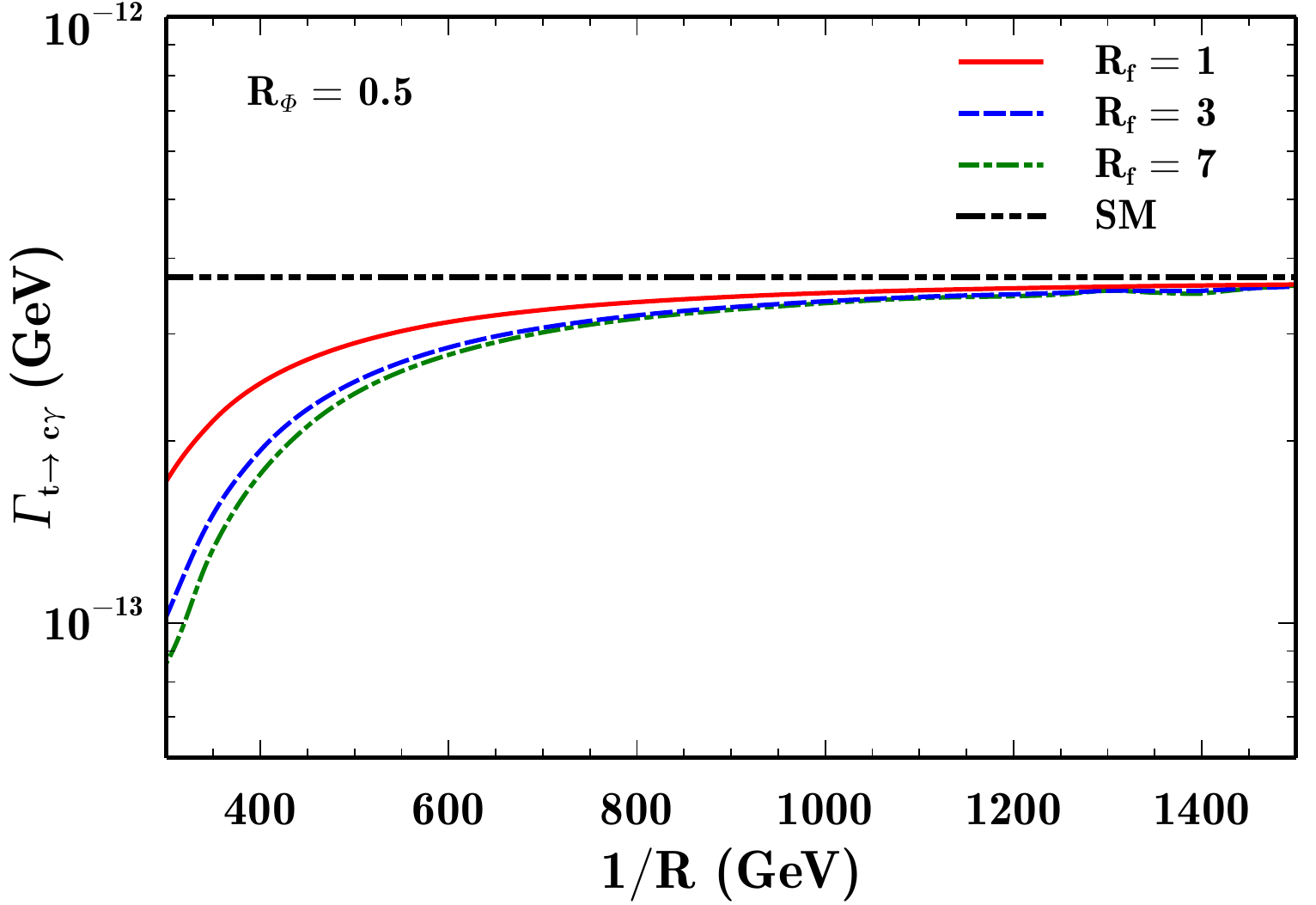}
~~~~
\includegraphics[scale=0.42]{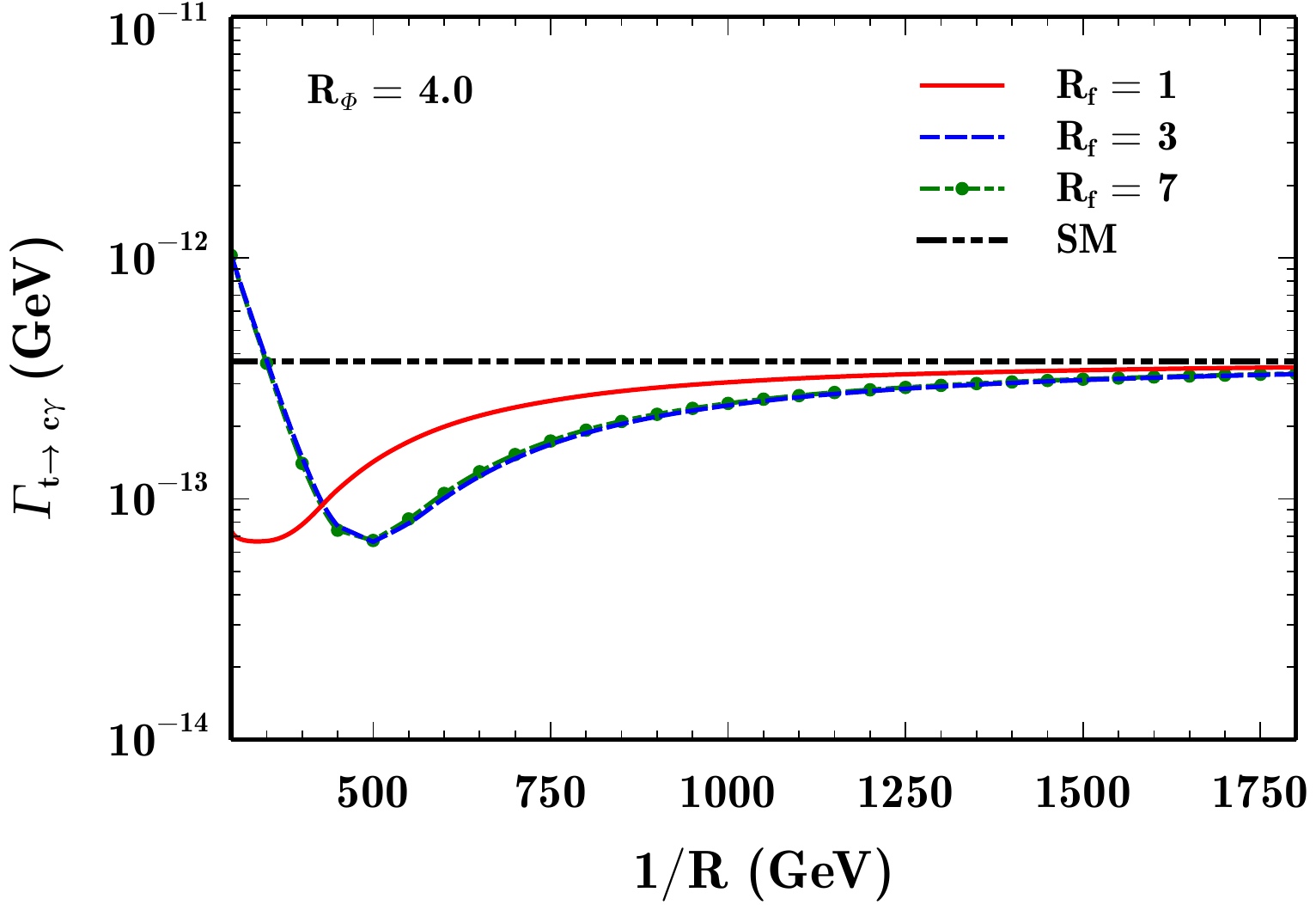} 
\\
\includegraphics[scale=0.42]{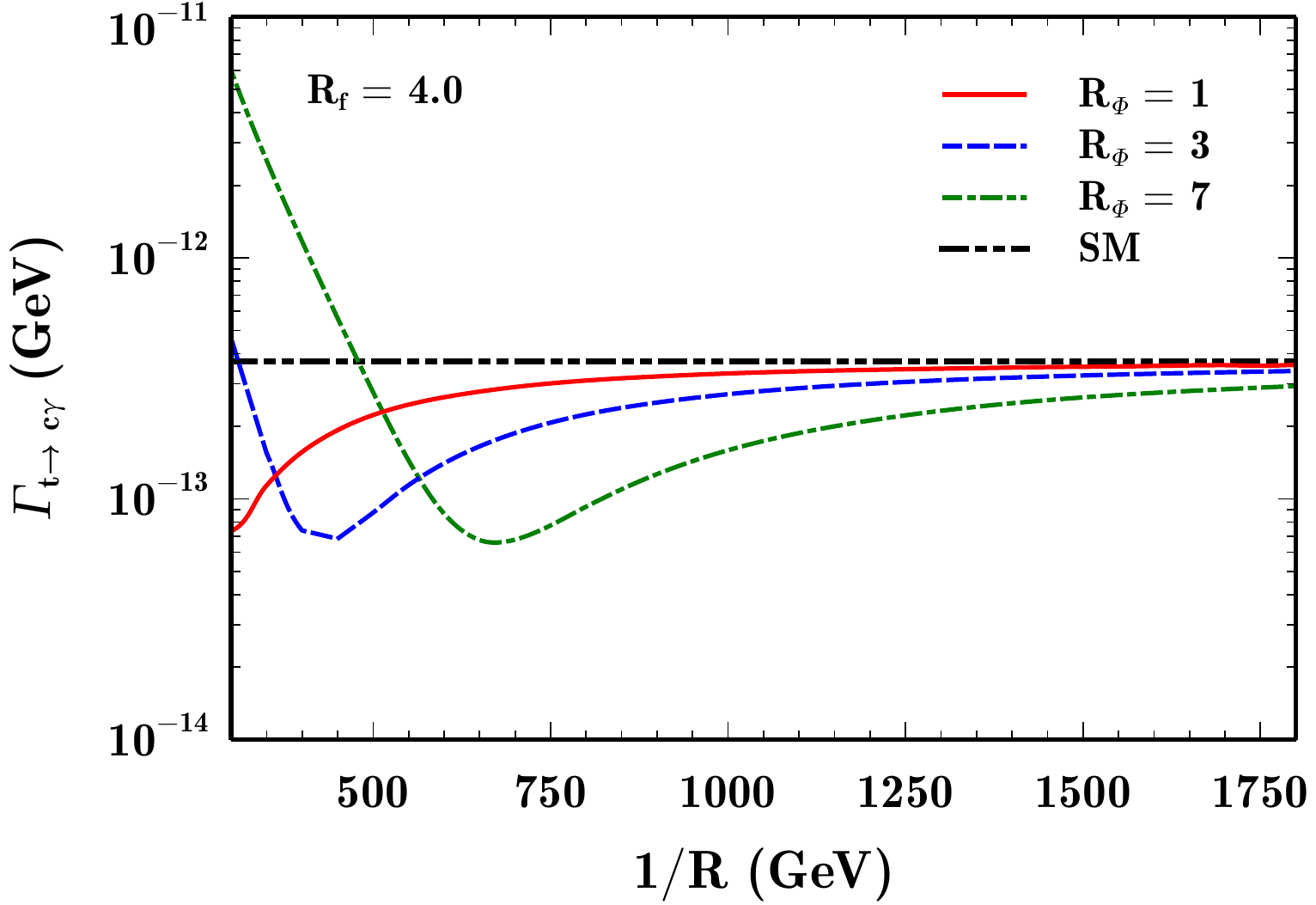}
~~~~
\includegraphics[scale=0.42]{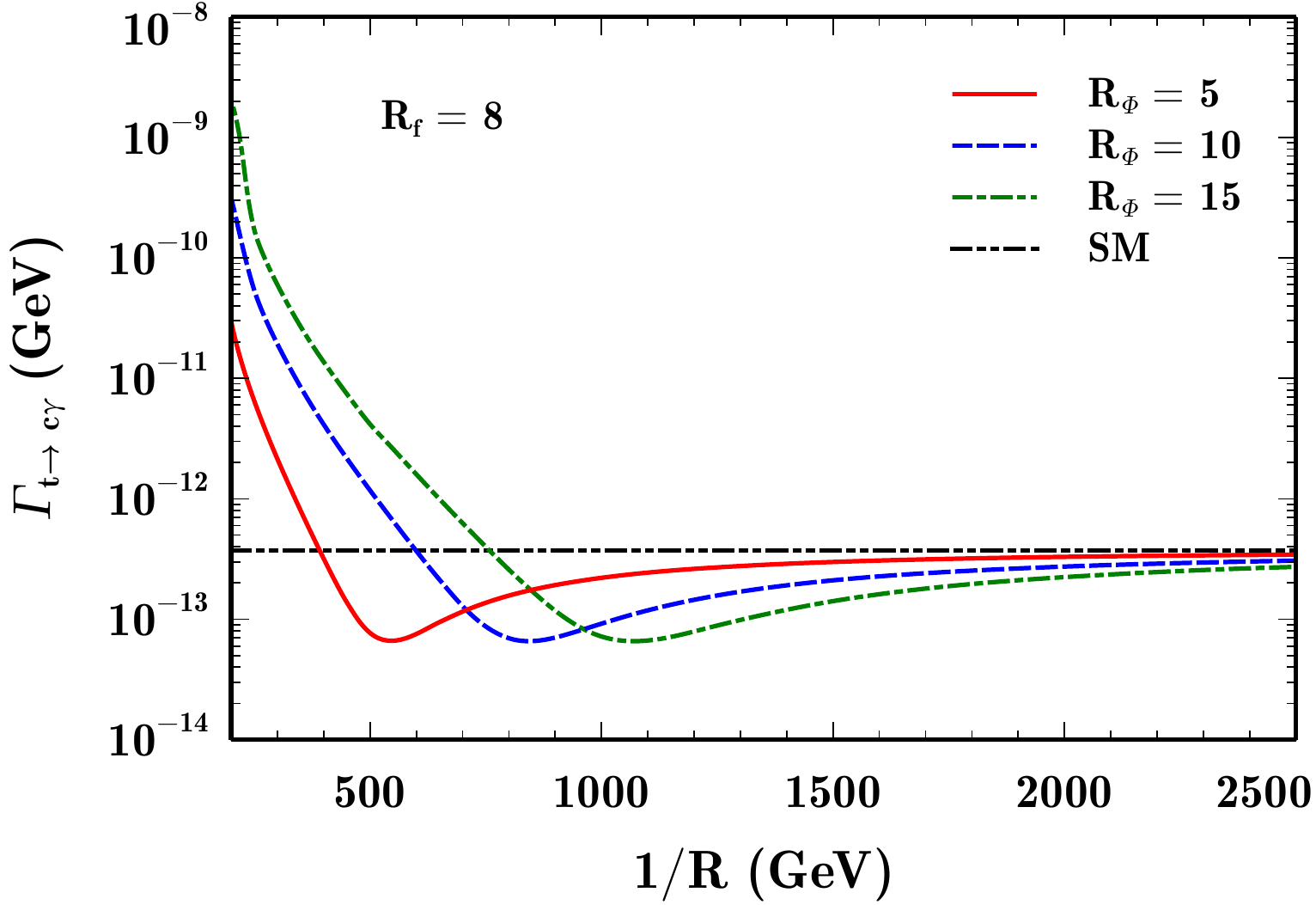} 
\caption{The decay width of the process $t\to c\gamma$ as a function of the inverse compactification radius $1/R$ in the case of nmUED for different values BLT parameters $R_{f}~(=r_{f}/R)$ and $R_{\Phi}~(=r_{\Phi}/R)$.}
\label{f:tcgam_nmueddiff}
\end{center}
\end{figure}
At this point it is worth mentioning that we find that the specific nature of the curves is mostly determined from the contribution from the Feynman diagrams that involve the KK excitations of the scalars. Moreover, as far as the nature of the curves are concerned, a na\"{i}ve interpolation to the same BLT scenario from the different BLT scenario is not possible. In the different BLT case there are overlap integrals that modify the appropriate couplings. These overlap integrals depend on the KK masses and, in some cases, on the difference of the KK masses between two different types of particles ({\it e.g.}, in our case some overlap integrals depend on the the difference between $M_{\Phi n}$ and $M_{Q n}$). Also, in the amplitude of the diagrams the differences in the physical masses (physical masses, $m_{X} = M_{X}/R$) play a significant role. Thus depending on the choice of different BLT parameters the difference between $m_{Q n}$ and $m_{\Phi n}$ can be positive or negative which, at the amplitude level, can positively or negatively contribute to the SM amplitude. So in the different BLT case, it is possible for some parameter region that the overall decay width be less than that of the SM prediction, leading to the specific nature of Fig. \ref{f:tcgam_nmueddiff}. The reverse is also possible, {\it e.g.}, we have checked that for $1/R = 600$ GeV, $R_{\Phi} = 6.0$, the KK contribution is always positive if $R_{f} < 0.58$ and in that case one can get the nature of Fig. \ref{f:tcgam_nmueddiff} similar to that of Fig. \ref{f:tcgam_nmuedsame}, {\it i.e.}, the curves show a monotonic increase as 1/R decreases. Clearly, if in future experiments the $t\to c\gamma$ decay width comes out to be a larger value than the SM calculations then the higher BLT scenario will be favored provided the constraints on the $1/R$ from other observations are met.
We end the discussion on the $t\to c\gamma$ decay width in nmUED with the caveat that in our study we take a common BLT parameter for fermions $R_{f}$ and for gauge/scalar fields a common $R_{\Phi}$ and thus the situation can be generalised by considering different types of BLT parameters for different fields and that eventually results in  richer details of this decay width.

\subsection{$t \to ch$} 
\label{sbsc:results_tch}
The loop-induced flavour changing top quark decay to charm quark and Higgs boson was first calculated in~\cite{Eilam:1990zc}. For $m_{t}\simeq 175$ GeV and $m_{h}\in [40~\text{GeV},2m_{W}]$, according to~\cite{Eilam:1990zc}, $\text{BR}(t\to ch) \simeq 10^{-7}-10^{-8}$. However, this result was erroneous, which was subsequently pointed out and corrected in~\cite{Mele:1998ag}. According to this, for $m_{t} = 175$ GeV, $m_{c} = 0$, $m_{b} = 5$ GeV, and $m_{h} = 120$ GeV,
\begin{align}
\text{Br}(t\to ch) = 4.605 \times 10^{-14}.
\end{align}
The SM prediction for the same process has recently been calculated in~\cite{AguilarSaavedra:2004wm, Abbas:2015cua, Bardhan:2016txk} and according to  these references,\footnote{The branching ratio is $3 \times 10^{-15}$ in~\cite{AguilarSaavedra:2004wm}, $(3.00\pm 0.17)\times 10^{-15}$ in~\cite{Abbas:2015cua}, $5.8 \times 10^{-15}$ in~\cite{Bardhan:2016txk} etc.} the branching ratio is $\sim 10^{-15}$. Again, the root of all these differences in the exact value of the branching ratio is that the choice of values of various SM parameters, most crucially the value of $m_{b}$, differs in each studies. For $m_{h} = 125$ GeV and taking the pole mass of the $b$-quark to be $4.18$ GeV our SM prediction for the $t\to ch$ branching ratio is $1.99\times 10^{-14}$ and for the running mass $\bar{m}_{b}(m_{t}) = 2.74$ GeV the branching ratio becomes $3.63\times 10^{-15}$. We again emphasize that the exact estimation of the decay width is highly sensitive to the $b$-quark mass, $m_{b}$. After this discussion we present our results in mUED and nmUED.

\subsubsection{mUED Results}
\label{sbsbsc:muedtch}
The relevant Feynman diagrams for the process $t\to ch$ in the case of mUED can be found in Fig.~\ref{fig:tch}. Again, the KK indices $m$ and $n$ have to be taken appropriately maintaining the conservation of KK number. Also, the other details, inherent to the model itself, remain the same as discussed in the case of $t\to c\gamma$; see Sec.~\ref{sbsbsc:muedtcgam}.    

\begin{figure}[H]
\begin{center}
\includegraphics[scale=0.5]{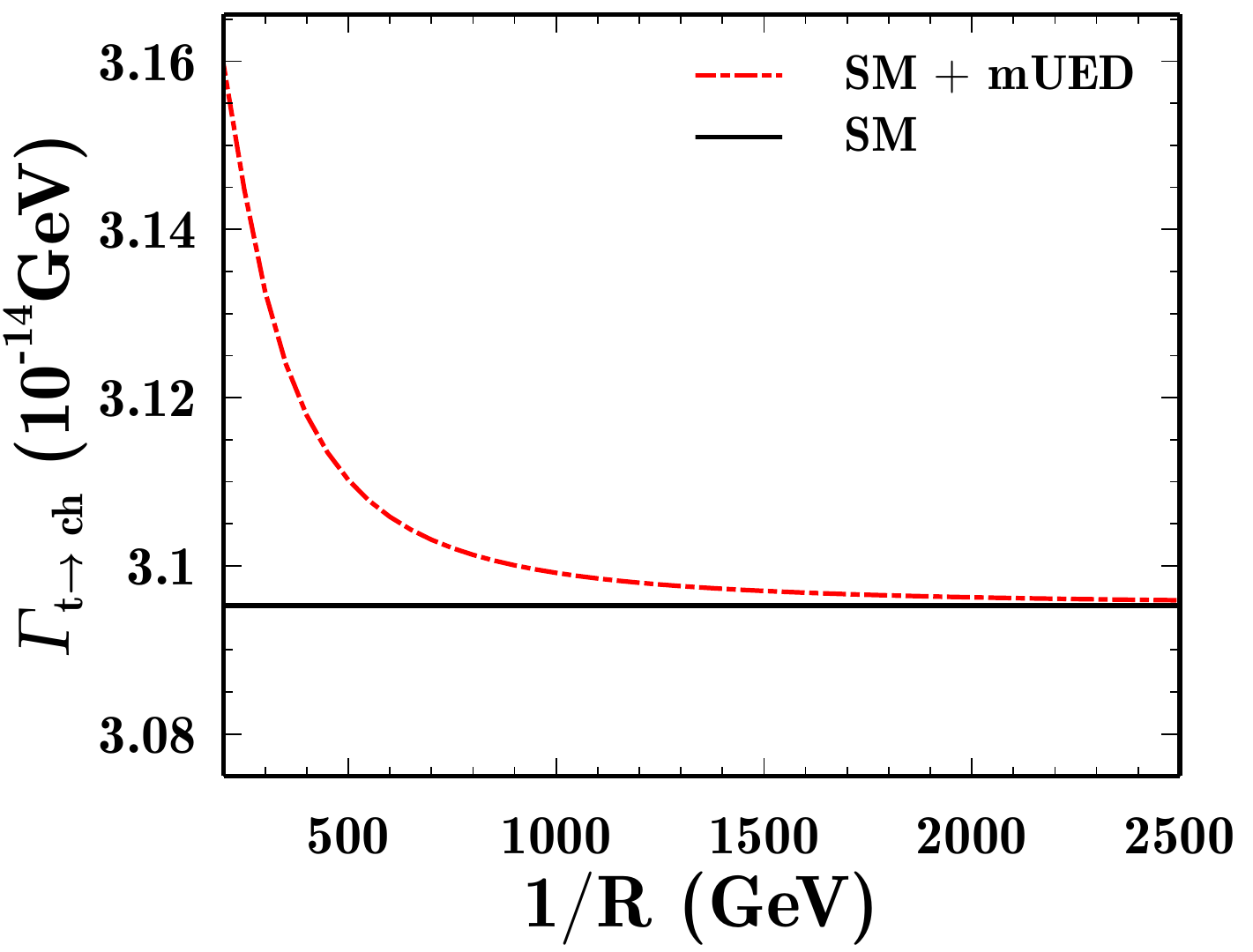}
\caption{The decay width of the process $t\to ch$ as a function of the inverse compactification radius $1/R$ in the case of mUED.}
\label{f:tch_mued}
\end{center}
\end{figure}

In Fig.~\ref{f:tch_mued} we present our results for the decay width of $t\to ch$ in mUED. Like the $t\to c\gamma$ case, here also we take the pole mass of the $b$-quark. In this figure, the black (solid) line represents our SM value of the decay width and the red (dash-dotted) curve is for the decay width in the case of SM combined with the mUED spectrum. In this case also we find no order of magnitude enhancement of the branching ratio for any reasonable values of $1/R$. The situation is almost similar in the nmUED scenario also, as we discuss in the next subsection.

\subsubsection{nmUED Results}
\label{sbsbsc:nmuedtch}
\begin{figure}[H]
\begin{center}
\includegraphics[scale=0.45]{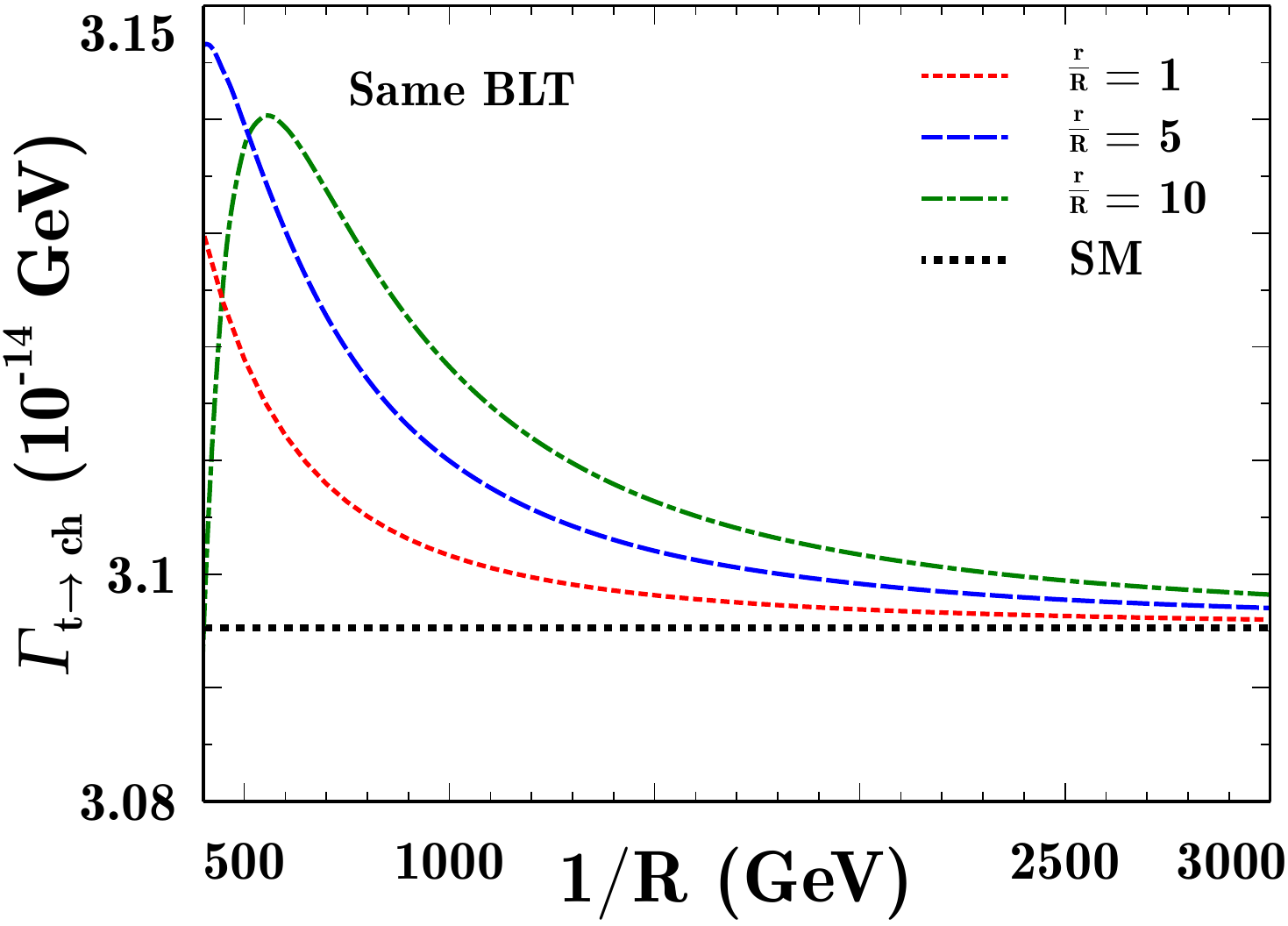}
\caption{The decay width of the process $t\to ch$ as a function of the inverse compactification radius $1/R$ in the case of nmUED for different BLT parameters. In this case we consider a universal BLT parameter $r$.}
\label{f:tch_nmuedsame}
\end{center}
\end{figure}

In Fig.~\ref{f:tch_nmuedsame} we present the results for the universal BLT scenario, \textit{i.e.,} $R_{\Phi} = R_{f}\equiv r/R$. For this case also we find that the value of decay width is of the same order as that of the SM for all choices of $r/R$. 

\begin{figure}[h]
\begin{center}
\includegraphics[scale=0.41]{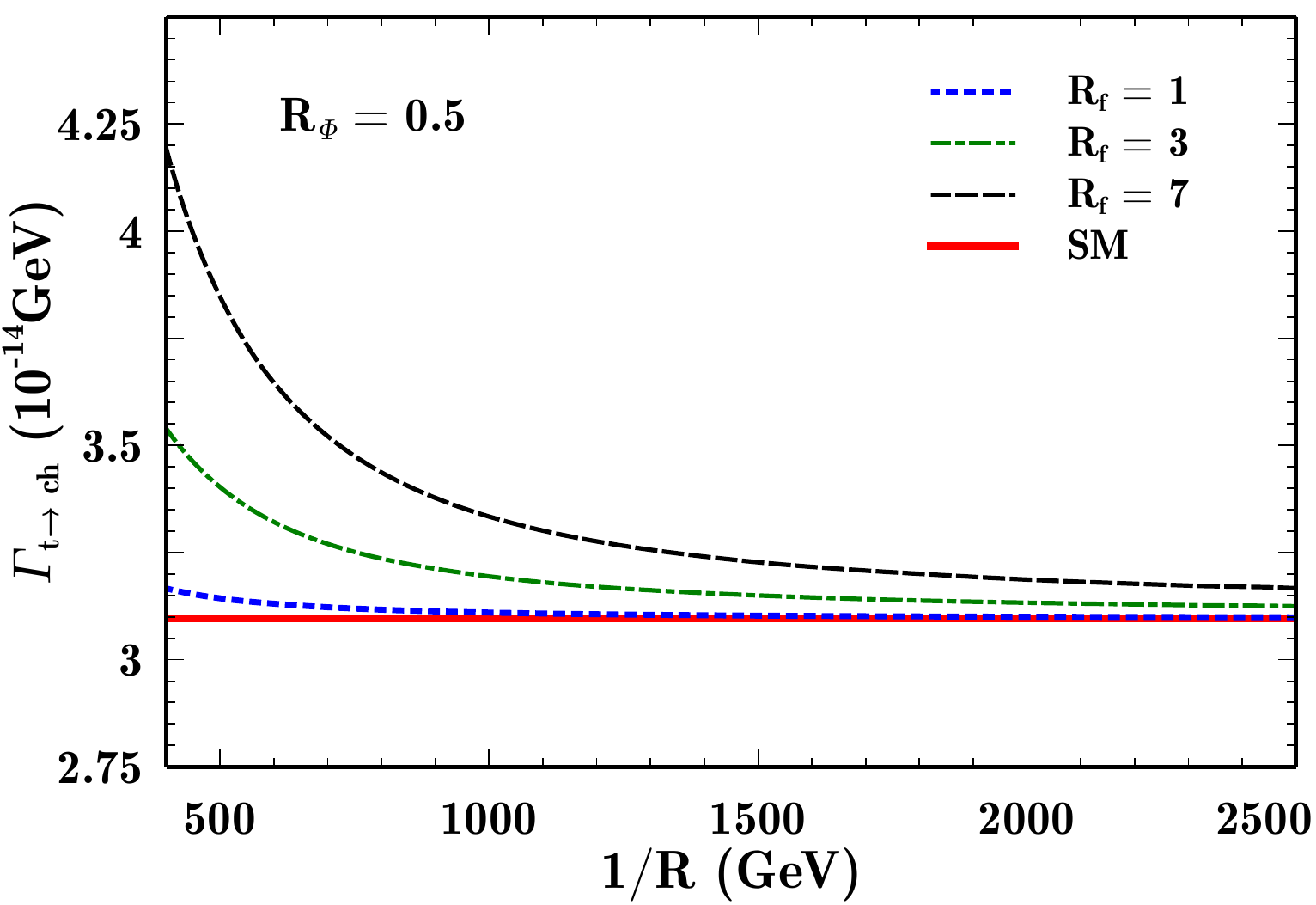}
~~~~
\includegraphics[scale=0.42]{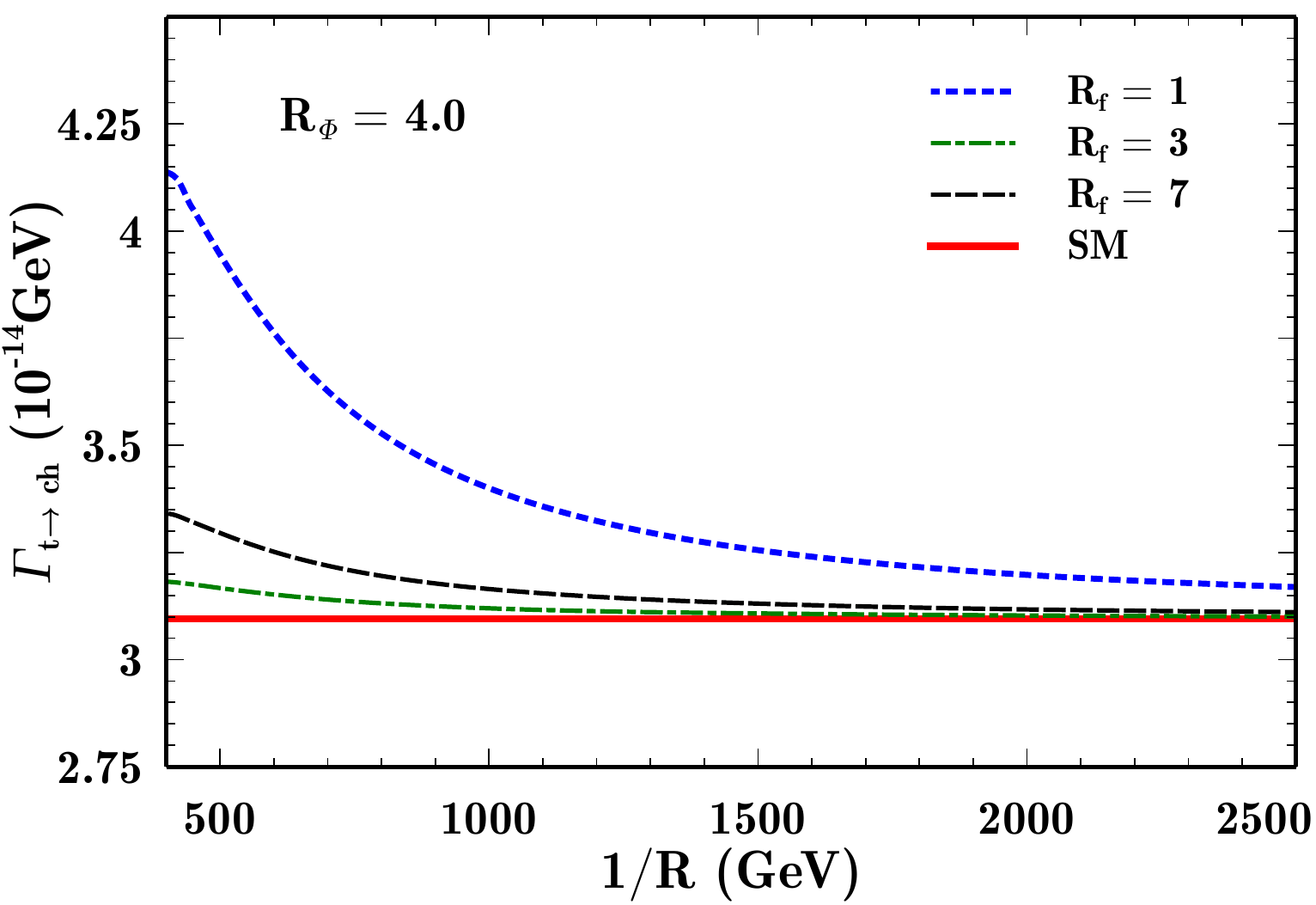} 
\\
\includegraphics[scale=0.42]{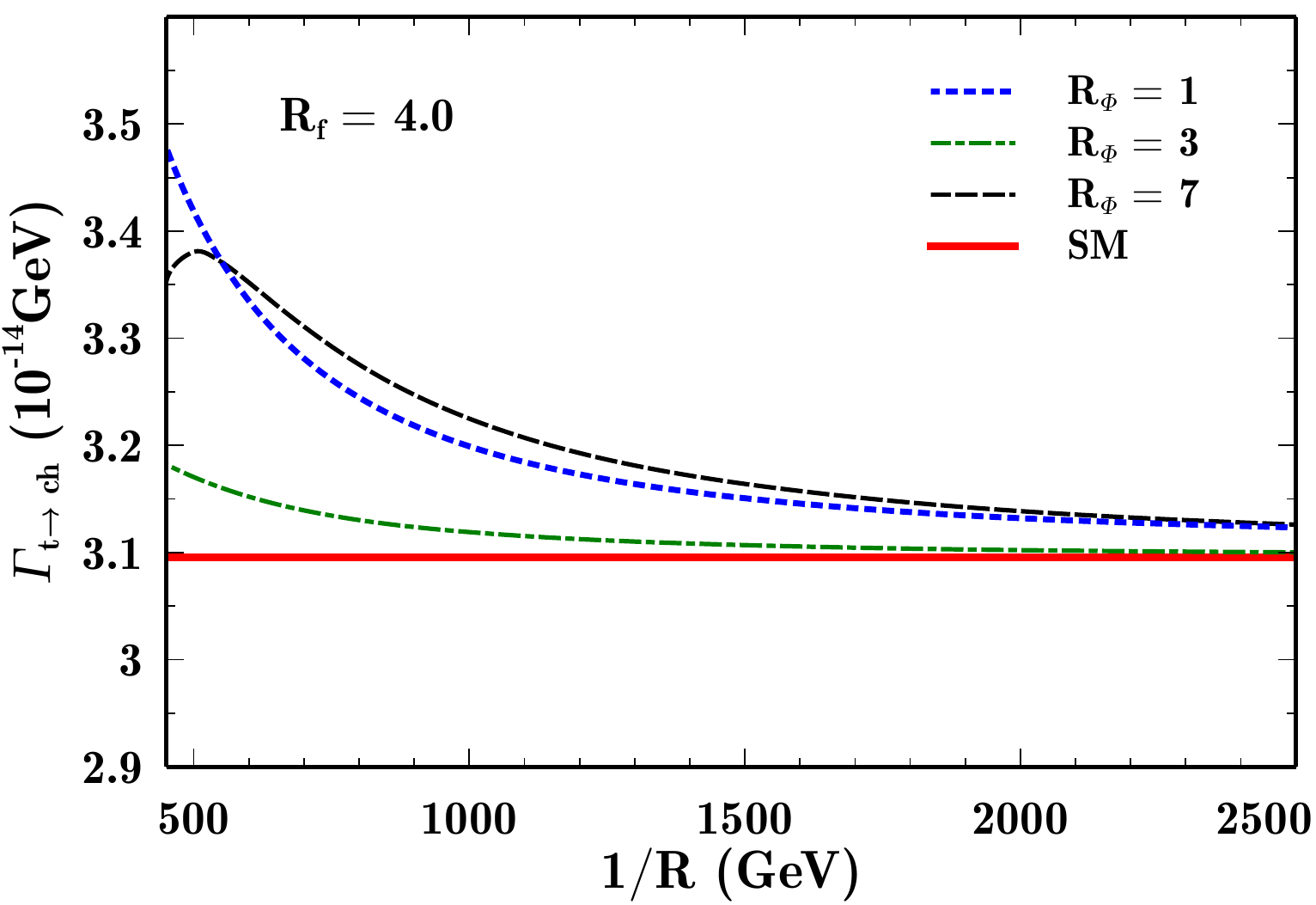}
~~~~
\includegraphics[scale=0.42]{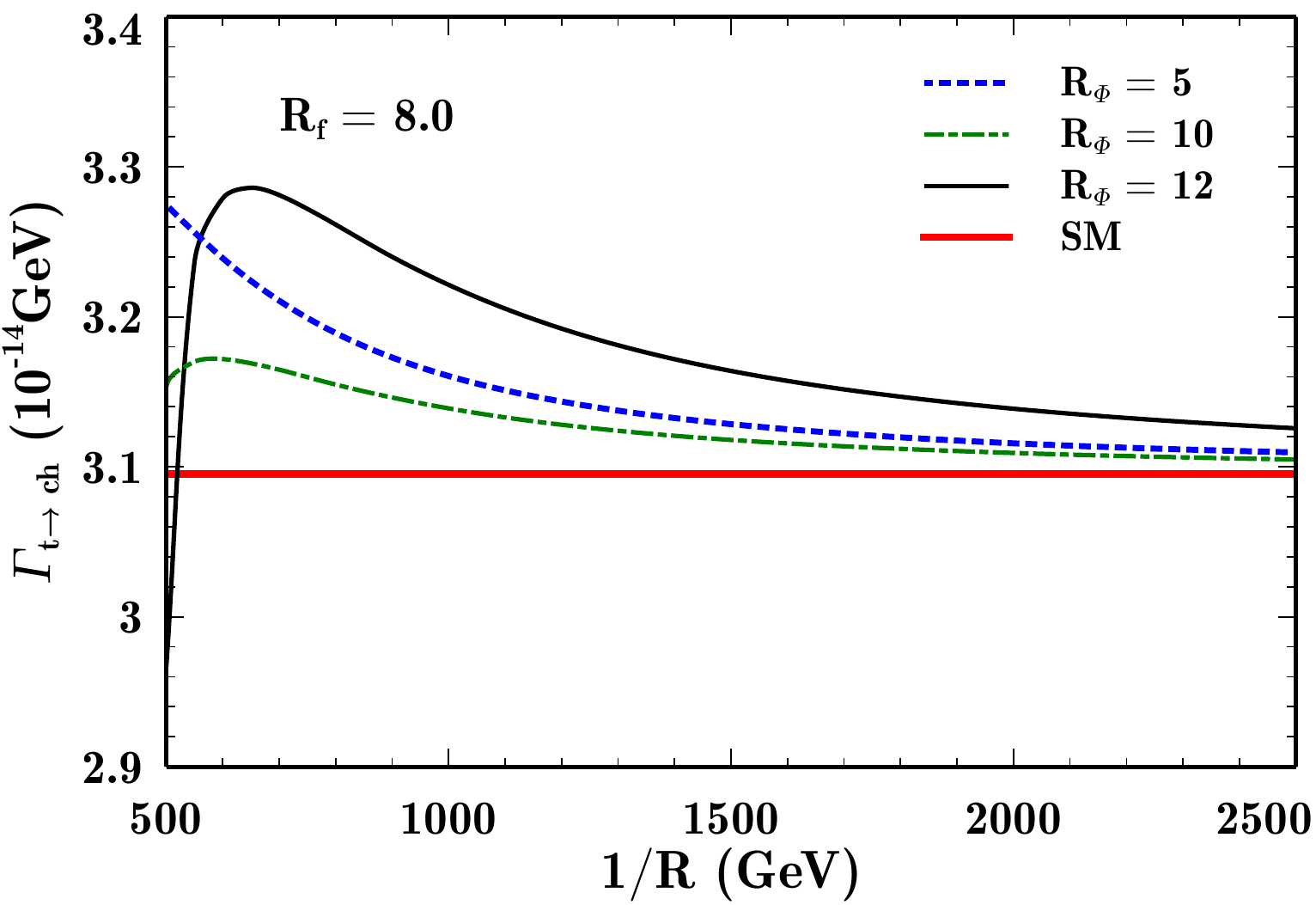} 
\caption{The decay width of the process $t\to ch$ as a function of the inverse compactification radius $1/R$ in the case of nmUED for different values of BLT parameters $R_{f}~(=r_{f}/R)$ and $R_{\Phi}~(=r_{\Phi}/R)$.}
\label{f:tch_nmueddiff}
\end{center}
\end{figure}
The same thing happens in the distinct BLT ($R_{\Phi} \neq R_{f}$) case also, as can be seen from Fig.~\ref{f:tch_nmueddiff}, where we plotted the $t\to ch$ decay width for various choices of BLT parameters. In this case also a very high value of BLT parameters can significantly enhance the decay rate.

\section{$S,T,U$ Parameters, FCNC and Other Issues}
\label{s:stufcnc}
\begin{figure}[!htbp]
\begin{center}
\includegraphics[scale=0.75]{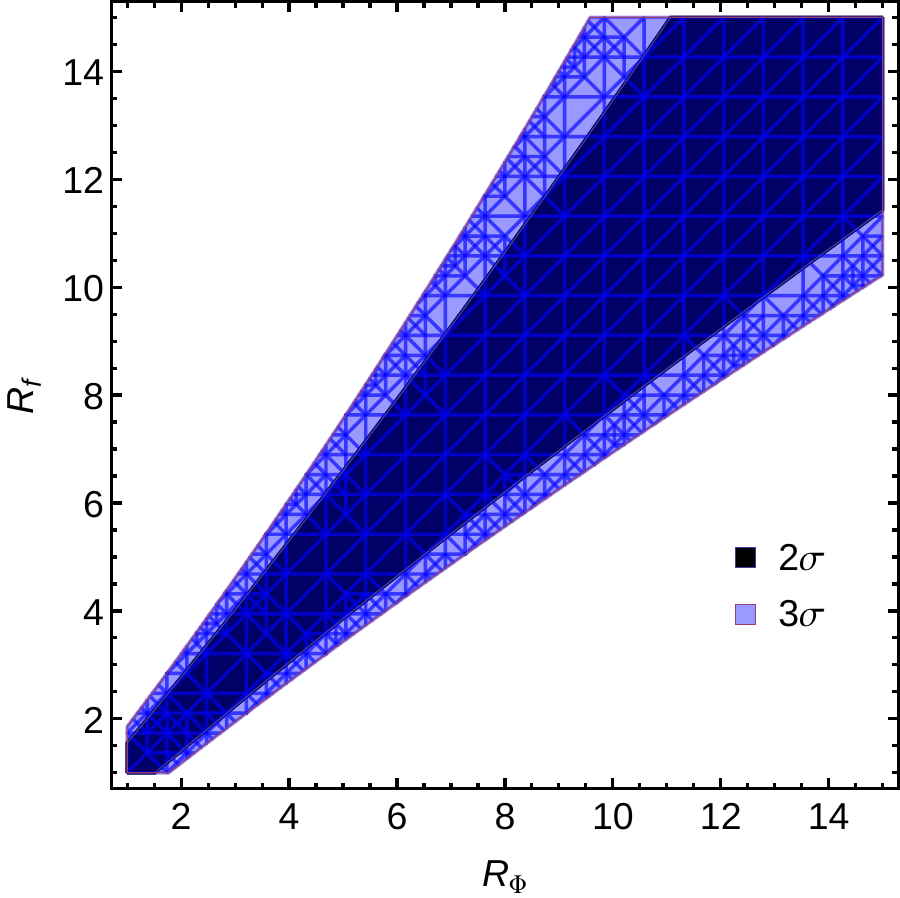}~~~~~
\includegraphics[scale=0.75]{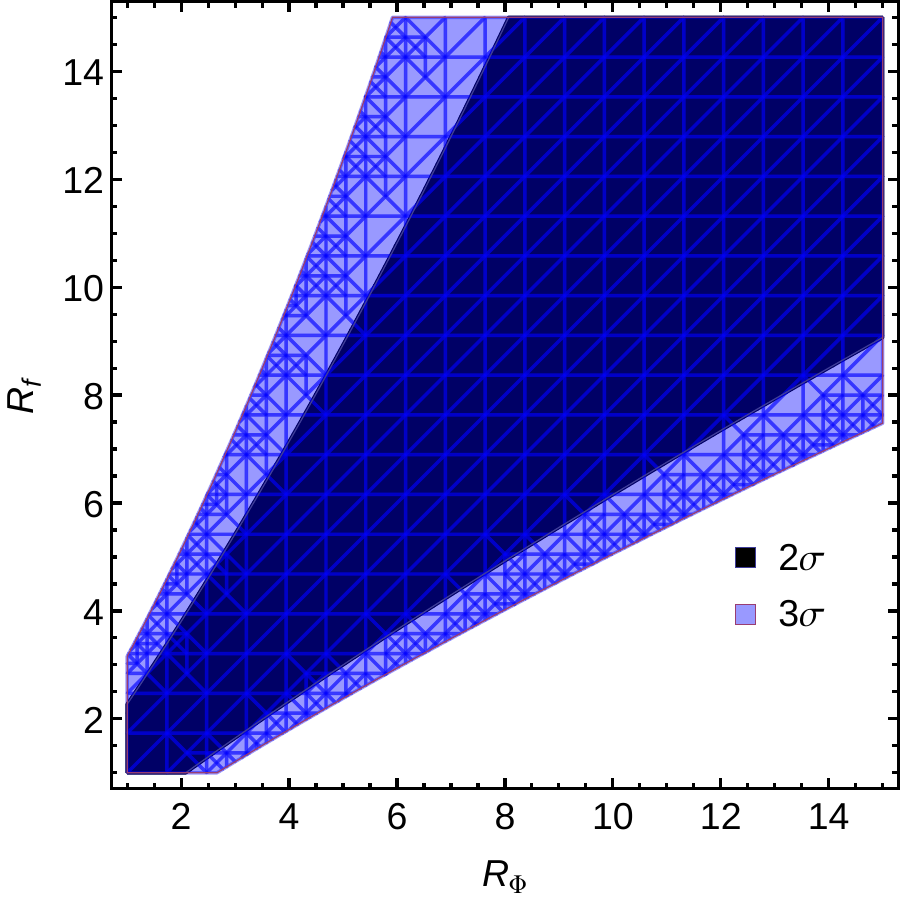}
\caption{The shaded region, in the $R_{\Phi}-R_{f}$ plane, represents the  $2\sigma$ (dark blue) and $3\sigma$ (light blue) deviation region satisfying the electroweak precision constraints for $1/R=$500 GeV (left) and 1000 GeV (right).}
\label{f:stu}
\end{center}
\end{figure}

The Peskin-Takeuchi parameters, \textit{i.e.,} $S,T$ and $U$ parameters, encode the oblique corrections to the electroweak gauge boson propagators~\cite{Peskin:1991sw}. These parameters put stringent constraints on many BSM physics scenarios. In nmUED these electroweak precision constraints are discussed in~\cite{Flacke:2008ne, Flacke:2013pla, Datta:2013yaa, Datta:2015aka}. Clearly in these cases the underlying action and some assumptions are different from our setup. For completeness we spell out these constraints in our case.
The effects on electroweak precision observables arise due to modifications to the Fermi constant $G_{F}$ at tree-level. Actually in nmUED second level KK gauge bosons have tree-level couplings with SM fermions, and this modifies the effective four Fermi interactions and thus the $G_{F}$. Note that this is in contrast with the mUED scenario where there is no 2-0-0 coupling at the tree-level. The corrected Fermi constant in the case of nmUED can be given as
\begin{align}
G_{F} = G_{F}^{0} + \delta G_{F},
\end{align} 
where $G_{F}^{0}$ ($\delta G_{F}$) comes from the $s$-channel SM (even KK) $W^{\pm}$-boson exchange. More precisely they can be written as~\cite{Datta:2015aka}
\begin{align}
G_{F}^{0} = \frac{g_{2}^{2}}{4\sqrt{2}M_{W}^{2}} ~~\mbox{and}~~
\delta G_{F} = \sum_{\substack{k\geq 2\\ \text{even}}} \frac{g_{2}^{2}}{4\sqrt{2}M_{W^{(k)}}^{2}} (\sqrt{r_{\Phi} + \pi R}~I_{3}^{k})^{2},
\end{align}
where $I_{3}^{k}$ is given in Eq.~\ref{eq:ovint3}. Now, in terms of these quantities the electroweak precision observables can be written as~\cite{Datta:2013yaa, Datta:2015aka}
\begin{align}
S_{nmUED} = 0,~~ T_{nmUED} = -\frac{1}{\alpha}\frac{\delta G_{F}}{G_{F}},
~~ U_{nmUED} = \frac{4\sin \theta_{W}^{2}}{\alpha}
                \frac{\delta G_{F}}{G_{F}}.
\end{align}
Now, the most recent fit to the electroweak precision data gives~\cite{Baak:2014ora}
\begin{align}
S = 0.05 \pm 0.11,~~~~T = 0.09 \pm 0.13, ~~~~ U = 0.01 \pm 0.11,
\end{align}
from which we write
\begin{align}
\hat{S} = 0.05,~~~\sigma_{S} = 0.11, \nonumber \\
\hat{T} = 0.09,~~~\sigma_{T} = 0.13, \nonumber \\
\hat{U} = 0.01,~~~\sigma_{U} = 0.11~. \nonumber
\end{align}
The $S,T$ and $U$ parameters are not independent parameters but are correlated. The correlation coefficients are given by~\cite{Baak:2014ora}
\begin{align}
\rho_{ST} = 0.90,~~~~\rho_{SU} = -0.59,~~~~\rho_{TU} = -0.83.
\end{align}
Now, constraints from $S,T,U$ parameters can be imposed by evaluating the $\chi^{2}$, given by
\begin{align}
\chi^{2} = \mathcal{X}^{T}\mathcal{C}^{-1}\mathcal{X},
\end{align} 
where $\mathcal{X}^{T} = (S_{nmUED}-\hat{S},T_{nmUED}-\hat{T},U_{nmUED}-\hat{U})$ and the covariance matrix $\mathcal{C}$ is given by
\begin{align}
\mathcal{C} = \begin{pmatrix}
\sigma_{S}^{2} & \sigma_{S}\sigma_{T}\rho_{ST} & \sigma_{S}\sigma_{U}\rho_{SU} \\
\sigma_{S}\sigma_{T}\rho_{ST} & \sigma_{T}^{2} & \sigma_{T}\sigma_{U}\rho_{TU} \\
\sigma_{U}\sigma_{S}\rho_{SU} & \sigma_{U}\sigma_{T}\rho_{TU} & \sigma_{U}^{2}
\end{pmatrix}.
\end{align}
For a maximal $2\sigma$ ($3\sigma$) deviation, given the two degrees of freedom, we need $\chi^{2}\leq 6.18$ (9.21)~\cite{Agashe:2014kda}. In Fig.~\ref{f:stu} we show the allowed region of parameter in the $R_{\Phi}-R_{f}$ plane consistent with electroweak precision data at  $2\sigma$ (and $3\sigma$) deviation for inverse compactification radius, $1/R$ = 500 and 1000 GeV. Note that the larger values of BLT parameters $R_{\Phi,f}$ lead to a larger allowed parameter space that is in agreement with the result shown in Fig.~11 of Ref.~\cite{Datta:2013yaa}. We mention that the dominant effect on the electoweak precision observables comes from the modification of the Fermi constant. From the one-loop contribution of KK particles another set of subdominant modification in the precision observables results. However, a detailed one-loop contribution from the KK particles in the case of nmUED is subject to further study. 

Normally BSM models suffer from the presence of tree-level FCNCs and appropriate symmetry etc. are imposed to get rid of them. In the most general setup of nmUED every field present in the model can have different BLT parameters. If there are different BLT parameters for fermions of different flavours then that leads to tree-level FCNCs. However, it has been shown in~\cite{Flacke:2013pla} that there exist no FCNCs if the fermion BLT parameters are flavour-blind, \textit{i.e.,} the BLT matrices are proportional to the unit matrix in flavour space. In our case we have used a universal BLT parameter for all fermions; clearly there is no tree-level FCNC in our present setup, no matter what the values of BLT parameters are. 
We have seen, at least in the case of $t\to c\gamma$, that somewhat large values of BLT parameters can result in an order of magnitude increment to the decay rate as compared to the SM. As it stands larger BLT parameters lead to smaller KK masses. In this regard a few points are in order. Presently the bounds on new physics particles are quite high as can be seen from the exotic particle searches of ATLAS and CMS~\cite{atlasExotic, cmsExotic}. In the case of mUED, LHC dilepton searches put constraints on second KK level particles to be $m_{KK^{(2)}}\geq 1.4$ TeV~\cite{Edelhauser:2013lia}. Now, for the BLT parameters and $R^{-1}$ that lead to a larger $\Gamma_{t\to c\gamma}$ compared to the SM the second level KK particles become much lighter. Clearly, a qualitative comparison with~\cite{Edelhauser:2013lia} shows that a lighter KK particle implies significant propagator enhancement leading the dilepton cross section to a degree that is ruled out by LHC dilepton data. We have checked that even the modification in couplings via the overlap integrals is not enough to evade these bounds. Thus the parameter space leading to an apparent enhancement in $\Gamma_{t\to c\gamma}$ is already ruled out from LHC dilepton searches. 
In passing we mention that the collider signatures of nmUED can mimic supersymmetry (SUSY). However, one must remember that $n=1$ KK masses in nmUED are more closely spaced than the masses of SUSY partners in any conventional SUSY models. This hinders one from directly translating available bounds on conventional SUSY models to nmUED models. So recasting the SUSY bounds on the nmUED parameters is an altogether different project and is beyond the scope of this work.   

\section{Summary and Conclusions} 
\label{s:concl}
We have performed a complete one-loop calculation of flavour-changing top quark decays ($t\to c\gamma$ and $t\to ch$) in the context of minimal and non-minimal UED. We have also verified , in the SM, the results of branching ratios of these decays with the existing literature. As far as the experimental searches are concerned, both ATLAS and CMS collaborations have performed some searches of FCNC top decays. For example, using the 19.6 fb$^{-1}$ data at $\sqrt{s} = 8$ TeV the CMS collaboration puts an upper bound on the rare decay to $c\gamma$ as Br$(t\to c\gamma) < 0.182\%$~\cite{CMS-PAS-TOP-14-003}. On the other hand for the $t\to ch$ channel, the ATLAS collaboration puts a bound of Br$(t\to ch) < 0.51\%$ using 4.7 fb$^{-1}$ data at $\sqrt{s} = 7$ TeV and 20.3 fb$^{-1}$ at $\sqrt{s} = 8$ TeV~\cite{Aad:2014dya}; whereas according to the CMS collaboration Br$(t\to ch) < 0.56\%$ by using 19.5 fb$^{-1}$ data at $\sqrt{s} = 8$ TeV~\cite{CMS-PAS-HIG-13-034}. Also see~\cite{Agashe:2013hma} for the projected limits for higher energies on top FCNCs at the LHC and ILC. From these numbers it is evident that even in the higher energetic Run-II of the LHC the sensitivity will not reach the limit to judge the small branching ratios as obtained from the theoretical calculations in the SM. However, there are many BSM scenarios in which these branching ratios are quite high and at the level that can be probed in the Run-II of LHC. The aim of this paper is to look into this issue of rare decays in one of the interesting BSM scenario, \textit{i.e.,} mUED and nmUED.     
We show that both the decay widths of $t\to c\gamma$ and $t\to ch$ do not change much from the SM value in mUED for any reasonable choice of the inverse compactification radius $1/R$. This result is somewhat contrary to what has been obtained in~\cite{GonzalezSprinberg:2007zz} by using various simplifying assumptions. In passing we note that a similar picture arises for the rare $B$- and $K$-decays in the case of mUED~\cite{Buras:2002ej}. On the other hand in the non-minimal case, \textit{i.e.,} in the presence of BLT parameters, if  $R_{f,\Phi}\geq 10$ and $R_{f}\neq R_{\Phi}$ the $t\to c\gamma$ decay width can, in principle, enhance up to four orders of magnitude from its SM value while respecting the electroweak precision data. But the required $R^{-1}$ along with such higher values of BLT parameters are not viable from the LHC data. For the case $R_{f} = R_{\Phi}$ we found no such enhancement. However in the case of $t\to ch$, irrespective of whether $R_{f}$ and $R_{\Phi}$ are equal or not the decay width does not get any enhancement compared to its SM result. Actually the GIM suppression is still at work, even though the KK particles are contributing the processes, and the decay widths remain, in most of the cases, almost the same level as SM. 
It is also worth-noting that in the KK parity conserving nmUED scenario the lightest KK particle is stable and can play the role of dark matter. But in this regard it is to be kept in mind that for some choice of BLT parameters it may so happen that some specific KK fermion becomes the LKP. Now, a KK fermion LKP is not preferable as it can have electric charge or if it is a KK neutrino it is excluded, as a viable dark matter candidate, by direct detection observations. KK parity can be broken if there are asymmetric BLTs. Thus taking asymmetric BLT parameters in a way that guarantees enough KK parity breaking to make the LKP unstable and sufficiently short-lived, the dark matter constraints can be relaxed.
Finally we add a few comments about the possible extensions in the ambit of this model. First, the effect of KK contribution on the CKM matrix elements can be systematically taken into account in the calculation of various rare top decays. Secondly, in nmUED by utilising the freedom of taking different BLT parameters for different types of fields one can, in principle, get a richer model set-up to look into these decays. Thirdly, and perhaps most importantly, by taking different BLT parameters at two orbifold fixed points for the same kind of field one can break KK-parity. The violation of KK-parity would lead to a host of new KK-parity violating vertices that can significantly modify the decay widths. It is imperative that these effects, which can lead to some interesting observations, be examined in detail.    
 

\paragraph*{Acknowledgements\,:} We thank Amitava Raychaudhuri and Anindya Datta for their insightful comments at different stages of this work. We also acknowledge Dr. Hiren H. Patel for numerous useful comments and suggestions regarding \texttt{Package-X}. We thank Triparno Bandyopadhyay and Nabanita Ganguly for useful discussions. TJ acknowledges  financial support from Council of Scientific and Industrial Research (CSIR) in terms of Senior Research Fellowship (SRF). UKD thanks Physical Research Laboratory (PRL), Department of Space (DoS), Government of India for financial support.

\appendix
\section*{Appendix} \label{s:appen}
\subsection*{Feynman rules}
All the momenta and fields are assumed to be incoming. To avoid cluttering we suppress the indices in the overlap integrals. Basically by $I_{3}$ we mean $I_{3}^{m}$ and by $I_{1,2}$ we imply $I_{1,2}^{n,m}$. Also $\beta = \left(\frac{\pi + R_{\Phi}}{\pi + R_{f}}\right)$.
Obviously in mUED the BLT parameters are vanishing and so the overlap integrals and $\beta$ will be unity and $\alpha_{n} = \frac12 \tan^{-1}\left(\frac{m_{j}}{n/R}\right)$ and $M_{\Phi k} = k/R$. Moreover for mUED, the conservation of KK number will ensure that there will be no 0-0-$n$ type coupling.
\begin{figure}[H]
  \includegraphics[scale=0.55]{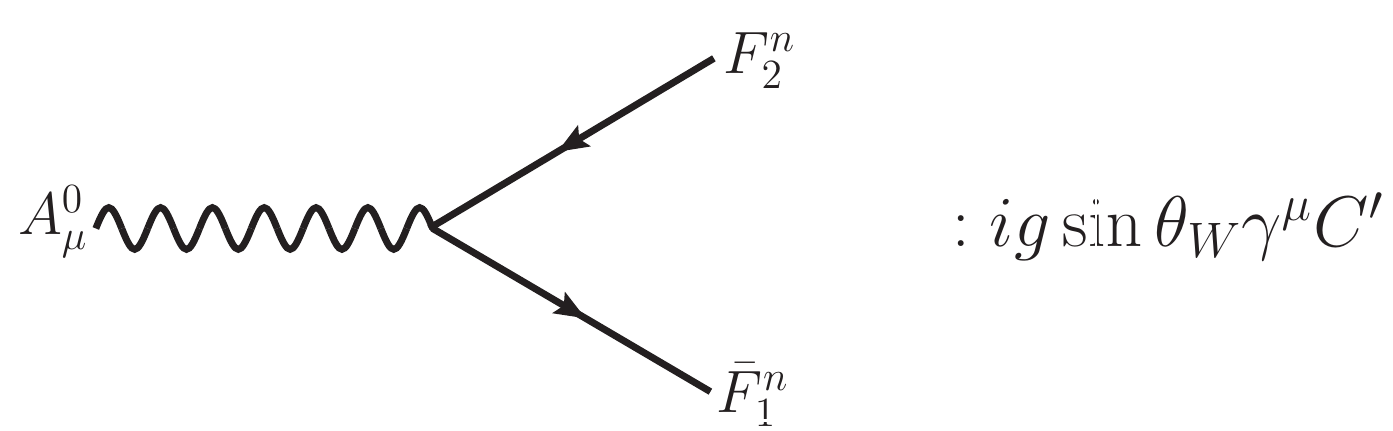}
\end{figure}
\begin{align*}
A_{\mu}^{0}\bar{Q}_{j}^{\prime n}Q_{j}^{\prime n}  &\colon   \begin{array}{rl}
                                      C' = -\frac{1}{3}
                                      \end{array} , ~~~~~~~~~ A_{\mu}^{0}\bar{D}^{\prime n}D^{\prime n}  \colon   \begin{array}{rl}
                                      C' = -\frac{1}{3}
                                     \end{array},\\
A_{\mu}^{0}\bar{Q}_{j}^{\prime n}D^{\prime n}  &\colon  \begin{array}{rl}
                                      C' =0
                                      \end{array} , ~~~~~~~~~~~~ A_{\mu}^{0}\bar{D}^{\prime n}Q_{j}^{\prime n}  \colon \begin{array}{rl}
                                      C' = 0
                                      \end{array} ,\\
A_{\mu}^{0}\bar{b}_{j}^{0}b_{j}^{0}  &\colon   \begin{array}{rl}
                                      C' = -\frac{1}{3}
                                      \end{array} , ~~~~~~~~~~~~ A_{\mu}^{0}\bar{t}_{i}^{0}t_{i}^{0}  \colon   \begin{array}{rl}
                                      C' = \frac{2}{3}
                                     \end{array},
\end{align*}
\begin{figure}[H]
  \includegraphics[scale=0.55]{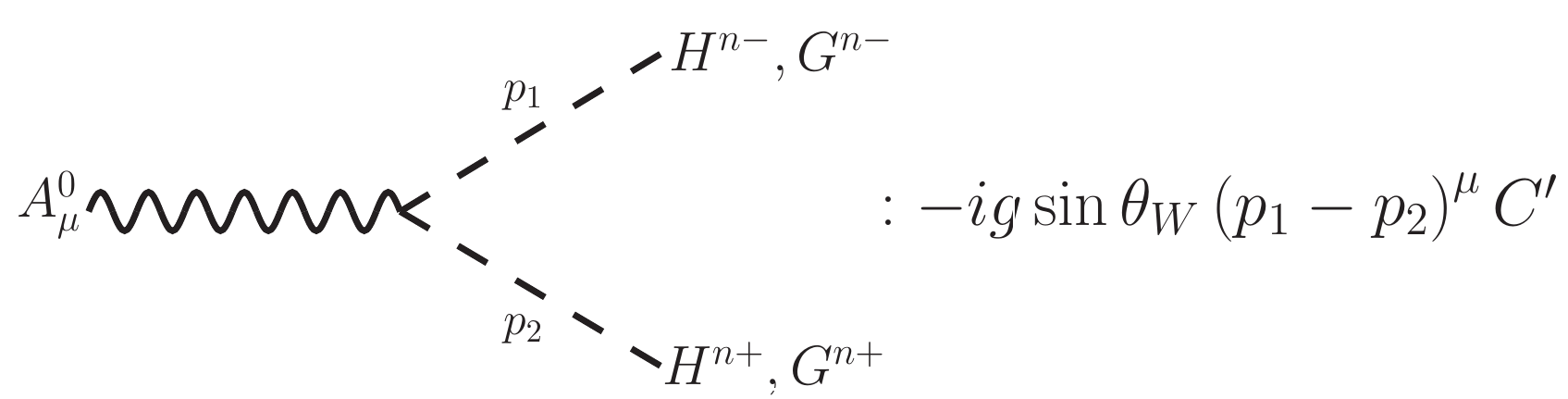}
\end{figure}
\begin{align*}
A_{\mu}^{0}{H}^{n +}H^{n-}  \colon   \begin{array}{rl}
                                      C' = 1
                                      \end{array} ,~~~~~~~~ 
A_{\mu}^{0}{G}^{n +}G^{n-}  \colon   \begin{array}{rl}
                                      C' = 1
                                     \end{array},~~~~~~~~
A_{\mu}^{0}{H}^{n \pm}G^{n \mp}   \colon  \begin{array}{rl}
                                      C' =0
                                      \end{array},
\end{align*}
\begin{figure}[H]
  \includegraphics[scale=0.55]{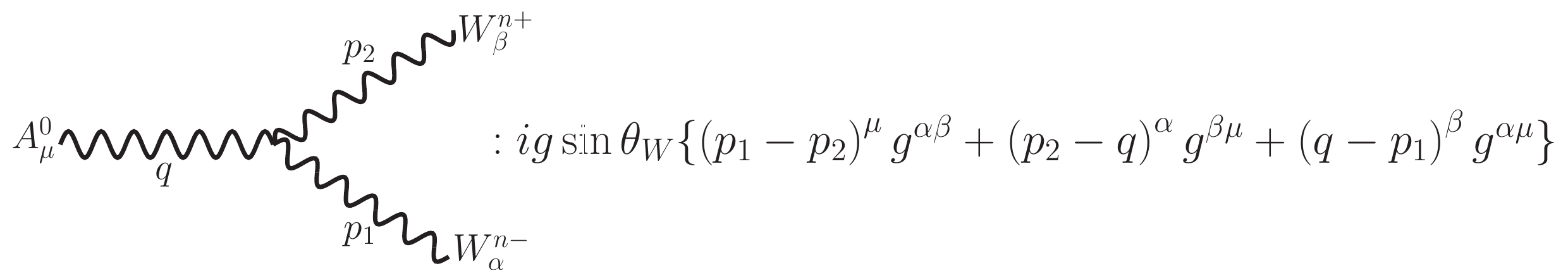}
  \includegraphics[scale=0.55]{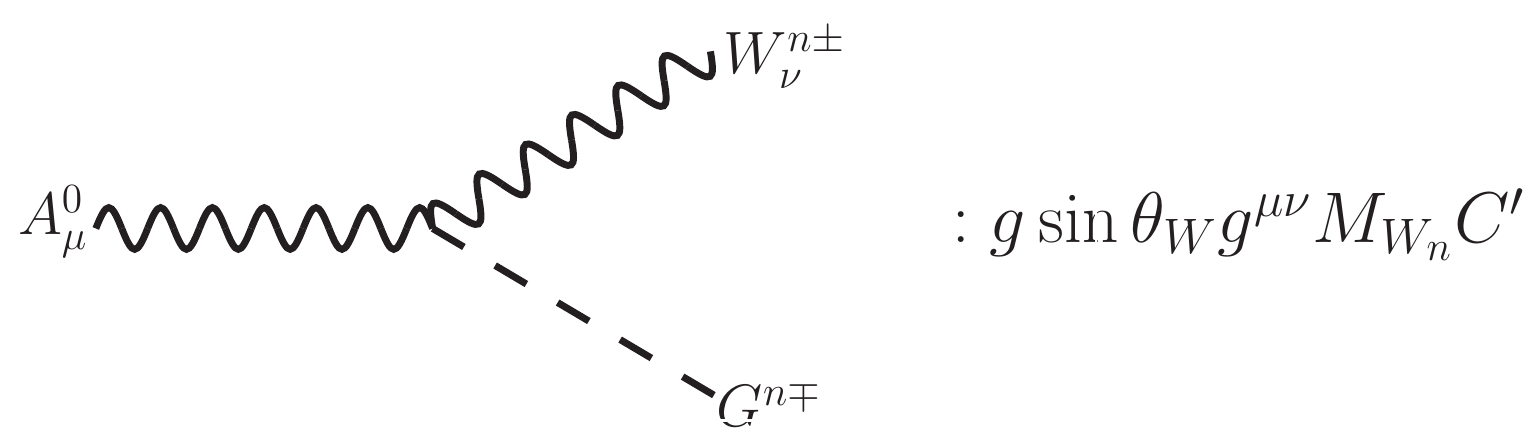}
\end{figure}
\begin{equation*}
A_{\mu}^{0}{G}^{n-}W_{\nu}^{n+}  \colon   \begin{array}{rl}
                                      C' = 1
                                      \end{array} , ~~~~~~~~~~~~~~~~~~~~ A_{\mu}^{0}{G}^{n+}W_{\nu}^{n-}  \colon   \begin{array}{rl}
                                      C' = -1
                                     \end{array},
\end{equation*}
\begin{figure}[H]
  \includegraphics[scale=0.55]{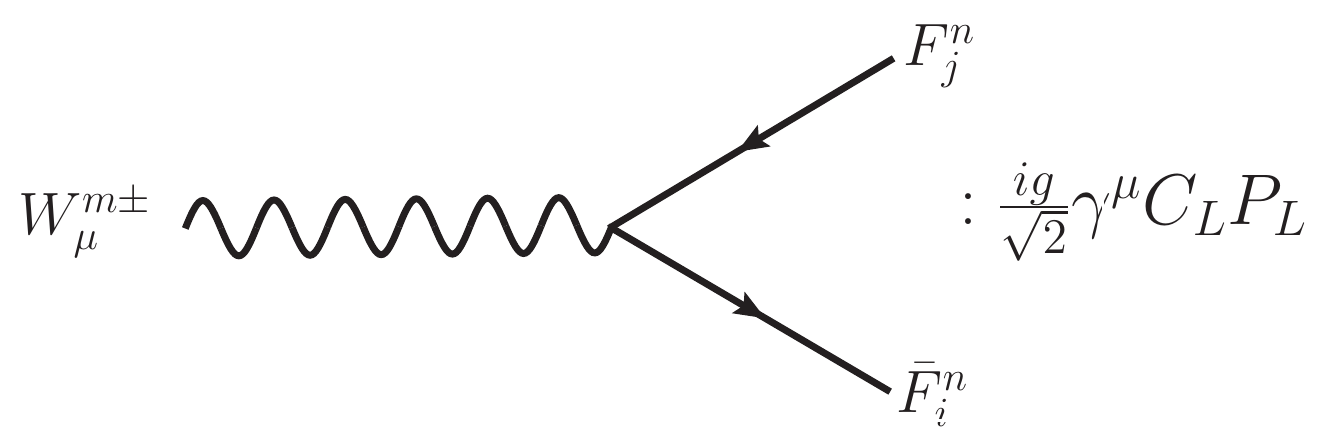}
\end{figure}
\begin{align*}
W^{m +}\bar{t}_{i}^{0}Q_{j}^{\prime n}  &\colon   C_{L} = -I_{1}\sqrt \beta\cos\alpha_{n} V_{ij},
                                       ~~~~~~~~~~~~~~~ W^{m -}\bar{Q}_{j}^{\prime n}t_{i}^{0}  \colon  
                                      C_{L} = -I_{1}\sqrt \beta\cos\alpha_{n}V_{ij}^{*},\\
W^{m +}\bar{t}_{i}^{0}D_{j}^{\prime n}  &\colon   C_{L} = I_{1}\sqrt \beta\sin\alpha_{n}V_{ij},
                                       ~~~~~~~~~~~~~~~~~ 
W^{m -}\bar{D}_{j}^{\prime n}t_{i}^{0}  \colon  
                                      C_{L} = I_{1}\sqrt \beta\sin\alpha_{n}V_{ij}^{*},\\
W^{m +}\bar{t}_{i}^{0}b_{j}^{0}  &\colon   C_{L} = I_{3}\sqrt {\frac{\beta}{\pi R + r_{f}}}V_{ij},
                                      ~~~~~~~~~~~~~~~~~ 
W^{m -}\bar{b}_{j}^{0}t_{i}^{0}  \colon  
                                     C_{L} = I_{3}\sqrt{\frac{\beta}{\pi R + r_{f}}} V_{ij}^{*}.
\end{align*}
\begin{figure}[H]
  \includegraphics[scale=0.55]{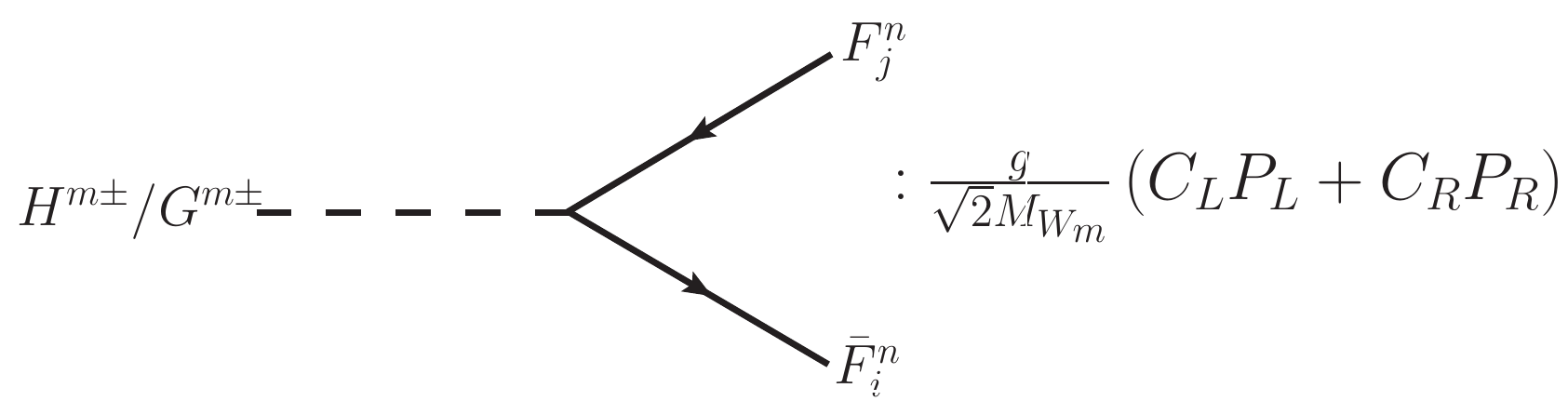}
\end{figure}
\begin{gather*}
\bar{t}_{i}^{0}G^{m +}Q_{j}^{\prime n}  \colon  \left\{ \begin{array}{rl}
                                       C_{L} &= \sqrt \beta I_{1}m_{i}\cos\alpha_{n} V_{ij} \\
                                       C_{R} &= -\sqrt \beta\left(I_{1}m_{j}\sin\alpha_{n} + I_{2}M_{\Phi m}\cos\alpha_{n}\right)V_{ij}
                                      \end{array} \right.,\\ \bar{Q}_{j}^{\prime n}G^{m -}t_{i}^{0}  \colon  \left\{ \begin{array}{rl}
                                      C_{L} &=  \sqrt \beta\left(I_{1}m_{j}\sin\alpha_{n} + I_{2}M_{\Phi m}\cos\alpha_{n}\right)V_{ij}^{*}\\
                                      C_{R} &= -\sqrt \beta I_{1}m_{i}\cos\alpha_{n} V_{ij}^{*}
                                     \end{array} \right.,\\
\bar{t}_{i}^{0}G^{m +}D_{j}^{\prime n}  \colon  \left\{ \begin{array}{rl}
                                      C_{L} &= -\sqrt \beta I_{1}m_{i}\sin\alpha_{n} V_{ij} \\
                                      C_{R} &= \sqrt \beta\left(I_{1}m_{j}\cos\alpha_{n} - I_{2}M_{\Phi m}\sin\alpha_{n}\right)V_{ij}
                                      \end{array} \right.,\\ \bar{D}_{j}^{\prime n}G^{m -}t_{i}^{0}  \colon  \left\{ \begin{array}{rl}
                                      C_{L} &=  -\sqrt \beta\left(I_{1}m_{j}\cos\alpha_{n} - I_{2}M_{\Phi m}\sin\alpha_{n}\right)V_{ij}^{*}\\
                                      C_{R} &= \sqrt \beta I_{1}m_{i}\sin\alpha_{n} V_{ij}^{*}
                                     \end{array} \right.,\\
\bar{t}_{i}^{0}H^{m +}Q_{j}^{\prime n}  \colon  \left\{ \begin{array}{rl}
                                      C_{L} &= -i\sqrt \beta I_{1}\frac{m_{i}M_{\Phi m}}{M_{W}}\cos\alpha_{n} V_{ij} \\
                                      C_{R} &= i\sqrt \beta\left(I_{1}\frac{m_{j}M_{\Phi m}}{M_{W}}\sin\alpha_{n} - I_{2}M_{W}\cos\alpha_{n}\right)V_{ij}
                                      \end{array} \right.,\\ \bar{Q}_{j}^{\prime n}H^{m -}t_{i}^{0}  \colon  \left\{ \begin{array}{rl}
                                      C_{L} &=  i\sqrt \beta\left(I_{1}\frac{m_{j}M_{\Phi m}}{M_{W}}\sin\alpha_{n} - I_{2}M_{W}\cos\alpha_{n}\right)V_{ij}^{*}\\
                                      C_{R} &= -i\sqrt \beta I_{1}\frac{m_{i}M_{\Phi m}}{M_{W}}\cos\alpha_{n} V_{ij}^{*}
                                     \end{array} \right.,\\
\bar{t}_{i}^{0}H^{m +}D_{j}^{\prime n}  \colon  \left\{ \begin{array}{rl}
                                      C_{L} &=  i\sqrt \beta I_{1}\frac{m_{i}M_{\Phi m}}{M_{W}}\sin\alpha_{n}V_{ij} \\
                                      C_{R} &= -i\sqrt \beta\left(I_{1}\frac{m_{j}M_{\Phi m}}{M_{W}}\cos\alpha_{n} + I_{2}M_{W}\sin\alpha_{n}\right)V_{ij}
                                      \end{array} \right.,\\ \bar{D}_{j}^{\prime n}G^{m -}t_{i}^{0}  \colon  \left\{ \begin{array}{rl}
                                      C_{L} &= -i\sqrt \beta\left(I_{1}\frac{m_{j}M_{\Phi m}}{M_{W}}\cos\alpha_{n} + I_{2}M_{W}\sin\alpha_{n}\right)V_{ij}^{*}\\
                                      C_{R} &= i\sqrt \beta I_{1}\frac{m_{i}M_{\Phi m}}{M_{W}}\sin\alpha_{n} V_{ij}^{*}
                                     \end{array} \right.,\\
\bar{t}_{i}^{0}G^{m +}b_{j}^{0}  \colon  \left\{ \begin{array}{rl}
                                      C_{L} &=  -\sqrt{\frac{\beta}{\pi R + r_{f}}} I_{3}m_{i}V_{ij} \\
                                      C_{R} &= \sqrt{\frac{\beta}{\pi R + r_{f}}} I_{3} m_{j} V_{ij}
                                      \end{array} \right., ~~~~~~~~~~~~~~~~
\bar{b}_{j}^{0}G^{m -}t_{i}^{0}  \colon  \left\{ \begin{array}{rl}
                                      C_{L} &= -\sqrt{\frac{\beta}{\pi R + r_{f}}} I_{3}m_{j}V_{ij}^{*}\\
                                      C_{R} &= \sqrt{\frac{\beta}{\pi R + r_{f}}} I_{3} m_{i}V_{ij}^{*}
                                     \end{array} \right.,\\
\bar{t}_{i}^{0}H^{m +}b_{j}^{0}  \colon  \left\{ \begin{array}{rl}
                                      C_{L} &=  i \sqrt{\frac{\beta}{\pi R + r_{f}}} I_{3} \frac{M_{\Phi m}}{M_{W}}m_{i}V_{ij} \\
                                      C_{R} &= -i \sqrt{\frac{\beta}{\pi R + r_{f}}} I_{3} \frac{M_{\Phi m}}{M_{W}} m_{j} V_{ij}
                                      \end{array} \right., ~~~~~
\bar{b}_{j}^{0}H^{m -}t_{i}^{0}  \colon  \left\{ \begin{array}{rl}
                                      C_{L} &= -i \sqrt{\frac{\beta}{\pi R + r_{f}}} I_{3} \frac{M_{\Phi m}}{M_{W}} m_{j}V_{ij}^{*}\\
                                      C_{R} &= i \sqrt{\frac{\beta}{\pi R + r_{f}}} I_{3} \frac{M_{\Phi m}}{M_{W}}m_{i}V_{ij}^{*}
                                     \end{array} \right..\\
\end{gather*}
\begin{figure}[H]
  \includegraphics[scale=0.55]{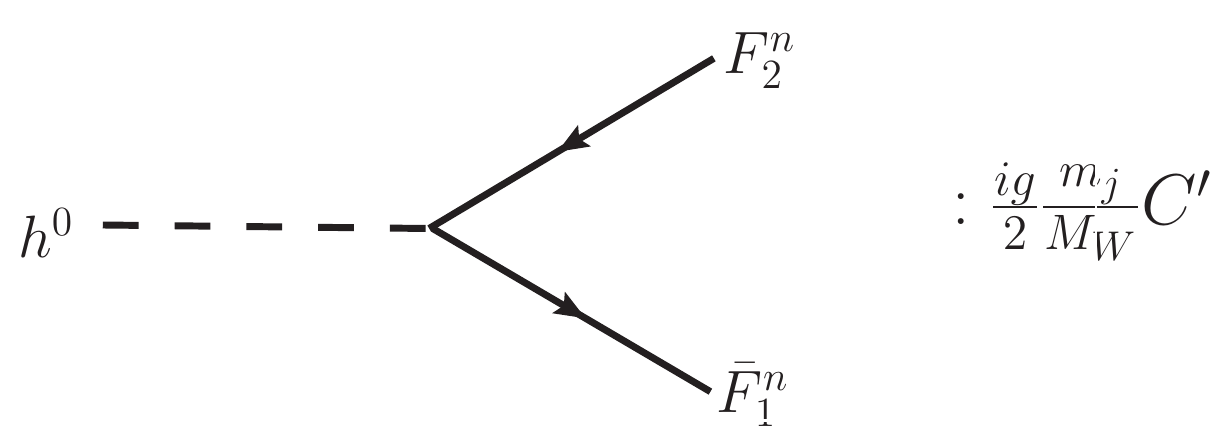}
\end{figure}
\begin{gather*}
h^{0}\bar{Q}_{j}^{\prime n}Q_{j}^{\prime n}  \colon   \begin{array}{rl}
                                      C' = -\sin 2\alpha_{n}
                                      \end{array} , ~~~~~~~~~~~~~~~~~~~~ h^{0}\bar{D}^{\prime n}D^{\prime n}  \colon   \begin{array}{rl}
                                      C' = -\sin 2\alpha_{n}
                                     \end{array},\\
h^{0}\bar{Q}_{j}^{\prime n}D^{\prime n}  \colon  \begin{array}{rl}
                                      C' = \cos 2\alpha_{n} \gamma_{5}
                                      \end{array} , ~~~~~~~~~~~~~~~~~~~~ h^{0}\bar{D}^{\prime n}Q_{j}^{\prime n}  \colon \begin{array}{rl}
                                      C' = -\cos 2\alpha_{n} \gamma_{5}
                                      \end{array} ,\\
h^{0}\bar{b}_{j}^{0}b_{j}^{0}  \colon   \begin{array}{rl}
                                      C' = -1 
                                      \end{array} , ~~~~~~~~~~~~~~~~~~~~ h^{0}\bar{t}_{i}^{0}t_{i}^{0}  \colon   \begin{array}{rl}
                                      C' = -\frac{m_{i}}{m_{j}}
                                     \end{array}.\\
\end{gather*}
\begin{figure}[H]
  \includegraphics[scale=0.55]{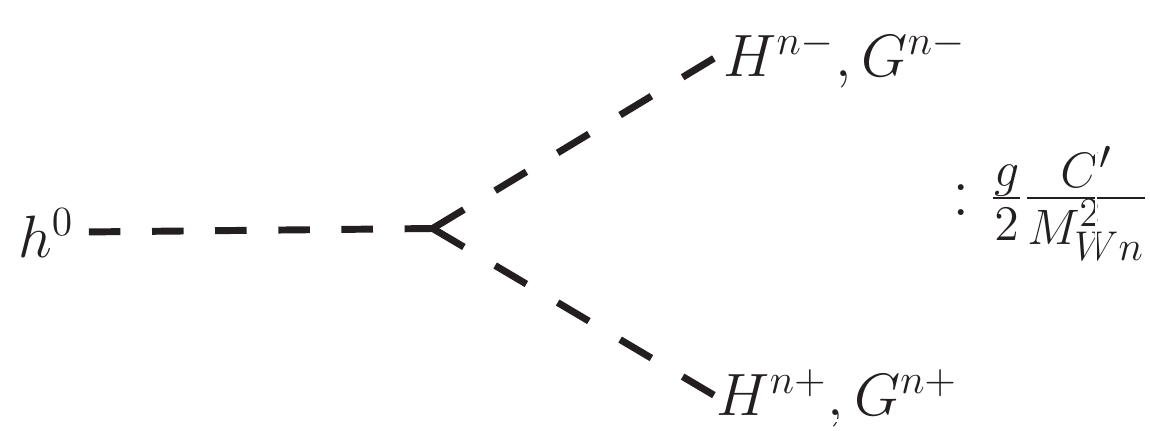}
\end{figure}
\begin{gather*}
h^{0}{H}^{n +}H^{n-}  \colon   \begin{array}{rl}
                                      C' = -i m_{h}^{2} M_{W}
                                      \end{array} , ~~~~~~~~~~~~~~~~ h^{0}{G}^{n +}G^{n-}  \colon   \begin{array}{rl}
                                      C' = -i \frac{m_{h}^{2} M_{\Phi n}^{2} + 2 M_{W}^{2} M_{Wn}^{2}}{M_{W}}
                                     \end{array},\\
h^{0}{H}^{n +}G^{n -}   \colon  \begin{array}{rl}
                                      C' =  - M_{\Phi n}(m_{h}^{2}-M_{Wn}^{2})
                                      \end{array},~~~~~~~~~~~~~~~~~~~ h^{0}{H}^{n -}G^{n+}   \colon  \begin{array}{rl}
                                      C' =  M_{\Phi n}(m_{h}^{2}-M_{Wn}^{2})
                                      \end{array}.
\end{gather*}
\begin{figure}[H]
  \includegraphics[scale=0.55]{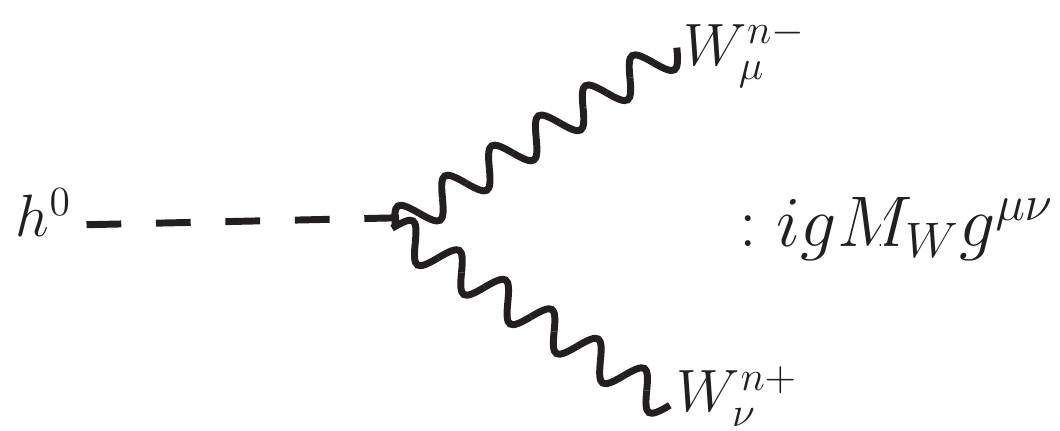}

  \includegraphics[scale=0.55]{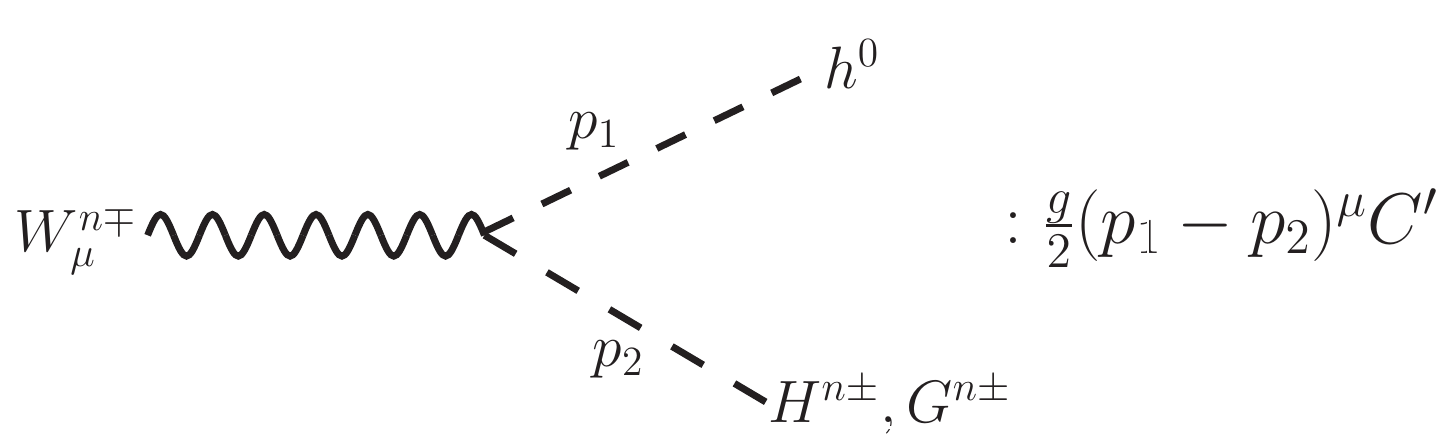}
\end{figure}
\begin{equation*}
h^{0}{H}^{n\pm}W^{n\mp}  \colon   \begin{array}{rl}
                                      C' = \mp i \frac{M_{\Phi n}}{M_{Wn}}
                                      \end{array} , ~~~~~~~~~~~~~~~~~~~~ h^{0}{G}^{n\pm}W^{n\mp}  \colon   \begin{array}{rl}
                                      C' = \frac{M_{W}}{M_{Wn}}
                                     \end{array}.\\
\end{equation*}


\bibliographystyle{JHEP}
\bibliography{raretop_ref.bib}

\providecommand{\href}[2]{#2}\begingroup\raggedright\begin{thebibliography}{10}

\bibitem{Aad:2012tfa}
{\bf ATLAS} Collaboration, G.~Aad et~al., {\it {Observation of a new particle
  in the search for the Standard Model Higgs boson with the ATLAS detector at
  the LHC}},  {\em Phys. Lett.} {\bf B716} (2012) 1--29,
  [\href{http://arxiv.org/abs/1207.7214}{{\tt arXiv:1207.7214}}].

\bibitem{Chatrchyan:2012xdj}
{\bf CMS} Collaboration, S.~Chatrchyan et~al., {\it {Observation of a new boson
  at a mass of 125 GeV with the CMS experiment at the LHC}},  {\em Phys. Lett.}
  {\bf B716} (2012) 30--61, [\href{http://arxiv.org/abs/1207.7235}{{\tt
  arXiv:1207.7235}}].

\bibitem{Antoniadis:1990ew}
I.~Antoniadis, {\it {A Possible new dimension at a few TeV}},  {\em Phys.
  Lett.} {\bf B246} (1990) 377--384.

\bibitem{Antoniadis:1998ig}
I.~Antoniadis, N.~Arkani-Hamed, S.~Dimopoulos, and G.~R. Dvali, {\it {New
  dimensions at a millimeter to a Fermi and superstrings at a TeV}},  {\em
  Phys. Lett.} {\bf B436} (1998) 257--263,
  [\href{http://arxiv.org/abs/hep-ph/9804398}{{\tt hep-ph/9804398}}].

\bibitem{Appelquist:2000nn}
T.~Appelquist, H.-C. Cheng, and B.~A. Dobrescu, {\it {Bounds on universal extra
  dimensions}},  {\em Phys. Rev.} {\bf D64} (2001) 035002,
  [\href{http://arxiv.org/abs/hep-ph/0012100}{{\tt hep-ph/0012100}}].

\bibitem{ArkaniHamed:1998rs}
N.~Arkani-Hamed, S.~Dimopoulos, and G.~R. Dvali, {\it {The Hierarchy problem
  and new dimensions at a millimeter}},  {\em Phys. Lett.} {\bf B429} (1998)
  263--272, [\href{http://arxiv.org/abs/hep-ph/9803315}{{\tt hep-ph/9803315}}].

\bibitem{Randall:1999ee}
L.~Randall and R.~Sundrum, {\it {A Large mass hierarchy from a small extra
  dimension}},  {\em Phys. Rev. Lett.} {\bf 83} (1999) 3370--3373,
  [\href{http://arxiv.org/abs/hep-ph/9905221}{{\tt hep-ph/9905221}}].

\bibitem{Randall:1999vf}
L.~Randall and R.~Sundrum, {\it {An Alternative to compactification}},  {\em
  Phys. Rev. Lett.} {\bf 83} (1999) 4690--4693,
  [\href{http://arxiv.org/abs/hep-th/9906064}{{\tt hep-th/9906064}}].

\bibitem{Servant:2002aq}
G.~Servant and T.~M.~P. Tait, {\it {Is the lightest Kaluza-Klein particle a
  viable dark matter candidate?}},  {\em Nucl. Phys.} {\bf B650} (2003)
  391--419, [\href{http://arxiv.org/abs/hep-ph/0206071}{{\tt hep-ph/0206071}}].

\bibitem{Servant:2002hb}
G.~Servant and T.~M.~P. Tait, {\it {Elastic scattering and direct detection of
  Kaluza-Klein dark matter}},  {\em New J. Phys.} {\bf 4} (2002) 99,
  [\href{http://arxiv.org/abs/hep-ph/0209262}{{\tt hep-ph/0209262}}].

\bibitem{Burnell:2005hm}
F.~Burnell and G.~D. Kribs, {\it {The Abundance of Kaluza-Klein dark matter
  with coannihilation}},  {\em Phys. Rev.} {\bf D73} (2006) 015001,
  [\href{http://arxiv.org/abs/hep-ph/0509118}{{\tt hep-ph/0509118}}].

\bibitem{Kong:2005hn}
K.~Kong and K.~T. Matchev, {\it {Precise calculation of the relic density of
  Kaluza-Klein dark matter in universal extra dimensions}},  {\em JHEP} {\bf
  01} (2006) 038, [\href{http://arxiv.org/abs/hep-ph/0509119}{{\tt
  hep-ph/0509119}}].

\bibitem{Cheng:2002iz}
H.-C. Cheng, K.~T. Matchev, and M.~Schmaltz, {\it {Radiative corrections to
  Kaluza-Klein masses}},  {\em Phys. Rev.} {\bf D66} (2002) 036005,
  [\href{http://arxiv.org/abs/hep-ph/0204342}{{\tt hep-ph/0204342}}].

\bibitem{Blennow:2011tb}
M.~Blennow, H.~Melbeus, T.~Ohlsson, and H.~Zhang, {\it {RG running in a minimal
  UED model in light of recent LHC Higgs mass bounds}},  {\em Phys. Lett.} {\bf
  B712} (2012) 419--424, [\href{http://arxiv.org/abs/1112.5339}{{\tt
  arXiv:1112.5339}}].

\bibitem{Datta:2012db}
A.~Datta and S.~Raychaudhuri, {\it {Vacuum Stability Constraints and LHC
  Searches for a Model with a Universal Extra Dimension}},  {\em Phys. Rev.}
  {\bf D87} (2013), no.~3 035018, [\href{http://arxiv.org/abs/1207.0476}{{\tt
  arXiv:1207.0476}}].

\bibitem{Flacke:2008ne}
T.~Flacke, A.~Menon, and D.~J. Phalen, {\it {Non-minimal universal extra
  dimensions}},  {\em Phys.Rev.} {\bf D79} (2009) 056009,
  [\href{http://arxiv.org/abs/0811.1598}{{\tt arXiv:0811.1598}}].

\bibitem{delAguila:2003bh}
F.~del Aguila, M.~Perez-Victoria, and J.~Santiago, {\it {Bulk fields with
  general brane kinetic terms}},  {\em JHEP} {\bf 0302} (2003) 051,
  [\href{http://arxiv.org/abs/hep-th/0302023}{{\tt hep-th/0302023}}].

\bibitem{delAguila:2003gu}
F.~del Aguila, M.~Perez-Victoria, and J.~Santiago, {\it {Bulk fields with brane
  terms}},  {\em Eur.Phys.J.} {\bf C33} (2004) S773--S775,
  [\href{http://arxiv.org/abs/hep-ph/0310352}{{\tt hep-ph/0310352}}].

\bibitem{delAguila:2003kd}
F.~del Aguila, M.~Perez-Victoria, and J.~Santiago, {\it {Some consequences of
  brane kinetic terms for bulk fermions}},
  \href{http://arxiv.org/abs/hep-ph/0305119}{{\tt hep-ph/0305119}}.

\bibitem{delAguila:2003gv}
F.~del Aguila, M.~Perez-Victoria, and J.~Santiago, {\it {Physics of brane
  kinetic terms}},  {\em Acta Phys.Polon.} {\bf B34} (2003) 5511--5522,
  [\href{http://arxiv.org/abs/hep-ph/0310353}{{\tt hep-ph/0310353}}].

\bibitem{delAguila:2006atw}
F.~del Aguila, M.~Perez-Victoria, and J.~Santiago, {\it {Effective description
  of brane terms in extra dimensions}},  {\em JHEP} {\bf 10} (2006) 056,
  [\href{http://arxiv.org/abs/hep-ph/0601222}{{\tt hep-ph/0601222}}].

\bibitem{Datta:2012tv}
A.~Datta, K.~Nishiwaki, and S.~Niyogi, {\it {Non-minimal Universal Extra
  Dimensions: The Strongly Interacting Sector at the Large Hadron Collider}},
  {\em JHEP} {\bf 1211} (2012) 154, [\href{http://arxiv.org/abs/1206.3987}{{\tt
  arXiv:1206.3987}}].

\bibitem{Datta:2013nua}
A.~Datta, U.~K. Dey, A.~Raychaudhuri, and A.~Shaw, {\it {Boundary Localized
  Terms in Universal Extra-Dimensional Models through a Dark Matter
  perspective}},  {\em Phys.Rev.} {\bf D88} (2013) 016011,
  [\href{http://arxiv.org/abs/1305.4507}{{\tt arXiv:1305.4507}}].

\bibitem{Datta:2013yaa}
A.~Datta, K.~Nishiwaki, and S.~Niyogi, {\it {Non-minimal Universal Extra
  Dimensions with Brane Local Terms: The Top Quark Sector}},  {\em JHEP} {\bf
  1401} (2014) 104, [\href{http://arxiv.org/abs/1310.6994}{{\tt
  arXiv:1310.6994}}].

\bibitem{Dey:2013cqa}
U.~K. Dey and T.~S. Ray, {\it {Constraining minimal and nonminimal universal
  extra dimension models with Higgs couplings}},  {\em Phys.Rev.} {\bf D88}
  (2013), no.~5 056016, [\href{http://arxiv.org/abs/1305.1016}{{\tt
  arXiv:1305.1016}}].

\bibitem{Flacke:2014jwa}
T.~Flacke, K.~Kong, and S.~C. Park, {\it {A Review on Non-Minimal Universal
  Extra Dimensions}},  {\em Mod. Phys. Lett.} {\bf A30} (2015), no.~05 1530003,
  [\href{http://arxiv.org/abs/1408.4024}{{\tt arXiv:1408.4024}}].

\bibitem{Jha:2014faa}
T.~Jha and A.~Datta, {\it {$ Z\to b\overline{b} $ in non-minimal Universal
  Extra Dimensional Model}},  {\em JHEP} {\bf 1503} (2015) 012,
  [\href{http://arxiv.org/abs/1410.5098}{{\tt arXiv:1410.5098}}].

\bibitem{Datta:2015aka}
A.~Datta and A.~Shaw, {\it {Nonminimal universal extra dimensional model
  confronts B$_s\to \mu^+\mu^-$}},  {\em Phys. Rev.} {\bf D93} (2016), no.~5
  055048, [\href{http://arxiv.org/abs/1506.08024}{{\tt arXiv:1506.08024}}].

\bibitem{ATLAS-CONF-2015-081}
{\bf ATLAS} Collaboration, {\it {Search for resonances decaying to photon pairs
  in 3.2 fb$^{-1}$ of $pp$ collisions at $\sqrt{s}$ = 13 TeV with the ATLAS
  detector}},  Tech. Rep. ATLAS-CONF-2015-081, CERN, Geneva, Dec, 2015.

\bibitem{CMS-PAS-EXO-15-004}
{\bf CMS} Collaboration, {\it {Search for new physics in high mass diphoton
  events in proton-proton collisions at $\sqrt{s} = 13$ TeV}},  Tech. Rep.
  CMS-PAS-EXO-15-004, CERN, Geneva, 2015.

\bibitem{Deshpande:1981zq}
N.~G. Deshpande and G.~Eilam, {\it {FLAVOR CHANGING ELECTROMAGNETIC
  TRANSITIONS}},  {\em Phys.Rev.} {\bf D26} (1982) 2463.

\bibitem{Deshpande:1982mi}
N.~G. Deshpande and M.~Nazerimonfared, {\it {Flavor Changing Electromagnetic
  Vertex in a Nonlinear $ R_{\xi} $ Gauge}},  {\em Nucl.Phys.} {\bf B213}
  (1983) 390--408.

\bibitem{DiazCruz:1989ub}
J.~Diaz-Cruz, R.~Martinez, M.~Perez, and A.~Rosado, {\it {Flavor Changing
  Radiative Decay of the $t$ Quark}},  {\em Phys.Rev.} {\bf D41} (1990)
  891--894.

\bibitem{Mele:1998ag}
B.~Mele, S.~Petrarca, and A.~Soddu, {\it {A New evaluation of the $t \to cH$
  decay width in the standard model}},  {\em Phys.Lett.} {\bf B435} (1998)
  401--406, [\href{http://arxiv.org/abs/hep-ph/9805498}{{\tt hep-ph/9805498}}].

\bibitem{Mele:1999zk}
B.~Mele, S.~Petrarca, and A.~Soddu, {\it {The $ t \to c H$ decay width in the
  standard model}},  \href{http://arxiv.org/abs/hep-ph/9912235}{{\tt
  hep-ph/9912235}}.

\bibitem{AguilarSaavedra:2002ns}
J.~Aguilar-Saavedra and B.~Nobre, {\it {Rare top decays $t \to c \gamma$, $t
  \to c g$ and CKM unitarity}},  {\em Phys.Lett.} {\bf B553} (2003) 251--260,
  [\href{http://arxiv.org/abs/hep-ph/0210360}{{\tt hep-ph/0210360}}].

\bibitem{AguilarSaavedra:2004wm}
J.~Aguilar-Saavedra, {\it {Top flavor-changing neutral interactions:
  Theoretical expectations and experimental detection}},  {\em Acta
  Phys.Polon.} {\bf B35} (2004) 2695--2710,
  [\href{http://arxiv.org/abs/hep-ph/0409342}{{\tt hep-ph/0409342}}].

\bibitem{Chen:2013qta}
K.-F. Chen, W.-S. Hou, C.~Kao, and M.~Kohda, {\it {When the Higgs meets the
  Top: Search for $t \to ch^{0}$ at the LHC}},  {\em Phys. Lett.} {\bf B725}
  (2013) 378--381, [\href{http://arxiv.org/abs/1304.8037}{{\tt
  arXiv:1304.8037}}].

\bibitem{Khanpour:2014xla}
H.~Khanpour, S.~Khatibi, M.~K. Yanehsari, and M.~M. Najafabadi, {\it {Single
  top quark production as a probe of anomalous $tq\gamma$ and $tqZ$ couplings
  at the FCC-ee}},  \href{http://arxiv.org/abs/1408.2090}{{\tt
  arXiv:1408.2090}}.

\bibitem{Hesari:2014eua}
H.~Hesari, H.~Khanpour, M.~K. Yanehsari, and M.~M. Najafabadi, {\it {Probing
  the Top Quark Flavour-Changing Neutral Current at a Future Electron-Positron
  Collider}},  {\em Adv. High Energy Phys.} {\bf 2014} (2014) 476490,
  [\href{http://arxiv.org/abs/1412.8572}{{\tt arXiv:1412.8572}}].

\bibitem{Kim:2015oua}
D.~Kim and M.~Park, {\it {Enhancement of new physics signal sensitivity with
  mistagged charm quarks}},  \href{http://arxiv.org/abs/1507.03990}{{\tt
  arXiv:1507.03990}}.

\bibitem{Hesari:2015oya}
H.~Hesari, H.~Khanpour, and M.~M. Najafabadi, {\it {Direct and Indirect
  Searches for Top-Higgs FCNC Couplings}},  {\em Phys. Rev.} {\bf D92} (2015),
  no.~11 113012, [\href{http://arxiv.org/abs/1508.07579}{{\tt
  arXiv:1508.07579}}].

\bibitem{Khatibi:2015aal}
S.~Khatibi and M.~M. Najafabadi, {\it {Top quark flavor changing via photon}},
  \href{http://arxiv.org/abs/1511.00220}{{\tt arXiv:1511.00220}}.

\bibitem{Guasch:1999jp}
J.~Guasch and J.~Sola, {\it {FCNC top quark decays: A Door to SUSY physics in
  high luminosity colliders?}},  {\em Nucl. Phys.} {\bf B562} (1999) 3--28,
  [\href{http://arxiv.org/abs/hep-ph/9906268}{{\tt hep-ph/9906268}}].

\bibitem{Eilam:2001dh}
G.~Eilam, A.~Gemintern, T.~Han, J.~M. Yang, and X.~Zhang, {\it {Top quark rare
  decay $t \to ch$ in R-parity violating SUSY}},  {\em Phys. Lett.} {\bf B510}
  (2001) 227--235, [\href{http://arxiv.org/abs/hep-ph/0102037}{{\tt
  hep-ph/0102037}}].

\bibitem{Frank:2005vd}
M.~Frank and I.~Turan, {\it {$t \to$ cg, $c \gamma$, cZ in the left-right
  supersymmetric model}},  {\em Phys.Rev.} {\bf D72} (2005) 035008,
  [\href{http://arxiv.org/abs/hep-ph/0506197}{{\tt hep-ph/0506197}}].

\bibitem{Cao:2007dk}
J.~J. Cao, G.~Eilam, M.~Frank, K.~Hikasa, G.~L. Liu, I.~Turan, and J.~M. Yang,
  {\it {SUSY-induced FCNC top-quark processes at the large hadron collider}},
  {\em Phys. Rev.} {\bf D75} (2007) 075021,
  [\href{http://arxiv.org/abs/hep-ph/0702264}{{\tt hep-ph/0702264}}].

\bibitem{Cao:2008vk}
J.~Cao, Z.~Heng, L.~Wu, and J.~M. Yang, {\it {R-parity violating effects in top
  quark FCNC productions at LHC}},  {\em Phys. Rev.} {\bf D79} (2009) 054003,
  [\href{http://arxiv.org/abs/0812.1698}{{\tt arXiv:0812.1698}}].

\bibitem{Cao:2014udj}
J.~Cao, C.~Han, L.~Wu, J.~M. Yang, and M.~Zhang, {\it {SUSY induced top quark
  FCNC decay $t \rightarrow { ch}$ after Run I of LHC}},  {\em Eur. Phys. J.}
  {\bf C74} (2014), no.~9 3058, [\href{http://arxiv.org/abs/1404.1241}{{\tt
  arXiv:1404.1241}}].

\bibitem{Dedes:2014asa}
A.~Dedes, M.~Paraskevas, J.~Rosiek, K.~Suxho, and K.~Tamvakis, {\it {Rare
  Top-quark Decays to Higgs boson in MSSM}},  {\em JHEP} {\bf 11} (2014) 137,
  [\href{http://arxiv.org/abs/1409.6546}{{\tt arXiv:1409.6546}}].

\bibitem{Bardhan:2016txk}
D.~Bardhan, G.~Bhattacharyya, D.~Ghosh, M.~Patra, and S.~Raychaudhuri, {\it {A
  Detailed Analysis of Flavour-changing Decays of Top Quarks as a Probe of New
  Physics at the LHC}},  \href{http://arxiv.org/abs/1601.04165}{{\tt
  arXiv:1601.04165}}.

\bibitem{Eilam:1990zc}
G.~Eilam, J.~Hewett, and A.~Soni, {\it {Rare decays of the top quark in the
  standard and two Higgs doublet models}},  {\em Phys.Rev.} {\bf D44} (1991)
  1473--1484.

\bibitem{Iltan:2001yt}
E.~O. Iltan, {\it {$t \to cH^{0}$ decay in the general two Higgs doublet
  model}},  {\em Phys. Rev.} {\bf D65} (2002) 075017,
  [\href{http://arxiv.org/abs/hep-ph/0111318}{{\tt hep-ph/0111318}}].

\bibitem{Arhrib:2005nx}
A.~Arhrib, {\it {Top and Higgs flavor changing neutral couplings in two Higgs
  doublets model}},  {\em Phys. Rev.} {\bf D72} (2005) 075016,
  [\href{http://arxiv.org/abs/hep-ph/0510107}{{\tt hep-ph/0510107}}].

\bibitem{Gaitan:2015hga}
R.~Gaitán, R.~Martinez, and J.~H.~M. de~Oca, {\it {Rare top decay $t
  \rightarrow c \gamma$ with flavor changing neutral scalar interactions in
  THDM}},  \href{http://arxiv.org/abs/1503.04391}{{\tt arXiv:1503.04391}}.

\bibitem{Abbas:2015cua}
G.~Abbas, A.~Celis, X.-Q. Li, J.~Lu, and A.~Pich, {\it {Flavour-changing top
  decays in the aligned two-Higgs-doublet model}},  {\em JHEP} {\bf 06} (2015)
  005, [\href{http://arxiv.org/abs/1503.06423}{{\tt arXiv:1503.06423}}].

\bibitem{Gao:2013fxa}
T.-J. Gao, T.-F. Feng, and J.-B. Chen, {\it {$t \to c \gamma$ and $t \to c g$
  in warped extra dimensions}},  {\em JHEP} {\bf 1302} (2013) 029,
  [\href{http://arxiv.org/abs/1303.0082}{{\tt arXiv:1303.0082}}].

\bibitem{GonzalezSprinberg:2007zz}
G.~Gonzalez-Sprinberg, R.~Martinez, and J.~A. Rodriguez, {\it {FCNC top quark
  decays in extra dimensions}},  {\em Eur.Phys.J.} {\bf C51} (2007) 919--926.

\bibitem{CorderoCid:2004vi}
A.~Cordero-Cid, M.~A. Perez, G.~Tavares-Velasco, and J.~J. Toscano, {\it
  {Effective Lagrangian approach to Higgs-mediated FCNC top quark decays}},
  {\em Phys. Rev.} {\bf D70} (2004) 074003,
  [\href{http://arxiv.org/abs/hep-ph/0407127}{{\tt hep-ph/0407127}}].

\bibitem{Datta:2009zb}
A.~Datta and M.~Duraisamy, {\it {Model Independent Predictions for Rare Top
  Decays with Weak Coupling}},  {\em Phys.Rev.} {\bf D81} (2010) 074008,
  [\href{http://arxiv.org/abs/0912.4785}{{\tt arXiv:0912.4785}}].

\bibitem{Datta:2014sha}
A.~Datta and A.~Shaw, {\it {A note on gauge-fixing in the electroweak sector of
  UED with BLKTs}},  \href{http://arxiv.org/abs/1408.0635}{{\tt
  arXiv:1408.0635}}.

\bibitem{Muck:2004zz}
A.~Muck, {\em {The standard model in 5D: Theoretical consistency and
  experimental constraints}}.
\newblock PhD thesis, Wurzburg U., 2004.

\bibitem{Buras:2002ej}
A.~J. Buras, M.~Spranger, and A.~Weiler, {\it {The Impact of universal extra
  dimensions on the unitarity triangle and rare K and B decays}},  {\em Nucl.
  Phys.} {\bf B660} (2003) 225--268,
  [\href{http://arxiv.org/abs/hep-ph/0212143}{{\tt hep-ph/0212143}}].

\bibitem{Abazov:2014dpa}
{\bf D0} Collaboration, V.~M. Abazov et~al., {\it {Precision measurement of the
  top-quark mass in lepton+jets final states}},  {\em Phys. Rev. Lett.} {\bf
  113} (2014) 032002, [\href{http://arxiv.org/abs/1405.1756}{{\tt
  arXiv:1405.1756}}].

\bibitem{Agashe:2014kda}
{\bf Particle Data Group} Collaboration, K.~A. Olive et~al., {\it {Review of
  Particle Physics}},  {\em Chin. Phys.} {\bf C38} (2014) 090001.

\bibitem{Patel:2015tea}
H.~H. Patel, {\it {Package-X: A Mathematica package for the analytic
  calculation of one-loop integrals}},  {\em Comput. Phys. Commun.} {\bf 197}
  (2015) 276--290, [\href{http://arxiv.org/abs/1503.01469}{{\tt
  arXiv:1503.01469}}].

\bibitem{Hahn:1998yk}
T.~Hahn and M.~Perez-Victoria, {\it {Automatized one loop calculations in
  four-dimensions and D-dimensions}},  {\em Comput. Phys. Commun.} {\bf 118}
  (1999) 153--165, [\href{http://arxiv.org/abs/hep-ph/9807565}{{\tt
  hep-ph/9807565}}].

\bibitem{Servant:2014lqa}
G.~Servant, {\it {Status Report on Universal Extra Dimensions After LHC8}},
  {\em Mod. Phys. Lett.} {\bf A30} (2015), no.~15 1540011,
  [\href{http://arxiv.org/abs/1401.4176}{{\tt arXiv:1401.4176}}].

\bibitem{Peskin:1991sw}
M.~E. Peskin and T.~Takeuchi, {\it {Estimation of oblique electroweak
  corrections}},  {\em Phys. Rev.} {\bf D46} (1992) 381--409.

\bibitem{Flacke:2013pla}
T.~Flacke, K.~Kong, and S.~C. Park, {\it {Phenomenology of Universal Extra
  Dimensions with Bulk-Masses and Brane-Localized Terms}},  {\em JHEP} {\bf 05}
  (2013) 111, [\href{http://arxiv.org/abs/1303.0872}{{\tt arXiv:1303.0872}}].

\bibitem{Baak:2014ora}
{\bf Gfitter Group} Collaboration, M.~Baak, J.~Cúth, J.~Haller, A.~Hoecker,
  R.~Kogler, K.~Mönig, M.~Schott, and J.~Stelzer, {\it {The global electroweak
  fit at NNLO and prospects for the LHC and ILC}},  {\em Eur. Phys. J.} {\bf
  C74} (2014) 3046, [\href{http://arxiv.org/abs/1407.3792}{{\tt
  arXiv:1407.3792}}].

\bibitem{atlasExotic}
\url{https://atlas.web.cern.ch/Atlas/GROUPS/PHYSICS/CombinedSummaryPlots/EXOTICS/ATLAS_Exotics_Summary/ATLAS_Exotics_Summary.pdf}.

\bibitem{cmsExotic}
\url{https://twiki.cern.ch/twiki/pub/CMSPublic/PhysicsResultsCombined/exo-limits_DecJamboree2015.pdf}.

\bibitem{Edelhauser:2013lia}
L.~Edelhäuser, T.~Flacke, and M.~Krämer, {\it {Constraints on models with
  universal extra dimensions from dilepton searches at the LHC}},  {\em JHEP}
  {\bf 08} (2013) 091, [\href{http://arxiv.org/abs/1302.6076}{{\tt
  arXiv:1302.6076}}].

\bibitem{CMS-PAS-TOP-14-003}
{\bf CMS} Collaboration, {\it { Search for anomalous single top quark
  production in association with a photon}},  Tech. Rep. CMS-PAS-TOP-14-003,
  CERN, Geneva, 2014.

\bibitem{Aad:2014dya}
{\bf ATLAS} Collaboration, G.~Aad et~al., {\it {Search for top quark decays $t
  \to qH$ with $H \to \gamma\gamma$ using the ATLAS detector}},  {\em JHEP}
  {\bf 1406} (2014) 008, [\href{http://arxiv.org/abs/1403.6293}{{\tt
  arXiv:1403.6293}}].

\bibitem{CMS-PAS-HIG-13-034}
{\bf CMS} Collaboration, {\it {Combined multilepton and diphoton limit on $t
  \to cH$}},  Tech. Rep. CMS-PAS-HIG-13-034, CERN, Geneva, 2014.

\bibitem{Agashe:2013hma}
{\bf Top Quark Working Group} Collaboration, K.~Agashe et~al., {\it {Working
  Group Report: Top Quark}},  \href{http://arxiv.org/abs/1311.2028}{{\tt
  arXiv:1311.2028}}.

\end{thebibliography}\endgroup
\end{document}